%% file: main.tex
\documentclass[10pt,reqno,twoside]{article}
\usepackage{natbib} 
\usepackage{mathtools} 
\usepackage{booktabs} 
\usepackage{tikz} 
\usepackage[utf8]{inputenc}
\usepackage{authblk}
\usepackage{enumitem}  
\usepackage{amsthm}
\usepackage{amsmath}
\usepackage{bm}

\usepackage{setspace} 
\usepackage{comment}

\usepackage[colorinlistoftodos,prependcaption,textsize=tiny]{todonotes}
\allowdisplaybreaks

\usepackage{amssymb}
\usepackage{amsfonts}
\usepackage{comment}
\usepackage{colortbl}
\usepackage{graphicx}
\usepackage{placeins}
\usepackage{caption}
\usepackage{subcaption}

\newtheorem{theorem}{Theorem}
\newtheorem{lemma}[theorem]{Lemma}
\newtheorem{defn}[theorem]{Definition}
\newtheorem{corollary}[theorem]{Corollary}
\newtheorem{proposition}[theorem]{Proposition}
\newtheorem{assumption}[theorem]{Assumption}

\newcommand{\inc}{\mathrm{inc}}
\newcommand{\cc}[1]{\mathcal{#1}}
\newcommand{\degree}{\mathrm{degree}}
\newcommand{\imec}{\cc{I}\text{-}\mathrm{MEC}}
\newcommand{\simec}{\cc{I}^\star\text{-}\mathrm{MEC}}
\newcommand{\himec}{\hat{\cc{I}}\text{-}\mathrm{MEC}}
\newcommand{\optimec}{\hat{\cc{I}}_\mathrm{opt}\text{-}\mathrm{MEC}}

\DeclareMathOperator*{\argmin}{argmin} 

\newcommand{\dagg}{\mathcal{D}}
\newcommand{\methName}{\emph{UT-LVCE}}
\setlist[enumerate]{nosep}
\usepackage{varwidth}

\newcommand{\R}{\mathbb{R}}
\newcommand{\id}{\mathrm{Id}}

\usepackage{tikz}
\usetikzlibrary{arrows, decorations.pathreplacing, positioning, shapes,arrows.meta}

\newcommand\independent{\protect\mathpalette{\protect\independenT}{\perp}}
\def\independenT#1#2{\mathrel{\rlap{$#1#2$}\mkern2mu{#1#2}}}
\usepackage[titletoc]{appendix}
\usepackage{algorithm,algorithmic}
\numberwithin{equation}{section}
\usepackage{amssymb}
\usepackage{amsfonts}
\usepackage{verbatim}

\newtheorem{innercustomgeneric}{\customgenericname}
\providecommand{\customgenericname}{}
\newcommand{\newcustomtheorem}[2]{%
  \newenvironment{#1}[1]
  {%
   \renewcommand\customgenericname{#2}%
   \renewcommand\theinnercustomgeneric{##1}%
   \innercustomgeneric
  }
  {\endinnercustomgeneric}
}

\newcustomtheorem{customthm}{Theorem}
\newcustomtheorem{customlemma}{Lemma}
\newcustomtheorem{customassump}{Assumption}

\usepackage[margin=1.2in]{geometry}
\newcommand{\Proj}{\mathcal{P}}
\newcommand{\suppmat}{Appendix}

\title{Learning and scoring Gaussian latent variable causal models \\ with unknown additive interventions
}

\author[1]{Armeen Taeb}
\author[2]{Juan L. Gamella}
\author[2]{Christina Heinze-Deml}
\author[2]{Peter B{\"u}hlmann}
\affil[1]{%
   Department of Statistics, University of Washington
}
\affil[2]{%
   Seminar for Statistics, ETH Z\"urich
}

\date{}

\usepackage{titlesec}
\titlespacing*{\section}{0pc}{4pt}{4pt}
\titlespacing{\section}{0pc}{4pt}{4pt}
\titlespacing{\subsection}{0pc}{4pt}{4pt}
\titlespacing{\subsubsection}{0pc}{4pt}{4pt}
\expandafter\def\expandafter\normalsize\expandafter{%
    \normalsize
    \setlength\abovedisplayskip{7pt}
    \setlength\belowdisplayskip{7pt}
    \setlength\abovedisplayshortskip{6pt}
    \setlength\belowdisplayshortskip{6pt}
}
\definecolor{darkgreen}{rgb}{0.0, 0.13, 0.8}
\usepackage[
            CJKbookmarks=true,
            bookmarksnumbered=true,
            bookmarksopen=true,
            colorlinks=true,
            citecolor=blue,
            linkcolor=black,
            anchorcolor=red,
            urlcolor=blue,
            ]{hyperref}
\usepackage{cleveref}

\begin{document}
\maketitle
\begin{abstract}
With observational data alone, causal structure learning is a challenging problem. The task becomes easier when having access to data collected from perturbations of the underlying system, even when the nature of these is unknown. Existing methods either do not allow for the presence of latent variables or assume that these remain unperturbed. However, these assumptions are hard to justify if the nature of the perturbations is unknown. We provide results that enable scoring {causal structures} in the setting with additive, but unknown interventions.
Specifically, we propose a maximum-likelihood estimator in a structural equation model that exploits system-wide invariances to {output an equivalence class of causal structures} from perturbation data. Furthermore, under certain structural assumptions on the population model, we provide a simple graphical characterization of all the DAGs in the interventional equivalence class. We illustrate the utility of our framework on synthetic data as well as real data involving California reservoirs and protein expressions. {The software implementation is available as the Python package \emph{utlvce}}.
\end{abstract}
\input{sections/introduction}
\input{sections/model}
\input{sections/theoretical}
\input{sections/algorithm}

\input{sections/experiments}
\input{sections/conclusion}

\bibliography{main.bib}
\newpage
\appendix
\input{sections/supp_mat}

\end{document}

%% file: sections/introduction.tex
\section{Introduction}
\label{section:intro}

Identifying causal relations from observational data alone is challenging. In the context of (acyclic) structural causal models \citep{Robins,Pearl}, one possibility is to find the \emph{Markov equivalence class} (MEC) 
 of the underlying directed acyclic graph (DAG) under the faithfulness assumption \citep{Verma} or the beta-min condition \citep{Sara}. Some of the well-known algorithms for structure learning of MECs with observational data include the constraint-based PC algorithm \citep{PC}, the score-based Greedy Equivalence Search (GES) algorithm \citep{chick02}, and hybrid methods that integrate constraint-based and score-based methods such as ARGES \citep{Arges}. 
 
 In contrast to the purely observational setting, randomized controlled experiments lie at the opposite pole \citep{Rubin}: they are the gold standard for causal inference but randomizing the treatment is often hindered by cost, feasibility, or ethical concerns. However, under some assumptions, it is possible to exploit {unspecific} interventions or perturbations in the underlying system of interest which may not have been explicitly designed and controlled by a human experimenter. Such interventions arise in many application domains. For example, in genomics, with the advance of gene editing technologies, high throughput interventional gene expression data is being produced \citep{Kemmeren2014LargeScaleGP,gene,nicolai_pnas}. While there is typically a particular gene that is targeted by an intervention in a particular experiment, there may be additional off-target effects whose nature is unknown. In this paper, we assume that we have access to interventional data from different so-called ``environments'' where the location and strength of the respective interventions do not have to be known. 
 
 Interventional data can be viewed as \emph{perturbations} to components of the system and can offer substantial gain in identifiability: \cite{GIES} demonstrated that combining interventional with observational data reduces ambiguity and enhances identifiability to a smaller equivalence class than the MEC, known as the I-MEC {(Interventional MEC)}. A variety of methods have been proposed for causal structure learning from observational and interventional data. This includes the modified GES algorithms by \cite{GIES,GamellaGnies2021}, permutation-based causal structure learning for observational data \citep{IGSP} and for interventional data \citep{caroline_unknown}, penalized maximum-likelihood procedure in Gaussian models \citep{jointly}, {the Joint Causal Inference framework based on conditional independence testing \citep{JCI},
 and methods based on a causal invariance framework \citep{nicolai_pnas,ICP,backshift,Dantzig,anchor_regress,Ghassami,christina,Huang2020CausalDF} {building on a concept of stability} \citep{Didelez,Dawid}}. For a more comprehensive list, see {also} \cite{marloes_summary,structureLearningHeinze} and the references therein. 

One reason why randomized controlled experiments are considered to be the gold standard for causal inference is that the randomization breaks the influence potential hidden confounders have on both the treatment as well as the response variable of interest. In less controlled settings, the presence of latent variables, which may be difficult to measure or are simply unknown, poses a major challenge as the causal graphical model structure is not closed under marginalization. Therefore, the graphical structure corresponding to the marginal distribution of the observed variables consists of potentially many confounding dependencies that are induced due to the marginalization over the latent variables. 

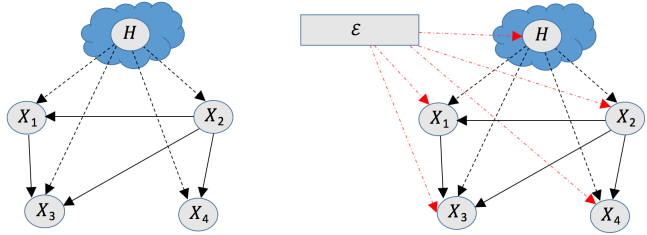
\begin{figure}
\centering
\begin{tabular}{cc}
\begin{tikzpicture}[every node/.style={scale=0.9},node distance={15mm},main/.style = {draw, circle, minimum size=1cm}]
  \tikzstyle{state}=[inner sep=1pt, minimum size=12pt]
    \tikzstyle{every edge}=[draw,>=stealth',->,line width = 0.25mm]	\node[main,inner sep=1.5pt,dotted](h_1) {$H$};
	\node[state](x_1)[below left = 0.7cm and 0.7cm of h_1]  {$X_1$};
	\node[state](x_2)[below right = 0.7cm and 0.7cm of h_1] {$X_2$};
 	\node[state](x_3)[below of=x_1] {$X_3$};
   	\node[state](x_4)[below of=x_2] {$X_4$};
	\draw (h_1) edge [->,dotted] (x_1);
 	\draw (h_1) edge [->,dotted] (x_2);
	\draw (h_1) edge [->,dotted] (x_3);
	\draw (h_1) edge [->,dotted] (x_4);
 	\draw (x_1) edge [->] (x_3);
 	\draw (x_2) edge [->] (x_3);
 	\draw (x_2) edge [->] (x_4);
   	\draw (x_2) edge [->] (x_1);
\end{tikzpicture}
&~~~~~~~~~~
\begin{tikzpicture}[every node/.style={scale=0.9},node distance={15mm},main/.style = {draw, circle, minimum size=1cm}]
  \tikzstyle{state}=[inner sep=1pt, minimum size=12pt]
      \tikzstyle{every edge}=[draw,>=stealth',->,line width = 0.25mm]	\node[main,inner sep=1.5pt,dotted](h_1) {$H$};
	\node[main,inner sep=1.5pt,dotted](h_1) {$H$};
 	\node[state](e)[left = 1.6cm of h_1] {${\color{red}\mathcal{E}}$};
	\node[state](x_1)[below left = 0.7cm and 0.7cm of h_1]  {$X_1$};
	\node[state](x_2)[below right = 0.7cm and 0.7cm of h_1] {$X_2$};
 	\node[state](x_3)[below of=x_1] {$X_3$};
   	\node[state](x_4)[below of=x_2] {$X_4$};

        \draw (e) edge [->,dotted,red] (h_1);
        \draw (e) edge [->,dotted,red] (x_1);
 	\draw (e) edge [->,dotted,red] (x_2);
	\draw (e) edge [->,dotted,red] (x_3);
	\draw (e) edge [->,dotted,red] (x_4);
 	
        \draw (h_1) edge [->,dotted] (x_1);
 	\draw (h_1) edge [->,dotted] (x_2);
	\draw (h_1) edge [->,dotted] (x_3);
	\draw (h_1) edge [->,dotted] (x_4);
 	\draw (x_1) edge [->] (x_3);
 	\draw (x_2) edge [->] (x_3);
 	\draw (x_2) edge [->] (x_4);
   	\draw (x_2) edge [->] (x_1);
\end{tikzpicture}
\end{tabular}
\caption{\small{Toy illustration of the setting considered in this paper where $X$ represent the observed variables and $H$ represent the latent variables where solid lines are connections among observed variables and dotted lines are connections between observed and latent variables; {\bf left}: without perturbations, {\bf right}: perturbations $\mathcal{E}$ on all components indicated with red dotted lines.}}
        \label{fig:toy}
\end{figure}

In this paper, we propose a modeling framework and estimator that allows for {unspecific} perturbations on some or all of the variables. Figure~\ref{fig:toy} demonstrates a toy example of our setup among $4$ observed variables $(X_1,X_2,X_3,X_4)$, latent variables $H$, and the environment variables $\mathcal{E}$ representing exogenous effects (to the graphical structure among observed and latent variables) that provide additive perturbations to the observed and latent variables.

Formally, we study a linear structural causal model (SCM) specifying the perturbation model and the relationship between $p$ observed variables $X \in \mathbb{R}^{p}$ and $h$ latent variables $H \in \mathbb{R}^h$. Here, the latent variables are assumed to act exogenously on the observed variables, but unless otherwise specified, no assumption is placed on the dependence structure among the latent variables. The SCM is parameterized by a connectivity matrix encoding the DAG structure among the observed variables, a coefficient matrix encoding the latent variable effects, and parameters involving the noise variances and unknown additive perturbation magnitudes and locations among all of the variables. {{Using data from this model, our objective is to estimate the DAG among the observed variables (e.g.\ the solid dotted lines in Figure~\ref{fig:toy}) or an equivalence class of DAGs when the underlying structure is not identifiable.}}



A key property of our modeling framework is that the connectivity matrix and the latent variable coefficient matrix remain \emph{invariant} across all the perturbation environments. With this insight, we propose a regularized maximum-likelihood procedure --  dubbed (U)nknown (T)arget (L)atent (V)ariable (C)ausal (E)stimator (\methName{}) -- to score any given DAG and estimate the associated parameters. Using the given DAG structure and the learned perturbation locations on the observed variables, we then provide a simple graphical approach to identify an equivalence class of DAGs that yield the same fit to the data. We show that to identify the population DAG among the observed variables, it is necessary to impose constraints on the latent effects. Otherwise, the problem is ill-posed. Under certain conditions on the population model, we demonstrate that applying this graphical procedure to the underlying DAG and the intervention locations of the observed variable fully characterizes, in the infinite data limit, the equivalence class of optimally scoring DAGs. Furthermore, under sufficiently many interventions, the optimally scoring DAG is uniquely the population DAG structure among the observed variables. Our characterization of the optimally scoring DAGs is valid under two types of structural assumptions on the latent effects: the first assumption is that the number of latent variables is small (compared to the observed variables) and they affect many observed variables; the second assumption substantially relaxes the first assumption and requires very mild conditions on the latent effects at the expense of approximately knowing the magnitude of the latent perturbations.

We envision several use cases for \methName{}. Firstly, in certain application domains, a DAG structure may be believed to approximate the underlying phenomenon (for example, protein expressions as in Section~\ref{sec:experiments}). \methName{} can be used to learn a latent variable causal model with respect to this DAG and return an equivalence class of DAGs that fit the data equally well. Secondly, along similar lines, a set of candidate DAGs, instead of only a singleton, may be available based on prior knowledge. In such settings, each DAG in this collection may be scored, and the best scoring ones as well as the DAGs in their respective equivalence class may be returned as output. Thirdly, the input candidate DAGs may be viewed as `starting points' that may contain spurious edges (obtained by domain expertise or by any structure learning algorithm) where the user aims to improve on these DAGs. Here, we propose to apply \methName{} on top of a greedy backward deletion approach to remove spurious dependencies due to latent confounding and identify an equivalence class of best scoring DAGs. 

\begin{figure}
\centering
\tikzstyle{every edge}=[draw,semithick]
\begin{tabular}{ccc}
\begin{tikzpicture}[every node/.style={scale=0.85},node distance={15mm},main/.style = {draw, circle, minimum size=1cm}]
  \tikzstyle{state}=[inner sep=1pt, minimum size=12pt]
    \tikzstyle{every edge}=[draw,>=stealth',->,line width = 0.25mm]
	\node[main,inner sep=1.5pt,dotted](h_1) {$H$};
 	\node[state](e)[left = 1.3cm of h_1] {$\mathcal{E}$};
	\node[state](x_1)[below left = 1.2cm and 1.4cm of h_1]  {$X_1$};
	\node[state](x_2)[right = 0.5cm of x_1] {$X_2$};
       \node[state](x_3)[right = 0.5cm of x_2] {$X_3$};
       \node[state](x_4)[right = 0.5cm of x_3] {$X_{4}$};
              \node[state](x_5)[right = 0.5cm of x_4] {$X_{5}$};
       \draw  (e) edge [->,dotted,red] (h_1);
       \draw  (e) edge [->,dotted,red] (x_1);
       \draw  (x_1) edge [->,line width = 0.2mm] (x_2);
       \draw  (x_2) edge [->] (x_3);
              \draw  (x_3) edge [->] (x_4);
                            \draw  (x_4) edge [->] (x_5);

              \draw  (h_1) edge [->,dotted] (x_1);
              \draw  (h_1) edge [->,dotted] (x_2);
              \draw  (h_1) edge [->,dotted] (x_3);
              \draw  (h_1) edge [->,dotted] (x_4);
            \draw  (h_1) edge [->,dotted] (x_5);

         \end{tikzpicture}
&~~~
\begin{tikzpicture}[every node/.style={scale=0.85},node distance={15mm},main/.style = {draw, circle, minimum size=1cm}]
  \tikzstyle{state}=[inner sep=1pt, minimum size=12pt]
  \tikzstyle{every edge}=[draw,>=stealth',->,line width = 0.25mm]
	\node[state](x_1) {$X_1$};
	\node[state](x_2)[right = 0.5cmof x_1] {$X_2$};
       \node[state](x_3)[right = 0.5cm of x_2] {$X_{3}$};
       \node[state](x_4)[right = 0.5cmof x_3] {$X_{4}$};
                     \node[state](x_5)[right = 0.5cm of x_4] {$X_{5}$};

         \draw (x_1) edge [->] (x_2);
         \draw (x_2) edge [->] (x_3);
         \draw (x_3) edge [->] (x_4);
                                     \draw  (x_4) edge [->] (x_5);
\end{tikzpicture}
&~~~
\begin{tikzpicture}[every node/.style={scale=0.85},node distance={15mm},main/.style = {draw, circle, minimum size=1cm}]
  \tikzstyle{state}=[inner sep=1pt, minimum size=12pt]
  \tikzstyle{every edge}=[draw,>=stealth',->,line width = 0.25mm]
	\node[state](x_1) {$X_1$};
	\node[state](x_2)[right = 0.5cmof x_1] {$X_2$};
       \node[state](x_3)[right = 0.5cmof x_2] {$X_{3}$};
       \node[state](x_4)[right = 0.5cmof x_3] {$X_{4}$};
                            \node[state](x_5)[right = 0.5cm of x_4] {$X_{5}$};

                   \draw    	(x_1) edge [->] (x_2);
          \draw    	(x_2) edge [o-o] (x_3);
          \draw    	(x_3) edge [o-o] (x_4);
                    \draw    	(x_4) edge [o-o] (x_5);

          \draw    	(x_1) edge [->,out=65,in=115,looseness=0.9] (x_3);
          \draw    	(x_1) edge [->,out=65,in=115,looseness=0.9] (x_4);
                    \draw    	(x_1) edge [->,out=65,in=115,looseness=0.9] (x_5);
           \draw    	(x_2) edge [o-o,out=-35,in=-145,looseness=0.9] (x_4);          
           \draw    	(x_2) edge [o-o,out=-65,in=-115,looseness=0.9] (x_5);        
                      \draw    	(x_3) edge [o-o,out=-35,in=-145,looseness=0.9] (x_5);

\end{tikzpicture}\\
a) &~~~ b) &~~~ c)
\end{tabular}
\caption{\small{An illustration comparing the output of our approach \methName{} and approaches that do not impose any constraints on the latent effects such as JCI \citep{JCI}. a) the setup consisting of five observed variables, one latent variable and interventions on the latent variable and observed variable $X_1$, b) the interventional equivalence class that \methName{} recovers (under some conditions) consisting uniquely of the subgraph among observed variables, c) the interventional equivalence class that \citep{JCI} recovers where many causal effects are not identified; here, directed edges denote ancestral relationships and circle marks represent uncertainty about edge marks.}}
        \label{fig:toy2}
\end{figure}

\subsection{Related work}
\label{section:related_work}

{A large body of causal structure learning methods with latent variables typically characterize 
a class of graphical independence models called maximal ancestral graphs (MAGs) \citep{PC,Richardson2002AncestralGM,jaber2020causal,differentiable_malinsky}. These methods, which allow for arbitrary hidden structure, tend to be overly conservative, recovering only a small subset of the causal effects. For example, suppose a latent variable influences many observed variables. Then, the underlying MAG tends to be dense where many edges cannot be directed. In this work, we take a middle-ground stance and place assumptions on the latent effects; these assumptions then enable us to direct edges and learn the sub-graph among observed variables (see the illustration in Figure~\ref{fig:toy2}). A similar perspective was taken in \cite{marloes} but without incorporating interventional data.}

In the {joint} observational and interventional setting with unspecified perturbations and latent confounders,
several methods exist in the literature for either learning the sub-graph among the observed variables or the causal parents (among the observed variables) of a target variable of interest. In particular, with unperturbed latent variables and only so-called shift interventions on the observed {covariates}, Causal Dantzig \citep{Dantzig} consistently estimates the causal effects on a response variable assuming that the interventions do not directly affect the response variable. {Such an assumption is relaxed in the} 
 backShift procedure \citep{backshift} {which still requires that the latent variables remain unperturbed for identifying the causal structure.} Both Causal Dantzig and backShift yield a single causal structure, even if the underlying model is not fully identifiable. On the other hand, in addition to allowing for perturbations on all the variables, \methName{} produces an equivalence class of DAGs. For a summary of the {assumptions for} \methName{} as compared to competing methods (including Instrumental Variable Regression (IV, \citealt{IV}), see Table~\ref{table:fd_illustration}. We will also provide more comparisons throughout the paper.

\begin{table}
\scalebox{1}{
    \begin{tabular}{ | c | c | c |c|}
    \hline
    Method &  Perturbed response & Unperturbed latent & Perturbed latent  \\\hline
    IV, ICP, Causal Dantzig & {\centering{{\color{red}x}}}& {{\color{blue}\checkmark} single DAG}& {{\color{red}x}}    \\ \hline
    backShift & {{\color{blue}\checkmark} single DAG}& {{\color{blue}\checkmark} single DAG}& {{\color{red}x}} \\ \hline
    \methName{} & {{\color{blue}\checkmark}} $\imec$&{{\color{blue}\checkmark}} $\imec$&\shortstack{{}\\{{\color{blue}\checkmark}} $\imec$} \\ \hline
\hline
\end{tabular}}
\caption{\small{Comparison of \methName{} with competing methods in the following settings: response variable is perturbed, latent variables are unperturbed, and the latent variables are perturbed. {The methods are Instrumental Variables IV \citep{IV}, Invariant Causal Predictions \citep{ICP}, Causal Dantzig \citep{Dantzig}, backShift \citep{backshift} and our proposal \methName{}. Here, we denote an interventional equivalence class of DAGs by $\imec$.}}}
\label{table:fd_illustration}
\end{table}
\subsection{Notation}
\label{sec:notation}
We denote the identity matrix by $\id$, with the size being clear from context. The collection of $d \times d$ symmetric matrices are denoted by $\mathbb{S}^d$ and positive-semidefinite matrices by $\mathbb{S}^d_+$ and the collection of strictly positive-definite matrices by $\mathbb{S}^d_{++}$. The collection of positive-definite diagonal matrices is denoted by $\mathbb{D}_{++}^d$. For a positive integer $a$, we denote the set $\{1,2,\dots,a\}$ by $[a]$. We denote the index set of the parents of a random variable $X_p$ by $\text{PA}(p)$. We denote $\mathrm{MEC}(\dagg)$ to be the Markov equivalence class of $\dagg$, namely DAGs that have the same skeleton and v-structures as $\dagg$. For a DAG $\dagg$ among $p$ variables, and a matrix $B \in \mathbb{R}^{p \times p}$, we use the notation $B \sim \dagg$ to denote that $B_{ij} \neq 0$ implies $i \to j$ in the DAG $\dagg$. Finally, for a set of diagonal and positive-definite matrices $\{\Omega_e\}_{e=1}^m \subseteq \mathbb{D}_{++}^p$ with positive integer $m \geq 2$, we let $\mathbb{I}(\{\Omega_e\}_{e=1}^m) := \{j: \exists~{e,f} \text{ such that }[\Omega_e]_{j,j} \neq [\Omega_f]_{j,j}\}$.



%% file: sections/model.tex
\section{Modeling framework and maximum-likelihood estimator}
In this section, we describe the data generation process associated with the perturbation model sketched in Figure~\ref{fig:toy}. 
Furthermore, we propose \methName, a regularized maximum-likelihood estimator. Given an input DAG, \methName{} identifies estimates of the unknown perturbation effects, the latent effects, and the causal relations among the observed variables. Finally, we describe a computationally efficient graphical procedure that uses the estimate obtained from \methName{} to find a set of equally scoring DAGs. 

\subsection{{Modeling framework}}
\label{section:modeling_framework}
We consider a directed acyclic graph whose $p+h$ nodes correspond to random variables $(X,H) \subseteq \R^p \times \R^h$, where $X$ are observable and $H$ are latent variables. We denote the induced subgraph DAG corresponding to the observed variables by $\dagg^\star$. We aim to learn $\dagg^\star$ or an equivalence class of DAGs when there are not enough interventions on the observed variables for full identifiability. Our methodology is also applicable in a setting where one is primarily interested in the causal effects on a particular response variable of interest. As such, we distinguish $X_p$ as the target {or response} variable. 

We assume that the observed and latent variables satisfy the following linear SCM:
\begin{equation}
    X = B^\star{X}+\Gamma^\star{H} + \epsilon. 
    \label{eq:SCM_comp}
\end{equation}
Here, the connectivity matrix $B^\star \in \mathbb{R}^{p \times p}$ contains zeros on the diagonal and is compatible with $\dagg^\star$: $B^\star_{ij} \neq 0$ if $X_j$ is a parent of $X_i$ in $\dagg^\star$. Thus, the $p$-th row vector $B^\star_{p,:}$ encodes the {(observable)} causal parents of the response variable and the magnitude of their effects. The matrix $\Gamma^\star$  in \eqref{eq:SCM_comp} encodes the effects of the latent variables on the observed variables where $\Gamma^\star_{k,j} \neq 0$ if the latent variable $H_j$ is a parent of the node $X_k$. Further, $\epsilon$ is a random vector with independent components. {We assume that the latent variables $H$ are exogenous to $X$, so that $\epsilon$ is independent of $H$. Unless otherwise specified, no assumptions are imposed on the causal structure among the latent variables $H$}.

The compact SCM \eqref{eq:SCM_comp} describes the generating process of $X$ in the observational setting when there are no external perturbations on the system. We next {describe} how the data generation process alters due to some type of perturbations to the variables $(X,H)$. We consider perturbations that directly shift the distributions of the random variables by some noise acting additively to the system. Specifically, the perturbations $\mathcal{E}$ generate the random pair $(X^e,H^e)$ for each environment $e \in \mathcal{E}$ satisfying the following SCM: 
\begin{equation}
\begin{gathered}
    X^e = B^\star{X^e} + \Gamma^\star{H}^e+\epsilon^e+\delta^e, 
\end{gathered}
\label{eq:SCM_env}
\end{equation}
where for every $e \in \mathcal{E}$, $\epsilon^e \stackrel{\text{dist}}{=} \epsilon$, $(H^e,\delta^e,\epsilon^e)$ are jointly independent, and the collection $(X^e,H^e,\delta^e,\epsilon^e)$ is independent across $e$. Further, $\delta^e \in \mathbb{R}^{p}$ is a vector that represents the additive perturbations {on the observed variables}. Some of the entries of $\delta^e$ could be identically zero indicating that no interventions occurred; the remaining entries are generated from a random distribution. The intervention targets are denoted by $\mathcal{I}^\star:=\{j \in [p]: \text{var}(\delta_j^e) \neq 0 \text{ for some }e\in\mathcal{E}\}$\footnote{Here, we consider interventions that vary the variance of the noise terms; see Section 2 of \cite{GamellaGnies2021} for why interventions on the means do not offer any identifiability in linear Gaussian SCMs.}. Importantly, the location of the nonzero components (i.e.\ variables that are intervened) is unknown. Finally, $H^e \in \mathbb{R}^{h}$ is a random vector that represents the perturbed latent variables across the environments. {That is, the perturbations on the latent variables are absorbed into $H^e$}. Without loss of generality, we assume that {all} variables are centered.

Given data of observed variables $X^e$ across environments $e \in \mathcal{E}$, our objective is to develop a procedure to estimate the unknown perturbation effects, the latent effects, and the causal relations among observed variables. To arrive at an estimator, we model the distribution of the random vectors $H^e,\epsilon^e$ and the nonzero components of $\delta^e$ to be Gaussian. Specifically, we model the random vectors $H^e$ as well as the sum $\epsilon^e+\delta^e$ as follows:
\begin{equation*}
\begin{aligned}
&H^e \sim \mathcal{N}(0,\Psi_e^\star),\ \Psi_e^\star \in \mathbb{S}_{++}^h, \\
&\epsilon^e + \delta^e \sim {\cal N}(0,\Omega^\star_e),\ \Omega^{\star}_e \in \mathbb{D}^{p}_{++},\ \mathbb{I}(\{\Omega_e^\star\}_{e=1}^m) = \cc{I}^\star. 
\end{aligned}
\end{equation*}
{The notations of $\mathbb{D}_{++}^{\cdot}$, $\mathbb{S}_{++}^{\cdot}$, $\mathbb{I}(\cdot)$ are defined in Section~\ref{sec:notation}}. We remark that non-Gaussian linear structural equation models are generally more identifiable than their Gaussian counterparts \citep{Shimizu2006ALN}. While we develop our procedure based on a Gaussian model, we will see that the output of our approach is conservative in the sense that the true set of equivalent DAGs is contained in the estimated set {in a non-Gaussian setting}. 

The compactified SCM \eqref{eq:SCM_env} characterizes the distribution among all of the observed variables and encodes system-wide invariances. Specifically, \eqref{eq:SCM_env} insists that for every $k = 1,2,\dots,p$, the regression coefficients when regressing $X^e_k$ on the parent sets $\{X_j^e: X_j \text{ parent }\allowbreak \text {of }X_k \}$ and  $\{H_l^e: H_l \text{ parent of }X_k\}$ remain invariant for all environments $e \in \mathcal{E}$. This is a point of departure from instrumental variable techniques \citep{IV} or {Invariant}
Causal Prediction \citep{ICP} in two significant ways: 1) such methods do not allow for perturbations on the latent variables or the response variable $X_p$ (i.e. they assume $H^e\stackrel{\text{dist}}{=}H$ and $\delta^e_p \equiv 0$ for all $e \in \mathcal{E}$) and 2) they only consider ``local" invariances arising from the distribution $X^e_p~|~\{(X_j^e,H_l^e) \text{ parents of } X_p\}$. The virtue of considering a joint model over all of the variables and exploiting system-wide invariances is that we can propose a maximum-likelihood estimator \methName{} {which} identifies the population DAG structure even under perturbations on the response variable and the latent variables. 

The SCM \eqref{eq:SCM_env} is similar in spirit to previous modeling frameworks in the literature. The authors \cite{jointly} consider jointly observational and interventional Gaussian data where the interventions are limited to do-interventions and there are no latent variables. In the context of \eqref{eq:SCM_env}, this means that $\delta^{e} \equiv 0$ and $\Gamma^\star \equiv 0$. As such, the framework considered in this paper is a substantial generalization of \cite{jointly}. Further, the backShift \citep{backshift} procedure considers the linear SCM \eqref{eq:SCM_env} with some modifications: i) there are no perturbations to the latent variables, i.e. $H^e \stackrel{\text{dist}}{=} H$ for all $e \in \mathcal{E}$, and ii) $B^\star$ may be a cyclic directed graph. {In addition, the backShift algorithm relies on exploiting invariances of differences of estimated covariance matrices across environments. Our \methName{} procedure is more in the "culture of likelihood modeling and inference" and has the advantage that it can cope well with having only a few observations per environment. This likelihood perspective also fits much more into the context of inference for mixed models as briefly discussed next.

The framework in \eqref{eq:SCM_env} bears some similarities to {standard} random effects mixed models \citep{mixed_models}. In particular, random effects mixed models are widely employed to model {grouped} data,  where some {parameter} components remain fixed and others are {random}.  In the context of our problem, the fixed parameters are the matrices $B^\star,\Gamma^\star$ and the random parameters are {the shift perturbations} $\delta^e$. However, a difference between our model in \eqref{eq:SCM_comp} and standard mixed models is that the effects of the random parameter $\delta^e$ propagate through the structural equations; and in practice, the order of propagation is usually unknown. 

\subsection{Scoring DAGs via \methName}
\label{section:directLikelihood}
In this section, we propose our method \methName, which scores a DAG $\dagg$ via regularized maximum likelihood estimation. As we will discuss, the scores of {a} candidate set of DAGs can then be obtained using this procedure to find the best scoring DAG(s). We suppose that there are $m$ environments $|\mathcal{E}| = m$, and for every environment $e = 1,2,\dots,m$, {we have samples of $X^e$:} $\{X^{e}_{i}\}_{i = 1}^{n_e}$ for some positive integer $n_e$ {which are independent and identically distributed (IID) for each $e$ and independent across $e$}. To obtain a score for a DAG $\dagg$, \methName{} identifies a causal model that best fits the data. This model is parameterized by $(B,\Gamma,\cc{I},\Omega_e,\Psi_e)$ for all $e = 1,2,\dots,m$, where $B$ is a connectivity matrix, $\Gamma$ encodes the latent effects, $\cc{I}$ is a subset representing the intervention locations, $\Omega_e$ represents the noise variances of the observed variables, and $\Psi_e$ encodes the perturbations on the latent variables (see \eqref{eq:SCM_env}). The quantities $(B,\Gamma,\cc{I},\Omega_e,\Psi_e)$ are unknown and {estimated} by solving the following regularized maximum-likelihood estimator for the DAG structure $\dagg$ with $\bar{h}$ latent variables:
\begin{equation}
    \begin{aligned}
 \argmin_{\substack{B \in \mathbb{R}^{p \times p},\Gamma \in \mathbb{R}^{p \times \bar{h}},\cc{I} \subseteq \{1,2,\dots,p\}\\\{\Omega_e ,\Psi_e\}_{e=1}^m \subseteq \mathbb{D}^{p}_{++} \times \mathbb{S}^{\bar{h}}_{++}}} \, &~~~\sum_{e = 1}^{m}\hat{\pi}_e\ell(B,\Gamma,\Omega_e,\Psi_e;\hat{\Sigma}_{e}) + \lambda\mathcal{R}_{\gamma}(\mathcal{D},\cc{I}).\\
\text{ subject-to:} \, &~~~B\sim\dagg~~;~~ \mathbb{I}(\{\Omega_e\}_{e=1}^m) \subseteq \cc{I}
    \end{aligned}
    \label{eqn:est2_marg}
\end{equation}
Here, $\ell(\cdot)$ is the negative Gaussian log-likelihood
\begin{gather*}
    \ell(\cdot) := \log\det\left(\Omega_e+\Gamma\Psi_e\Gamma^T\right)
    +\mathrm{trace}\left(\left[\Omega_e+\Gamma\Psi_e\Gamma^T\right]^{-1}(\id-B)\hat{\Sigma}_{e}(\id-B)^T\right),
    \end{gather*}
where the matrix $\hat{\Sigma}_{e}$ is the sample covariance of the data $\{X^{e}_{i}\}_{i = 1}^{n_e}$. The quantity $\hat{\pi}_e = {n_e}/{\sum_{e = 1}^{m}n_e}$ represents the estimated mixture components. The constraint $B \sim \dagg$ ensures that the connectivity matrix $B$ satisfies the sparsity pattern of the graph $\dagg$. Further, the constraint on the matrices $\{\Omega_e\}_{e = 1}^m$ ensures that the variances corresponding to unperturbed coordinates are the same across all environments. {The notations of $\mathbb{D}_{++}^{\cdot}$, $\mathbb{S}_{++}^{\cdot}$, $\cdot\sim \cdot$, $\mathbb{I}(\cdot)$ are defined in Section~\ref{sec:notation}}. Finally, $\lambda\mathcal{R}_\gamma(\cdot,\cdot)$ represents a regularization term with $\lambda,\gamma \geq 0$ and $\mathcal{R}_\gamma(\cdot,\cdot)$ given by:
\begin{equation*}
    \begin{aligned}
    \mathcal{R}_{\gamma}(\dagg,\cc{I}) :=\left(\|\dagg\|_{\ell_0} +  p~\degree[\text{moral}(\dagg)]\right)+\gamma|\cc{I}|.
    \end{aligned}
\end{equation*}
Here, $\|\dagg\|_{\ell_0}$ denotes the number of edges in $\dagg$. Further, $\text{moral}(\dagg)$ denotes the moralization of $\dagg$ which forms an undirected graph of $\dagg$ by adding edges between nodes that have common children, and $\text{degree}[\cdot]$ computes the maximal degree of the undirected graph. The sum $\|\dagg\|_{\ell_0} +  p~\degree[\text{moral}(\dagg)]$\footnote{{One can also add an extra tuning parameter, e.g. $\|\dagg\|_{\ell_0} +  \kappa~\degree[\text{moral}(\dagg)]$ for some $\kappa \geq 0$; for simplicity, we use a fixed value $\kappa = p$.}} regularizes the complexity of $\dagg$; although this term is a constant in the \methName{} estimator \eqref{eqn:est2_marg}, it will play a crucial role for comparing different DAGs that are scored via the \methName{} estimator \eqref{eqn:est2_marg}. The quantity $|\cc{I}|$ penalizes the number of interventions on the observed variables. Furthermore, the regularization parameter $\lambda$ provides overall control of the trade-off between the fidelity of the model to the data and the complexity of the model. Additionally, the regularization parameter $\gamma$ provides a trade-off between the complexity of the DAG and the number of intervention targets. Overall, $\mathcal{R}_\gamma(\dagg,\cc{I})$ is akin to the {Akaike Information Criterion (AIC) or} Bayesian Information Criterion (BIC) score as it prevents overfitting by incorporating the denseness of the DAG $\dagg$ as well as the number of interventions in the likelihood score.

We note that regularization terms controlling for the complexity of estimated DAGs are commonly employed in causal structural learning (see \citealt{marloes_summary} and the references therein). Previous work on penalized likelihood scores only contain the regularization term $\|\dagg\|_{\ell_0}$. Thus, our regularization penalty $\mathcal{R}_\gamma(\dagg,\cc{I})$ contains the novel terms $\degree[\text{moral}(\dagg)]$ and $|\cc{I}|$. The quantity $\degree[\text{moral}(\dagg)]$ is an important addition in our context to ensure identifiability of the underlying DAG among observed variables in the presence of latent variables as described in Section~\ref{sec:dense_sparse}. The quantity 
$|\cc{I}|$ is motivated by the following observation: if the size of the intervention set is not penalized, for any finite sample size, \eqref{eqn:est2_marg} returns $\hat{\cc{I}} =[p]$. Intuitively, a model that contains interventions on all of the variables is in its own equivalence class (we formalize this in Section~\ref{sec:theo_ident}). Thus, without the penalty on the size of the intervention target, the optimum of \eqref{eqn:est2_marg} may be unique even if there are multiple DAGs in the equivalence class of the population model.

In summary, the estimator \eqref{eqn:est2_marg} takes as input the DAG $\mathcal{D}$, the {tuning} parameters $(\lambda,\gamma,\bar{h})$ and the {observed empirical} covariance matrices $\hat{\Sigma}_e$ to obtain a causal model with the following parameters:
\begin{equation*}
\begin{aligned}
    \hat{\Theta}(\dagg,\bar{h}):= \text{ any minimizer } (\hat{B},\hat{\Gamma},\hat{\cc{I}},\{\hat{\Omega}_e,\hat{\Psi}_e\}_{e=1}^m) \text{ of }\eqref{eqn:est2_marg}.
\end{aligned}
\end{equation*}
Then the score for the DAG $\dagg$ given parameters $\hat{\Theta}(\dagg,\bar{h})$ is computed as:
\begin{equation}
    \begin{aligned}
        \texttt{score}_{\lambda,\gamma}(\dagg,\hat{\Theta}(\dagg,\bar{h})) := \sum_{e = 1}^{m}\hat{\pi}_e\ell(\hat{B},\hat{\Gamma},\hat{\Omega}_e,\hat{\Psi}_e;\hat{\Sigma}_{e}) + \lambda\mathcal{R}_\gamma(\mathcal{D},\hat{\cc{I}}).
    \end{aligned}
    \label{eqn:score}
\end{equation}
We select the parameters $(\lambda,\gamma,\bar{h})$ via cross-validation; see Section~\ref{sec:optimization} for more discussion. {Note that while the minimizer of \eqref{eqn:est2_marg} is not unique, the associated score \eqref{eqn:score} is the same for all minimizers of \eqref{eqn:est2_marg}.}

 In comparison to the \methName{} procedure, backShift \citep{backshift} fits the SCM \eqref{eq:SCM_env} (with some restrictions outlined in Section~\ref{section:modeling_framework}) by performing joint diagonalization to the difference of sample covariance matrices. \methName{} allows for much more modeling flexibility. First, in contrast to backShift where the latent effects are subtracted by computing the difference of covariances, \methName{} explicitly models these effects. This feature of \methName{} enables the possibility of perturbations to the latent variables and a manner to control the number of estimated latent variables (as opposed to an arbitrary number of latent variables with backShift). We discuss in Section~\ref{sec:theo_ident} that controlling the number of latent variables may lead to identifiability using \methName{} with two environments, whereas backShift is guaranteed to fail. {Furthermore, \methName{} allows to pool information over different environments $e$ for the parameter $B$ of interest: this enables \methName{} to be used with only a few sample points per environment.} Finally, \methName{} explicitly models the intervention structure among the observed variables (via the set $\cc{I}$). We will see in the next section that encoding the perturbation structure allows for outputting a set of equally scoring DAGs. The procedure backShift on the other hand returns a single DAG, even if the underlying model is not identifiable. 

\subsection{Score equivalent DAGs}
The estimator \eqref{eqn:est2_marg} can be used in conjunction with \eqref{eqn:score} to score a collection of DAGs and find the ones that best fit the data. Such an approach raises the following question: are there multiple DAGs that fit the data equally well? In this section, we answer in the affirmative and provide a {procedure} to obtain {a set of score equivalent DAGs.}
The set of equally scoring DAGs is closely related to an \emph{interventional equivalence class}, which was first introduced {for this model without latent variables} in \cite{GamellaGnies2021}, and we present {it} below. 
\begin{defn}[$\imec$, \cite{GamellaGnies2021}] Let $\mathcal{D}$ be a DAG and $\cc{I} \subseteq [p]$ denotes an intervention set. Furthermore, let $\mathrm{MEC}(\dagg)$ be the standard {observational} Markov equivalence class of $\dagg$. Then, we define the following interventional equivalence class:
\begin{equation}
   \imec(\dagg) := \left\{\tilde{\dagg} \in \mathrm{MEC}(\dagg)~|~ \mathrm{PA}_{\tilde{\dagg}}(i) = \mathrm{PA}_\dagg(i) \text{ for all }i \in \cc{I}\right\}.
\end{equation}
\label{defn:equiv_class}
\end{defn}
\vspace{-0.2in}
The interventional equivalence class $\imec(\dagg)$ is closely related to the notion of \emph{transition pair equivalence}, introduced by \citet{tian2001causal} (see \citealt{GamellaGnies2021} for more discussion). This class is a subset of the standard {observational} Markov equivalence class and consists of DAGs that have the same parents as $\dagg$ for variables in the intervention target set $\cc{I}$. Thus, the intervention target set $\cc{I}$ controls the cardinality of $\imec(\dagg)$, i.e. for any $\cc{I}_1 \subseteq \cc{I}_2$:
${\{\dagg \}}\subseteq \cc{I}_2\text{-}\mathrm{MEC}(\dagg)\subseteq \cc{I}_1\text{-}\mathrm{MEC}(\dagg) \subseteq \text{MEC}(\dagg)$. As noted in \cite{GamellaGnies2021}, the set $\imec(\dagg)$ can be computed efficiently given $(\dagg,\cc{I})$ using Meek's rules with background knowledge \citep{Meek1995CausalIA}. The following theorem statement
formally relates the interventional equivalence class $\imec(\dagg)$ to the set of equally scoring DAGs.
\begin{theorem}[score equivalent DAGs] Consider an SCM \eqref{eq:SCM_env} with structure given by a DAG $\dagg$ and parameter set $\Theta(\dagg,\bar{h}) := (B,\Gamma,\cc{I},\{\Omega_e,\Psi_e\}_{e=1}^m)$. Then, for any DAG $\tilde{D} \in \imec({\dagg})$, there exists a parameter set $\tilde{\Theta}(\tilde{\dagg},\bar{h}) := (\tilde{B},\tilde{\Gamma},\tilde{\cc{I}},\{\tilde{\Omega}_e,\tilde{\Psi}_e\}_{e=1}^m)$ with $\tilde{B}\sim\tilde{D}$ such that $\texttt{score}_{\lambda,\gamma}(\dagg,\Theta(\dagg,\bar{h})) = \texttt{score}_{\lambda,\gamma}(\tilde{\dagg},\tilde{\Theta}(\tilde{\dagg},\bar{h}))$ for all $\lambda,\gamma \geq 0$. 
\label{thm:IMEC}
\end{theorem}
The proof of Theorem~\ref{thm:IMEC} is shown in \suppmat{} Section~\ref{sec:supp_thm_equiv} and extends the analysis provided in \cite{GamellaGnies2021} to the setting with latent variables. The results of Theorem~\ref{thm:IMEC} enable the characterization of the set of best scoring DAGs. Specifically, let $\hat{\dagg}_\text{opt}$ be an optimal scoring DAG with a corresponding intervention set $\hat{\cc{I}}_\mathrm{opt}$, i.e. $(\hat{\dagg}_\mathrm{opt},\bar{h}_\text{opt}) \in \argmin_{\dagg,\bar{h}} \texttt{score}_{\lambda,\gamma}(\dagg,\hat{\Theta}(\dagg,\bar{h}))$ and $\hat{\cc{I}}_\text{opt}$ is an intervention set encoded in $\hat{\Theta}(\dagg_\text{opt},\bar{h}_\text{opt})$. Then, all the DAGs inside $\optimec(\hat{\dagg}_\mathrm{opt})$ are also optimal.



%% file: sections/theoretical.tex
\section{{Identifiability guarantees with \methName}}
\label{sec:theo_ident}
\begin{figure}[t]
\centering
\begin{tabular}{cc}
\begin{tikzpicture}[every node/.style={scale=0.8},node distance={15mm},main/.style = {draw, circle, minimum size=1cm}]
  \tikzstyle{state}=[inner sep=1pt, minimum size=12pt]
    \tikzstyle{every edge}=[draw,>=stealth',->,line width = 0.25mm]	\node[main,inner sep=1.5pt,dotted](h_1) {$H$};
     	\node[state](e)[left = 2.6cm of h_1] {${\color{red}\mathcal{E}}$};
	\node[state](x_1)[below left = 0.7cm and 1.2cm of h_1]  {$X_1$};
	\node[state](x_2)[below = 0.8cm of x_1] {$X_2$};
 	\node[state](x_3)[below left= 2.2cm and 0.1cm of h_1] {$X_3$};
 	\node[state](x_4)[below right= 2.2cm and 0.1cm of h_1] {$X_4$};
  	\node[state](x_p)[below right = 0.7cm and 1.2cm of h_1]  {$X_p$};
   	\node[state](x_5)[below = 0.8cm of x_p] {$X_5$};
        \draw node[fill,circle,minimum size=0.1cm,inner sep=0pt](dot1)[above=0.25cm of x_5] {};
        \draw node[fill,circle,minimum size=0.1cm,inner sep=0pt](dot2)[above=0.1cm of dot1] {};
	\draw (h_1) edge [->,dotted] (x_1);
 	\draw (h_1) edge [->,dotted] (x_2);
	\draw (h_1) edge [->,dotted] (x_3);
	\draw (h_1) edge [->,dotted] (x_4);
 	\draw (h_1) edge [->,dotted] (x_5);
	\draw (h_1) edge [->,dotted] (x_p);
 	\draw (x_4) edge [->] (x_1);
 	\draw (x_3) edge [->] (x_2);
 	\draw (x_3) edge [->] (x_p);
   	\draw (x_2) edge [->] (x_5);
   	\draw (x_4) edge [->] (x_5);
        \draw (e) edge [->,dotted,red] (h_1);
        \draw (e) edge [->,dotted,red] (x_4);
        \draw (e) edge [->,dotted,red] (x_5);
                \draw (e) edge [->,dotted,red] (x_1);
        \draw (e) edge [->,dotted,red] (x_2);
        \draw (e) edge [->,dotted,red] (x_3);
                \draw (e) edge [->,dotted,red] (x_p);
\end{tikzpicture}
&~~~~~~~~~~
\begin{tikzpicture}[every node/.style={scale=0.8},node distance={15mm},main/.style = {draw, circle, minimum size=1cm}]
  \tikzstyle{state}=[inner sep=1pt, minimum size=12pt]
    \tikzstyle{every edge}=[draw,>=stealth',->,line width = 0.25mm]	\node[main,inner sep=1.5pt,dotted](h_1) {$H$};
     	\node[state](e)[left = 2.6cm of h_1] {${\color{red}\mathcal{E}}$};
	\node[state](x_1)[below left = 0.7cm and 1.2cm of h_1]  {$X_1$};
	\node[state](x_2)[below = 0.8cm of x_1] {$X_2$};
 	\node[state](x_3)[below left= 2.2cm and 0.1cm of h_1] {$X_3$};
 	\node[state](x_4)[below right= 2.2cm and 0.1cm of h_1] {$X_4$};
  	\node[state](x_p)[below right = 0.7cm and 1.2cm of h_1]  {$X_p$};
   	\node[state](x_5)[below = 0.8cm of x_p] {$X_5$};
        \draw node[fill,circle,minimum size=0.1cm,inner sep=0pt](dot1)[above=0.25cm of x_5] {};
        \draw node[fill,circle,minimum size=0.1cm,inner sep=0pt](dot2)[above=0.1cm of dot1] {};
	\draw (h_1) edge [->,dotted] (x_3);
	\draw (h_1) edge [->,dotted] (x_4);
 	\draw (x_4) edge [->] (x_1);
 	\draw (x_3) edge [->] (x_2);
 	\draw (x_3) edge [->] (x_p);
   	\draw (x_2) edge [->] (x_5);
   	\draw (x_4) edge [->] (x_5);
        \draw (e) edge [->,dashed,blue] (h_1);
        \draw (e) edge [->,dotted,red] (x_4);
        \draw (e) edge [->,dotted,red] (x_5);
            \draw (e) edge [->,dotted,red] (x_1);
        \draw (e) edge [->,dotted,red] (x_2);
            \draw (e) edge [->,dotted,red] (x_3);
        \draw (e) edge [->,dotted,red] (x_p);
\end{tikzpicture}\\
\small{(a) dense latent effects with $\text{dim}(H) \ll p$} &~~~~~~~~~~ \small{(b) approximately known latent perturbations}
\end{tabular}
\caption{\small{Structural assumptions needed for equivalence class characterization: a) dense latent effects with a small number of latent variables as compared to the ambient dimension (Section~\ref{sec:dense_sparse}) and b) known perturbations on the latent variables (indicated by blue dashed line) and some confounding dependencies induced by the latent variables (Section~\ref{sec:known_latent_perturb}).}}
\label{fig:sec3_assumptions}
\end{figure}
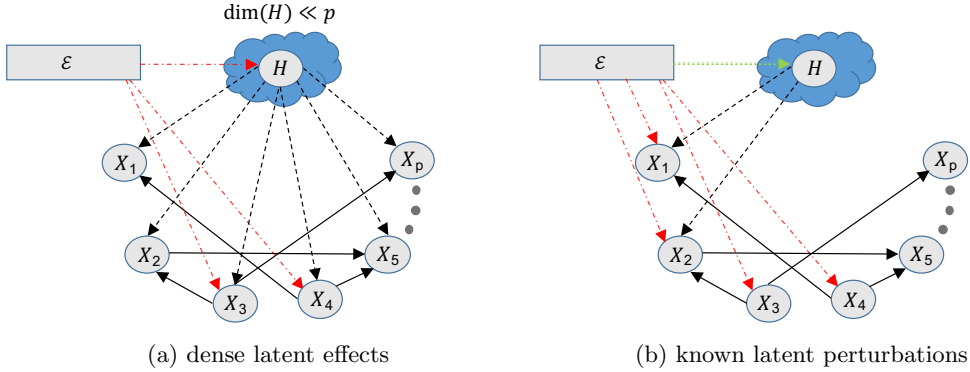
In this section, we analyze the identifiability guarantees of \methName{} in the infinite data limit. Specifically, we analyze the DAGs that minimize the score \eqref{eqn:score}, i.e. are solutions to $\argmin_{\dagg,\bar{h}} \texttt{score}_{\lambda,\gamma}\allowbreak(\dagg,\hat{\Theta}(\dagg,\bar{h}))$ when $\lambda \to 0$ and the sample size $n_e \to \infty$ for every $e\in[m]$.

In Section~\ref{sec:impossible}, we show that without imposing any additional structure on the problem, an optimal scoring DAG may be very different than the population DAG $\dagg^\star$; we will describe how this result is related to the violation of the faithfulness assumption (over the graph of observed and latent variables) that is typically made in the literature of latent variable causal discovery. Hence, since identifying optimal DAGs is meaningless without any conditions, in Sections~\ref{sec:dense_sparse} and \ref{sec:known_latent_perturb}, we impose structural assumptions on the latent effects and constrain the causal parameters appropriately; these conditions ensure that an estimated DAG is inside the interventional equivalence class of the population DAG in the infinite data limit, i.e. for any optimally scoring DAG $\hat{\dagg}_\mathrm{opt}$, we have $\hat{\dagg}_\mathrm{opt} \in \simec(\dagg^\star)$ with probability tending to one as the sample size in every environment tends to infinity. In Section~\ref{sec:dense_sparse}, we assume that the number of latent variables is small and they affect many observed variables (visualized in Figure~\ref{fig:sec3_assumptions}a). In Section~\ref{sec:known_latent_perturb}, we assume that the latent perturbations are approximately known and the latent variables induce some confounding dependencies (visualized in Figure~\ref{fig:sec3_assumptions}b). We extend our theoretical guarantees to the non-Gaussian setting in Section~\ref{sec:non-Gaussian}. Throughout, we describe quantitative measures that our software outputs to indicate when deviations from assumptions may have occurred.

We assume that the data is generated according to the perturbation model in \eqref{eq:SCM_env} with population parameters $B^\star,\Gamma^\star,\cc{I}^\star,\{(\Omega_e^\star,\Psi_e^\star)\}_{e=1}^m$ (see Section~\ref{section:modeling_framework}). We let \begin{equation}\hat{\dagg}_\mathrm{all.opt},\hat{B}_\mathrm{all.opt},\hat{\cc{I}}_\mathrm{all.opt}
\label{eqn:estimates_theory}
\end{equation}
be the optimally scoring DAGs and associated connectivity matrice(s) and set(s) of intervention targets, which are all solutions to:
\begin{equation}
    \begin{aligned}
 \argmin_{\dagg,\bar{h}}\argmin_{\substack{B,\Gamma,\cc{I} \\\{\Omega_e ,\Psi_e\}_{e=1}^m }} \, &~~~\sum_{e = 1}^{m}\hat{\pi}_e\ell(B,\Gamma,\Omega_e,\Psi_e;\hat{\Sigma}_{e}) + \lambda\mathcal{R}_{\gamma}(\mathcal{D},\cc{I})\\
\text{ subject-to:} \, &~~~B\sim\dagg~~;~~ \mathbb{I}(\{\Omega_e\}_{e=1}^m) \subseteq \cc{I}.
    \end{aligned}
    \label{eqn:estimator_theory}
\end{equation}
Here, the dimension of the matrices $\Gamma$ and $\Psi_e$ are $\mathbb{R}^{p \times \bar{h}}$ and $\mathbb{S}^{\bar{h}}$, respectively. Compared to the estimator \eqref{eqn:est2_marg}, the estimator \eqref{eqn:estimator_theory} searches for the optimal DAG and number of latent variables; thus, $\hat{\dagg}_\mathrm{all.opt} = \argmin_{\dagg,\bar{h}} \texttt{score}_{\lambda,\gamma}(\dagg,\hat{\Theta}(\dagg,\bar{h}))$ and $\hat{B}_\mathrm{all.opt},\hat{\cc{I}}_\mathrm{all.opt}$ are the associated model parameters. Throughout, we choose $\lambda \to 0$, with details on the specific rate described in Appendix~\ref{appendix:infinite_characterization}. We assume that ${\lim_{\mbox{all}\ n_e \to \infty}} \frac{n^e}{\sum_{e=1}^m n^e} > 0$ for all $e\in[m]$. Finally, for technical reasons, to prove consistency, the parameter space $(B,\Gamma,\{(\Omega_e,\Psi_e)\}_{e=1}^m)$ in \eqref{eqn:estimator_theory} is assumed to be compact. Assuming such a compactness constraint enables uniform convergence of M-estimators \citep{van2000empirical}; see again Appendix~\ref{appendix:infinite_characterization} for more details.

\subsection{Impossibility results without imposing structural assumptions}
\label{sec:impossible} 


The problem of identifying the underlying DAG is ill-posed if no assumption is placed on the latent effects. {Specifically, the distribution among the observed variables in every environment can be expressed as one generated according to an SCM \eqref{eq:SCM_comp} where the graph among the observed variables is arbitrary.  We formalize this next.
\begin{proposition}[Equivalent SCMs] Regardless of the intervention set $\cc{I}^\star$ and the perturbation magnitudes encoded in $\Omega_e^\star$ on the observed variables, for any DAG $\dagg$ and any connectivity matrix $B \sim \dagg$, there exists parameters $(\Gamma,\{(\Omega_e,\Psi_e)\}_{e=1}^m)$ such that the associated SCM specified by these parameters is compatible with the data distribution. 
\label{prop:all_dags_scm}
\end{proposition}

\begin{figure}[t]
\centering
\begin{tabular}{cc}
\begin{tikzpicture}[every node/.style={scale=0.8},node distance={20mm},main/.style = {draw, circle, minimum size=1cm}]
 \tikzstyle{state}=[inner sep=1pt, minimum size=12pt]
    \tikzstyle{every edge}=[draw,>=stealth',->,line width = 0.25mm]	
	\node[main,inner sep=1.5pt,dotted](h_1) {$H_1$};
 	\node[state](e)[left of=h_1] {$\mathcal{E}$};
	\node[state](x_1)[below left of=h_1]  {$X_1$};
	\node[state](x_2)[below of=h_1] {$X_2$};
 	\node[state](x_3)[below right of=h_1] {$X_3$};
  	\draw (e) edge [->,dotted,red] (h_1);
	\draw (h_1) edge [->,dotted] (x_2);
	\draw (h_1) edge[->,dotted](x_3);
	\draw (x_1) edge[->](x_2);
 	\draw (x_2) edge[->] (x_3);
    	\draw (e) edge[->,dotted,red] (x_1);
  	\draw (e)  edge [->,dotted,red] (x_2);
  	\draw (e)  edge [->,dotted,red] (x_3);

\end{tikzpicture}
&~~~~~~~~~~~~~
\begin{tikzpicture}[every node/.style={scale=0.8},node distance={20mm},main/.style = {draw, circle, minimum size=1cm}]
 \tikzstyle{state}=[inner sep=1pt, minimum size=12pt]
    \tikzstyle{every edge}=[draw,>=stealth',->,line width = 0.25mm]	
	\node[main,inner sep=1.5pt,dotted](h_1) {$\tilde{H}_1$};
  	\node[state](e)[left of=h_1] {$\mathcal{E}$};
 	\node[main,inner sep=1.5pt,dotted](h_2)[right of=h_1]  {$\tilde{H}_2$};
	\node[main,inner sep=1.5pt,dotted](h_3)[right of=h_2]  {$\tilde{H}_3$};
	\node[state](x_1)[below of=h_1]  {$X_1$};
	\node[state](x_2)[below of=h_2] {$X_2$};
 	\node[state](x_3)[below of=h_3] {$X_3$};
  	\draw  (e) edge[->,dotted,red] (h_1);
	\draw (e) edge[->,out=45,in=135,looseness=0.5,dotted,red] (h_2);
	\draw (e) edge[->,out=45,in=135,looseness=0.5,dotted,red](h_3);
	\draw (e) edge[->,dotted,red] (x_1);
	\draw (e) edge[->,dotted,red] (x_2);
	\draw (e) edge[->,dotted,red] (x_3);
	\draw (h_1) edge[->,dotted] (h_2);
        \draw(h_1) edge[->,dotted,out=45,in=135,looseness=0.5] (h_3);
        \draw (h_2) edge[->,dotted] (h_3);
	\draw  (h_1) edge[->,dotted] (x_1);
 	\draw (h_1) edge[->,dotted] (x_2);
	\draw (h_1)edge[->,dotted] (x_3);
\draw (h_2) edge[->,dotted] (x_1);
 	\draw (h_2) edge[->,dotted] (x_2);
	\draw (h_2) edge[->,dotted] (x_3);
\draw (h_3) edge[->,dotted] (x_1);
 	\draw (h_3) edge[->,dotted] (x_2);
	\draw (h_3) edge[->,dotted] (x_3);

\end{tikzpicture}\\
\small{(a)} &~~~~~~~~~~~~~
\small{(b)}
\end{tabular}
\caption{Two equivalent models with respect to the distribution among the observed variables; see text.}
\label{fig:illustration_faithfulness_main}
\end{figure}
The proof of Proposition~\ref{prop:all_dags_scm} is presented in Appendix \ref{proof_prop_all_dags_scm}. Figure~\ref{fig:illustration_faithfulness_main} provides an illustration of this result among three observed variables. Here, Figure~\ref{fig:illustration_faithfulness_main}(a) and Figure~\ref{fig:illustration_faithfulness_main}(b) represent equivalent models, although the subgraphs among the observed variables are different; see Appendix~\ref{sec:illutration_faithfulness} for formal description of the parameters of these models. Note that a typical assumption in causal structure learning with latent variables is that the joint distribution of the observed and latent variables is faithful to the graph among these variables. Indeed, many algorithms such as FCI \citep{PC}, RFCI \citep{RFCI}, and JCI \citep{JCI} rely on this condition for characterizing and learning an equivalence class of models. The result of Proposition~\ref{prop:all_dags_scm} provides a construction of non-faithful models; see Appendix~\ref{sec:illutration_faithfulness}.
\begin{corollary}[Optimally scoring DAGs without structural assumptions]Regardless of the intervention set $\cc{I}^\star$ and the perturbation magnitudes encoded in $\Omega_e^\star$ on the observed variables, 
$\hat{\dagg}_\text{all.opt}$ is the empty graph.
\label{prop:all_dags}
\end{corollary}
The proof of Corollary~\ref{prop:all_dags} is presented in Appendix \ref{proof_prop_all_dags}. The results of Proposition~\ref{prop:all_dags} and Corollary~\ref{prop:all_dags} state that searching for the highest scoring DAG is meaningless if no structure is imposed on the problem.

\subsection{Equivalence class characterization under a small number of latent variables with dense effects}
\label{sec:dense_sparse}
We now analyze the estimates \eqref{eqn:estimates_theory}
 under structural assumptions on the denseness of the latent effects and sparsity of the underlying DAG where the perturbations on the latent variables may be arbitrary. We substantially relax this assumption in Section~\ref{sec:known_latent_perturb} at the expense of approximate knowledge of the perturbations on the latent variables.

{Before proceeding, we present some real-world applications where the assumed structure may be reasonable. For example, \cite{Chand2012} showed that a large fraction of the conditional dependencies among stick returns can be explained by a few latent variables. In a similar spirit, \cite{water} demonstrated that the California reservoir network is influenced by a few external latent factors (correlated with environmental variables), and these have a system-wide effect; we explore this application further in our experiments. Finally, in an analysis of gene expression data, \cite{sparse_dense} find that a small number of dense latent factors explain much more variability than sparse latent factors, and these dense factors correlated well with some known biological and technical covariates (e.g. batch effects).} 

How are the aforementioned structural assumptions useful for identifiability? To motivate the utility of these structural assumptions, we first note that the structural equation model \eqref{eq:SCM_env} yields the covariance model of observed variables $\Sigma_e^\star = (\id-B^\star)^{-1}(\Omega^\star_e+\Gamma^\star\Psi_e^\star{\Gamma^\star}^T)(\id-B^\star)^{-T}$ for every $e\in[m]$. By the Woodbury Inversion lemma, we obtain the following decomposition  ${\Sigma^\star_e}^{-1} = S^\star_e - L^\star_e$ of the precision matrix ${\Sigma_e^\star}^{-1}$ for every $e\in[m]$. Here, the matrix $S^\star_e = (\id-B^\star)^T{\Omega^\star_{e}}^{-1}(\id-B^\star)$ is the inverse of the conditional covariance of the observed variables conditioned on the latent variables. The matrix $L^\star_e$ is the rank-$h$ matrix $(\id-B^\star)^T{\Omega^\star_{e}}^{-1}\Gamma^\star({\Psi_e^\star}^{-1}+{\Gamma^\star}^T{\Omega^\star_{e}}^{-1}\Gamma^\star)^{-1}{\Gamma^\star}^T{\Omega^\star_{e}}^{-1}(\id-B^\star)$ that summarizes the effect of marginalization over latent variables.

Without assuming any additional structure on the population model, the matrices $S^\star_e$ and $L^\star_e$ are not identifiable from ${\Sigma_e^\star}^{-1}$. This lack of identifiability implies ${\Sigma_e^\star}^{-1}$ can be modeled by a different DAG ${\dagg} \in \hat{\dagg}_\mathrm{all.opt}$ that may be arbitrarily different from $\dagg^\star$. 
Appealing to the previous literature on sparse-plus-low rank decompositions, the matrices $S_e^\star$ and $L_e^\star$ are identifiable from their sum if the matrix $S_e^\star$ is sparse and the matrix $L_e^\star$ is low-rank with its energy spread across the coordinates \citep{Recht2010GuaranteedMS,Chandrasekaran2011RankSparsityIF,Candes11RobustPCA}. It is straightforward to check that the entry $[S_e^\star]_{i,j}$ is nonzero if the variables $X_i$ and $X_j$ are connected in the moral graph of $\dagg^\star$. Thus, a sparse moral graph of $\dagg^\star$ implies that the matrix $S_e^\star$ is sparse. The assumption on $L_e^\star$ can be interpreted as the number of latent variables being small (as compared to the ambient dimension {$p$}) with their effects spread across all the observed variables. We measure the sparsity of the moral graph of a DAG $\dagg$ by the maximal degree of the moral graph, denoted by $\degree[\mathrm{moral}(\dagg)]$. Thus, we require $\degree[\mathrm{moral}(\dagg^\star)]$ to be small so that no observed variable is directly connected to ``many'' other observed variables in the moral graph of $\dagg^\star$. To measure the "diffuseness" of the latent effects, we consider the following quantity for any linear subspace $T \subseteq \mathbb{R}^p$ \citep{Candes11RobustPCA,rechtCandesMatrix,Chandrasekaran2011RankSparsityIF,Chand2012}:
\begin{equation*}
\inc[T] :=\max_i \|\mathcal{P}_{T}({\mathbf{e}}_i)\|_2,
\label{eqn:incoherence}
\end{equation*}
where $\mathcal{P}_{T}$ is the projection onto the subspace $T$ and $\mathbf{e}_i$ is a standard coordinate basis. The quantity $\inc[T]$ is also known as the ``incoherence parameter'' \citep{rechtCandesMatrix,Chandrasekaran2011RankSparsityIF}. It measures how aligned the subspace $T$ is with respect to standard basis elements and is lower-bounded by $\sqrt{\frac{\text{dim}(T)}{p}}$ and upper-bounded by one. In our setting, the relevant subspace is $\mathrm{col}\text{-}\mathrm{space}((\id-B^\star)^T{\Omega_e^\star}^{-1}\Gamma^\star))$ which is the column-space of $L_e^\star$. A small value of $\inc[\mathrm{col}\text{-}\mathrm{space}((\id-B^\star)^T{\Omega_e^\star}^{-1}\Gamma^\star))]$ ensures the matrix $L^\star_e$ has small rank and cannot have its support concentrated in a few locations.

In summary, to enable identifiability, the population quantities $\degree[\mathrm{moral}(\dagg^\star)]$ and $\inc[\mathrm{col}\text{-}\mathrm{space}((\id-B^\star)^T\allowbreak{\Omega_e^\star}^{-1}\Gamma^\star)]$ are assumed to be sufficiently small. We will first analyze the estimates \eqref{eqn:estimates_theory} under these assumptions as well as faithfulness and access to an observational environment. For notational simplicity, we define $d^\star :=\degree[\mathrm{moral}(\dagg^\star)]$ and $\inc^\star_e :=\inc[\mathrm{col}\text{-}\mathrm{space}((\id-B^\star)^T{\Omega_e^\star}^{-1}\Gamma^\star)]$. Formally, we assume:
\begin{assumption}
sparse DAG and incoherent {(dense)} latent effects across all environments: $32d^\star{\inc^\star_e}^2 < 1 \text{ for all }e\in[m]$.
\label{assum:dense_incoherence}
\end{assumption}
\begin{assumption}{the distribution }$X^e|H^e$ {is faithful with respect to} $\dagg^\star$ {for all} $e \in[m]$.
\label{assum:faithful}
\end{assumption}
\begin{assumption}
observational environment $e=1$ with no perturbations on the observed variables: $\Omega_{e}^\star \succeq \Omega_{1}^\star \text{ for all }e \in[m]$.
\label{assump:observational}
\end{assumption}
Assumption \ref{assum:dense_incoherence} ensures that the population model consists of a sufficiently sparse moral graph and dense latent effects with a small number of latent variables as compared to a relatively large ambient dimension $p$. This assumption bears resemblance to conditions for identifiability in sparse-plus-low rank decompositions \citep{Chandrasekaran2011RankSparsityIF,Chand2012,marloes}, although we demonstrate in Appendix~\ref{sec:discussion_dense} that our condition is weaker (in terms of high-dimensional scaling) than those imposed in these previous works. We also provide in Appendix~\ref{sec:discussion_dense} examples of SCMs \eqref{eq:SCM_env} that satisfy Assumption~\ref{assum:dense_incoherence}. Further, Assumptions \ref{assum:faithful}-\ref{assump:observational} are standard conditions for identifiability of an equivalence class of DAGs both in observational and interventional settings \citep{chick02,jointly,IGSP}. We will see later that Assumption~\ref{assump:observational} can be replaced by conditions on the informativeness of the interventions.
 
Recall that Proposition~\ref{prop:all_dags} and Corollary~\ref{prop:all_dags} tells us that solving \eqref{eqn:estimator_theory} is meaningless unless the parameters of the estimated causal models are also appropriately constrained. Thus, under Assumptions \ref{assum:dense_incoherence}-\ref{assump:observational}, we consider the theoretical properties of \eqref{eqn:estimates_theory} when the incoherence of the estimated latent effects is controlled.
\begin{proposition}[Equivalence class characterization under incoherent latent effects]Consider the estimator \eqref{eqn:estimator_theory} with the additional constraint that $\inc[\mathrm{col}\text{-}\mathrm{space}((\id-B)^T\Omega_e^{-1}\Gamma)] \leq 2\inc^\star_e$ for all $e\in[m]$. {Then}, under Assumptions \ref{assum:dense_incoherence}-\ref{assump:observational} and $\gamma$ selected so that $d^\star\allowbreak\geq\gamma>0$, we have that the estimates \eqref{eqn:estimates_theory} satisfy: $\{\dagg^\star\} \subseteq \hat{\dagg}_\mathrm{all.opt} \subseteq \mathrm{MEC}(\dagg^\star)$ with probability tending to one as the sample size in every environment tends to infinity.
\label{thm:sparse_dense_1}
\end{proposition}
We present the proof of Proposition~\ref{thm:sparse_dense_1} in Appendix~\ref{proof_prop_dense_without_interventions}. This result states that if the latent effects of the estimated causal models are constrained to be low-dimensional and dense, in infinite data limit and with probability tending to one, the set of equally scoring latent variable causal models {$\hat{\dagg}_\mathrm{all.opt}$ } are a subset of the Markov equivalence class $\mathrm{MEC}(\dagg^\star)$ and contain the population {DAG} $\dagg^\star$. In Proposition~\ref{thm:sparse_dense_1} the interventions on the variables indexed by $\cc{I}^\star$ do not appear to directly constrain the optimal set of solutions. Indeed, we show in Appendix~\ref{necessary_intervention} that under certain ``worst-case configurations" of intervention strengths, the set $\hat{\dagg}_\mathrm{all.opt}$ is precisely equal to the Markov equivalence class $\mathrm{MEC}(\dagg^\star)$. Thus, to improve identifiability, additional assumptions on the informativeness of the interventions are needed, which we state below: 
\begin{assumption}interventions on the observed variables are heterogeneous: for every $i \in \cc{I}^\star, \text{ there }\text{exists }\\e \text{ such that } [\Omega^\star_e]_{i,i} >[\Omega^\star_1]_{i,i} \text{ and }[\Omega^\star_e{\Omega^\star_{1}}^{-1}]_{i,i} \neq [\Omega^\star_e{\Omega^\star_{1}}^{-1}]_{j,j} \text{ for all }j \neq i$.
\label{assump:heterogenous}
\end{assumption}
\begin{assumption}
interventions on the observed variables are "truthful": for every Markov equivalent connectivity \footnote{A Markov equivalent connectivity matrix $B$ w.r.t $X^e|H^e$ in Assumption 4 satisfies: compatibility with a DAG $\dagg\in \mathrm{MEC}(\dagg^\star)$ and the relation $\Sigma^\star_{X^e|H^e}=(\id-B)^{-1}\Omega(\id-B)^{-T}$ for some $\Omega \in \mathbb{D}_{++}^p$.} $B \text{ to }B^\star \text{ w.r.t } X^1|H^1, \text{ and }(i,j) \text{ where } i \in \cc{I}^\star, i \leadsto j \text{ in }\dagg^\star$, $[\id-B]_{j,:} \not\propto [\id-B^\star]_{i,:}$. 
\label{assump:truthful}
\end{assumption}
Here, the notation $v_1 \not\propto v_2$ for vectors $v_1$ and $v_2$ means that the vectors $v_1$ and $v_2$ are not proportional. Further, the notation $i \leadsto j \text{ in }\dagg^\star$ means that the variable $X_i$ is an ancestor of the variable $X_j$ in the DAG $\dagg^\star$. While Assumption \ref{assump:heterogenous} ensures that the interventions on the observed variables are sufficiently diverse \footnote{Assumption \ref{assump:heterogenous} can be satisfied even if the interventions occur in different environments.}, Assumption \ref{assump:truthful} excludes a pathological configuration of the intervention strengths and the connectivity matrix $B^\star$ and is similar in spirit to "interventional faithfulness" assumptions imposed in previous work \citep{AICP_2021,caroline_unknown,GamellaGnies2021}. In summary, Assumptions \ref{assump:heterogenous}-\ref{assump:truthful} are both rather weak and ensure that interventions on the observed variables improve identifiability: 
\begin{theorem}[Equivalence class characterization under incoherent latent effects, and truthful and heterogeneous interventions]Consider the estimator \eqref{eqn:estimator_theory} with the additional constraint that $\inc[\mathrm{col}\text{-}\mathrm{space}\allowbreak((\id-B)^T\Omega_e^{-1}\Gamma)] \leq 2\inc^\star_e$ for all $e \in[m]$.  {Then}, under Assumptions \ref{assum:dense_incoherence}-\ref{assum:faithful} and \ref{assump:heterogenous}-\ref{assump:truthful}, and if the parameter $d^\star\geq \gamma > 0$: $\{\dagg^\star\} \subseteq \hat{\dagg}_\mathrm{all.opt} = \simec(\dagg^\star)$, $B^\star \in \hat{B}_\mathrm{all.opt}$ and {$\hat{\cc{I}}_\mathrm{all.opt} = \{\cc{I}^\star\}$}, all with probability tending to one as the sample size in every environment tends to infinity.
\label{thm:sparse_dense}
\end{theorem}
We present the proof of Theorem~\ref{thm:sparse_dense} in Appendix~\ref{proof_thm_dense}. Notice that Assumptions \ref{assump:heterogenous}-\ref{assump:truthful} replace the need for access to an observational environment in Assumption~\ref{assump:observational}, as $\imec(\dagg^\star) \subseteq \mathrm{MEC}(\dagg^\star)$. Theorem~\ref{thm:sparse_dense} states that $\simec(\dagg^\star)$ contains the set of optimally scoring DAGs when the latent effects of the estimated causal models are constrained to be low-dimensional and dense and the interventions are informative. Importantly, our proposed procedure \methName{} cannot directly control the incoherence of the latent effects. Instead, it can only constrain the number of latent variables. In the following corollary, we provide sufficient conditions for when the estimator \eqref{eqn:estimator_theory} can obtain a DAG from the set $\simec(\dagg^\star)$.
\begin{corollary}Suppose that Assumptions \ref{assum:dense_incoherence}-\ref{assum:faithful} and \ref{assump:heterogenous}-\ref{assump:truthful} are satisfied. Let $d^\star\geq \gamma > 0$.  Consider the estimator \eqref{eqn:estimator_theory} with the constraint that $\bar{h} \leq h_\text{max}$ for some non-negative integer $h_\text{max} \geq \text{dim}(H)$. Let $\nu^\star$ be a positive integer with $\nu^\star \geq d^\star$. Suppose there exists an estimate $(\hat{\dagg},\hat{B},\hat{\Gamma},\hat{\cc{I}},\{{\Omega}_e,\hat{\Psi}_e\}_{e=1}^m)$ that satisfies $48\nu^\star\inc[\mathrm{col}\text{-}\mathrm{space}((\id-\hat{B})^T\hat{\Omega}_e^{-1}\hat{\Gamma})]<1$ for all $e\in[m]$. {Then, $\hat{\cc{I}} = \cc{I}^\star$ and  $\himec(\hat{\dagg}) = \simec({\dagg}^\star)$} with probability tending to one as the sample size in every environment tends to infinity.
\label{cor:incoherent}
\end{corollary}
The proof of Corollary~\ref{cor:incoherent} is presented in Appendix~\ref{proof_corr_sufficient}. The result in Corollary~\ref{cor:incoherent} suggests the following procedure when Assumptions \ref{assum:dense_incoherence}-\ref{assump:truthful} are believed to be satisfied and the user has access to $\nu^\star$ that serves as an upper-bound for the maximal degree of the moral graph of the underlying DAG: obtain the best latent variable causal model(s) based on the likelihood score on test data when the regularization parameters $\lambda,\bar{h}$ are varied with $\bar{h}$ smaller than a pre-specified value $h_\text{max}$. Then, compute the incoherence of the latent effects of these best scoring models. If for an optimal DAG $\hat{\dagg}$, the incoherence parameter multiplied by $\nu^\star$ is sufficiently small for all environments, in large data settings and with probability tending to one, $\dagg^\star$ lies inside $\himec(\hat{\dagg})$. To highlight when such an assumption is far from being satisfied, our software outputs the following quantitative indicator: $\max_e\degree[\mathrm{moral}(\hat{\dagg})]\inc[\mathrm{col}\text{-}\mathrm{space}((\id-\hat{B})^T\hat{\Omega}_e^{-1}\hat{\Gamma})]$. Here, large values (e.g. far above 1) indicate strong deviations from our assumptions.

\subsection{Equivalence class characterization under approximately known latent perturbations}
\label{sec:known_latent_perturb}
In the previous discussion, a central assumption was that the number of latent variables is small and their effects are dense. We next consider a setting where the perturbations on the latent variables are approximately known. These assumptions enable an equivalence class characterization of DAGs without needing Assumption~\ref{assum:dense_incoherence}, i.e. without imposing conditions on the number of latent variables and the denseness {(incoherence)} of their effects. For technical simplicity, we analyze the setting where the latent variables are independent and identically distributed, i.e. $\Psi^{\star}_e = \psi_e^\star\id$ with $\psi^{\star}_e\in \mathbb{R}_+$.

{For illustrative purposes, we first start with an extreme setting where the perturbations are exactly known (although the number of latent variables remains unknown) and demonstrate that identifiability is possible with relatively mild assumptions. We then deviate from this extreme setting by assuming that the latent perturbations are approximately known and show once again that identifiability is possible under some conditions whose severity depends on the level of the approximation.}

\textbf{Illustrative setting: known latent perturbations} Without loss of generality, we can take $\psi_1^\star = 1$ and $\psi_e^\star$ to be known positive values that may be different than $\psi_1^\star$. Our theoretical guarantees require Assumptions 2 and 5 as well as modifications to Assumptions 3 and 4 (dubbed 3' - 4'). In particular, we assume that there are two observational environments ($e = 1$ and $e = 2$ without loss of generality) with no interventions on the observed variables and interventional environments (so that $m \geq 3$) with sufficiently heterogeneous perturbations on the observed variables:
\begin{customassump}{3'}
environments $e = 1,2$ with no perturbations on the observed variables: $\Omega^\star_{1}=\Omega^\star_{2} \text{ and } \Omega_e^\star \succeq \Omega_1^\star \text{ for all }e=3,\dots,m$.
\end{customassump}
\begin{customassump}{4'}
heterogeneous interventions on observed and latent variables: $\text{ for every }i \in \cc{I}^\star, j \in \cc{I}^\star \text{ with }i\neq j,  \text{ there exists } e  \text{ such that the collection }(\psi_1^\star,\psi_2^\star,\psi_e^\star) \text{ are distinct }, [\Omega_e^\star]_{i,i} > [\Omega_1^\star]_{i,i} \text{ and}\\
\left[(\Omega_{e}^\star-\psi_{e}^\star\Omega_{1}^\star\right){\Omega_{1}^\star}^{-1}]_{i,i} \neq \left[(\Omega_{e}^\star-\psi_{e}^\star\Omega_{1}^\star\right){\Omega_{1}^\star}^{-1}]_{j,j}$.
\end{customassump}
Assumption 3' (analogous to Assumption 3) ensures that there are environments where no perturbations act on the observed variables. Assumption 4' (analogous to Assumption 4) ensures that the interventions on the latent variables and observed variables are informative for additional identifiability. One can show that if the parameters $\Omega^{\star}_e,\Omega^{\star}_1$ and $\psi^\star_1,\psi^{\star}_2,\psi^{\star}_e$ are drawn from continuous distributions, Assumption 4' is satisfied almost surely. 

\begin{theorem}[Equivalence class characterization under known perturbations on the latent variables] Consider the estimator \eqref{eqn:estimator_theory} with the additional constraint $\psi_e\id = \psi_e^\star\id$ for all $e \in [m]$. Suppose Assumptions 2, 5 and Assumptions 3'-4' are satisfied.  Letting $\frac{1}{|\cc{I}^\star|}>\gamma > 0$, then $\{\dagg^\star\} \subseteq \hat{\dagg}_\mathrm{all.opt} = \simec(\dagg^\star)$ with probability tending to one as the sample size in every environment tends to infinity.
\label{thm:perturbed_latent_known}
\end{theorem}
The proof of Theorem~\ref{thm:perturbed_latent_known} is presented in Appendix~\ref{sec:known_latent_proof}. This result highlights that at the expense of knowing the latent perturbations, no assumptions on the incoherence (denseness) of the latent variables or their number is required for characterizing the equivalence class of optimally scoring DAGs. We note that when the number of latent variables is unconstrained, three environments are necessary for improved identifiability. Indeed, in Appendix \ref{sec:two_environments}, we show that two environments (regardless of the number of interventions and their strengths) only offer identifiability up to the Markov equivalence class of $\dagg^\star$. 

\textbf{Approximately known latent perturbations} Knowing the perturbations on the latent variables can be a stringent condition in practice. One can relax this to approximately knowing the perturbations at a pre-specified level $C_\psi$, e.g. $|\tilde{\psi}_e - \psi_e^\star| \leq C_\psi$ where $\tilde{\psi}_e$ is the (approximate) {known} perturbation on the latent variables. A natural choice for the approximate perturbations $\tilde{\psi}_e$ would be $\tilde{\psi}_e = 1$ for all $e$, encoding no perturbations on the latent variables: the level $C_{\psi}$ then describes the deviation from no perturbations on the latent variables. This and versions thereof will be discussed in the remarks below.

\noindent{\emph Remark 2:} To account for the latent perturbation approximation, the following two assumptions ensure equivalence class characterization. The first assumption is that the latent variables induce some confounding dependencies among the observed variables; this condition becomes more stringent with larger $C_\psi$ (e.g. weaker knowledge of the latent perturbations) although we demonstrate in Appendix~\ref{sec:approximately_known} that it is generally \emph{far weaker} than the incoherence condition in Assumption 1. The second assumption is that the observed variables in the set $\cc{I}^\star$ receive strong enough interventions, Under these two conditions (as well as assumptions 2,3',5 and an assumption similar in spirit to 4'), the estimator \eqref{eqn:estimator_theory} obtains (in the infinite data limit) $\simec({\dagg}^\star)$ as the set of optimally scoring DAGs. For a formal description of the assumptions and the result, we refer the reader to Appendix~\ref{sec:approximately_known}. Finally, as a quantitative indicator of deviations from assumptions, our software displays the strength of interventions on each variable, i.e. $\xi_j := \frac{1}{m}\sum_{e=1}^m([\hat{\Omega}_e]_{j,j}-\frac{1}{m}\sum_{e=1}^m [\hat{\Omega}_e]_{j,j})^2$ for each $j \in \hat{\cc{I}}_\mathrm{opt}$. Here, $\xi_j$ being small for any $j \in \hat{\cc{I}}_\mathrm{opt}$ indicates that the perturbations on the corresponding variable are weak.


\noindent{\emph Remark 3:} Assuming that the latent variables remain unperturbed across all environments is a special case of knowing the latent perturbations. In such settings, the equivalence class -- when $\cc{I}^\star \subset [p]$ -- can, in general, be very different than $\simec({\dagg}^\star)$ (see Appendix~\ref{sec:no_latent_perturb} for a simple illustration), highlighting that interventions on the latent variables may be beneficial for improved identifiability. Nevertheless, when $\cc{I}^\star = [p]$, similar to the backShift procedure, \methName{} attains full identifiability of the population DAG.

\subsection{Identifiability guarantees for non-Gaussian models}
\label{sec:non-Gaussian}
The \methName{} estimator \eqref{eqn:estimator_theory} fits a Gaussian perturbation model \eqref{eq:SCM_env} to the data. However, the perturbation data may be non-Gaussian but satisfy the linear SCM \eqref{eq:SCM_env}. In such settings, deploying \methName{} yields the same equivalence class characterization (outlined previously) -- but due to the non-Gaussianity, better identifiability is possible using a different method. In particular, since the estimator \eqref{eqn:estimator_theory} operates on covariance models (matching the second moments of the underlying distribution), it provides \emph{conservative estimates} in the sense that $\{\dagg^\star\}\subseteq \hat{\dagg}_\text{all.opt}$ with probability tending to one. On the other hand, tailored methods that match additional moments obtain an equivalent set of DAGs $\dagg^\text{non-Gaussian}_\text{all.opt}$ satisfying ${\{\dagg^\star\}} \subseteq \hat{\dagg}^{\text{non-Gaussian}}_\text{all.opt}\subseteq \hat{\dagg}_\text{all.opt}$ with probability tending to one. {How to design such tailored methods in the current modeling context is beyond the scope of this work.} 

%% file: sections/algorithm.tex
\section{Practical use cases of \methName{}}
\label{sec:optimization}
We next describe how \methName{} can be used in practice to account for latent effects and obtain a set of DAGs that fit the data well. In Section~\ref{sec:alternating}, we propose an alternating minimization strategy to solve \eqref{eqn:est2_marg} with the DAG, hence also the support of $B$, {being} pre-specified. Building on this, in Section~\ref{sec:alg_candidate_dags}, we consider the setting where a candidate set of DAGs are available (for example as for the protein expressions dataset in Section~\ref{sec:experiments}) and describe how \methName{} can be used to obtain an optimally scoring equivalence class of DAGs. Finally, in Section~\ref{sec:alg_starting_point}, we extend our algorithmic framework to the setting when a set of DAGs represent starting points, and we deploy \methName{} to improve on these DAGs by removing any spurious dependencies. A python package containing the implementation of all components of \methName{} is available at \url{https://github.com/juangamella/ut-lvce}.

We remark here that searching for optimally scoring DAGs (according to the score \eqref{eqn:score}) is a very difficult computational task. In particular, an immediate approach that comes to mind is to develop a greedy DAG search over the space of equivalent models akin to GES \cite{chick02}. Indeed, \cite{GamellaGnies2021} develops a greedy algorithm to move in the space of interventionally equivalent DAGs for the model \eqref{eq:SCM_env} \emph{without} latent variables. By employing the \methName{} score function \eqref{eqn:score}, one may adapt the method of \cite{GamellaGnies2021} to incorporate latent effects. However, a significant conceptual challenge is that the likelihood score \eqref{eqn:score} is not decomposable according to the DAG structure due to latent confounding; the lack of score decomposability renders greedy-based techniques computationally expensive. Thus, our focus in this paper is to demonstrate the utility of \methName{} on the use cases described in the previous paragraph.

\subsection{Alternating minimization strategy to compute the \methName{} estimator}
\label{sec:alternating}
We first describe an optimization approach for solving \methName{} given an input DAG $\dagg$ and a fixed intervention target set $\cc{I}$. Our optimization algorithm is based on the following alternating minimization strategy: starting with an initialization of all of the model parameters, we fix $B$ and perform gradient updates to find updated estimates for the parameters $(\Gamma,\{\Omega_e,\Psi_e\}_{e=1}^m$ where the noise variances $\{\Omega_e\}_{e=1}^m$ are compatible with the input intervention set $\cc{I}$, and then update $B$ by solving a convex program to optimality with the remaining parameters fixed. We find that the alternating method described above is relatively robust to the initialization scheme, but we nonetheless propose the following concrete strategy:
\begin{equation}
\begin{aligned}
1)~&\text{fit }B^{(0)}  \text{ via linear regression with pooled data}, \\
2)~&\Gamma^{(0)} = UD^{1/2} \text{ where } UDU^T \text{ is SVD of }(\id-B^{(0)})\hat{\Sigma}_\text{pooled}(\id-B^{(0)})^T, \\
3)~& \Omega_e^{(0)} = \texttt{diag}\left\{(\id-B^{(0)})\hat{\Sigma}_\text{pooled}(\id-B^{(0)})^T\right\} \text{ for every }e\in[m], \\
4)~& \Psi_e^{(0)} = \mathrm{Id} \text{ for every } e \in [m],
\end{aligned}
\label{eqn:init}
\end{equation}
 where $\hat{\Sigma}_{\text{pooled}}$ is the covariance matrix of the pooled data. The first step follows since the DAG structure is known. The entire procedure, involving the initialization step and the parameter updates, is presented in Algorithm \ref{algo:dag_score}.
 \setcounter{algorithm}{-1}
\begin{algorithm}[]
\caption{Solving \methName{} to score a given DAG $\dagg$ for a fixed $(
\lambda,\gamma,\bar{h})$ and targets $\cc{I}$}
\begin{algorithmic}[1]
\vspace{0.1in}
\STATE {\bf Input}: DAG $\dagg$; intervention targets $\cc{I}$; data $\hat{\Sigma}_e$, $\hat{\pi}_e$ for $e \in [m]$; regularization params. $\lambda,\gamma \geq 0$; $\#$ of latent vars. $\bar{h}$
\vspace{.05in}
\STATE{\bf Initialize parameters}: via relation \eqref{eqn:init}
\vspace{0.04in}
\STATE{\bf Alternating minimization}: solve for causal parameters
\begin{itemize}
	\item[(a)] fixing $(\Gamma^{(t)}, \{(\Omega_e^{(t)},\Psi_e^{(t)})\}_{e = 1}^m)$, update $B^{(t+1)}$ by solving {the} convex optimization program \eqref{eqn:est2_marg} where $B^{(t+1)} \sim \dagg$. Fixing $B^{(t+1)}$, perform gradient updates until convergence to find $(\Gamma^{(t+1)}, \{(\Omega_{e}^{(t+1)},\Psi_e^{(t+1)})\}_{e = 1}^m)$ where $\mathbb{I}(\{\Omega_e^{(t+1)}\}_{e=1}^m)\subseteq \cc{I}$
    \item[(b)] apply alternating iterates for positive integers $t$ until convergence at iteration $T$
    \item[(c)] obtain estimates $\hat{\Theta}(\dagg,\bar{h}):= (\hat{B}^{(T)},\hat{\Gamma}^{(T)},\cc{I},\{(\hat{\Omega}_e^{(T)},\hat{\Psi}_e^{(T)})\}_{e=1}^m)$ 
     \end{itemize}
\vspace{0.04in}
\STATE{\bf Output:} causal parameters $\hat{\Theta}(\dagg,\bar{h})$ and regularized likelihood $\texttt{score}_{\lambda,\gamma}(\dagg,\hat{\Theta}(\dagg,\bar{h}))$
\end{algorithmic} \label{algo:dag_score}
\end{algorithm}

Step 3(b) involves two convergence criteria: the convergence of the gradient steps for the parameters $(\Gamma^{(t)},\{(\Omega_{e}^{(t)},\Psi_e^{(t)})\}_{e = 1}^m)$ as well as the convergence of the alternating procedure. For the first criterion, we terminate the gradient descent when the relative change in the likelihood score is below $\epsilon_1$. For the second criterion, we terminate the alternating minimization at step $T$ when $\|B^{(T)}-B^{(T-1)}\|_\infty \leq \epsilon_2$, where $\|\cdot\|_{\infty}$ computes the maximum entry in magnitude of an input matrix. In our experiments, we set $\epsilon_1 = 10^{-6}$ and $\epsilon_2 = 10^{-2}$.

\subsection{Using \methName{} to identify the best scoring DAGs from a candidate set}
\label{sec:alg_candidate_dags}
Let $\dagg_\text{cand}$ be a candidate set of DAGs (potentially a singleton). Building on Algorithm~\ref{algo:dag_score}, we present an algorithm to identify an optimally scoring DAG as well as DAGs in its equivalence class. First, using Algorithm~\ref{algo:dag_score}, we score each DAG in the candidate set with $\cc{I} = [p]$, and obtain an optimally scoring DAG $\hat{\dagg}_\text{opt}$ with noise variances $\{\hat{\Omega}_e\}_{e=1}^m$. To estimate the intervention targets $\hat{\cc{I}}_\mathrm{opt}$, we measure the variation in each coordinate of $\hat{\Omega}_e$ across the environments as large variations indicate that the corresponding variable has received an intervention. To quantify the degree of variation, we compute a ``variance like" metric $\xi_j := \frac{1}{m}\sum_{e=1}^m([\hat{\Omega}_e]_{j,j}-\frac{1}{m}\sum_{e=1}^m [\hat{\Omega}_e]_{j,j})^2$ for each $j \in [p]$ (also defined in Section~\ref{sec:known_latent_perturb}), where large values of $\xi_j$ provide stronger evidence for the presence of an intervention on variable $X_j$. We propose a systematic approach to estimate an intervention set $\hat{\cc{I}}_\mathrm{opt}$ using the values $\xi_j$: we greedily remove the variable with the smallest variation $\xi_j$ and compute the regularized likelihood with the resulting intervention set. We repeat this process until the likelihood score can no longer be improved. Thus, as output, we return the equivalence class of optimally scoring DAGs $\optimec(\hat{\dagg}_\text{opt})$. A detailed summary of our procedure is presented in Algorithm~\ref{algo:equiv_class}.

\begin{algorithm}
\caption{Equivalence class of best scoring DAGs from a candidate set via \methName{}} 
\begin{algorithmic}[1]
\vspace{0.1in}
\STATE {\bf Input}: candidate DAG(s) $\dagg_\text{cand}$; data $\hat{\Sigma}_e,\hat{\pi}_e$ for $e \in[m]$; regularization params. $\lambda,\gamma \geq 0$; $\#$ of latent vars. $\bar{h}$
\vspace{.05in}
\STATE{\bf Obtain likelihood score for each DAG}: for each $\dagg \in \dagg_\text{cand}$, supply $(\dagg,\cc{I} = [p])$ to Algorithm~\ref{algo:dag_score} to obtain the causal parameters $\hat{\Theta}(\dagg,\bar{h})$ and $\texttt{score}_{\lambda,\gamma}(\dagg,\hat{\Theta}(\dagg,\bar{h}))$
\vspace{0.05in}
\STATE{\bf Find an optimal scoring DAG}: obtain $\hat{\dagg}_\text{opt} = \argmin_{\dagg \in \dagg_\text{cand}} \texttt{score}_{\lambda,\gamma}(\dagg,\hat{\Theta}(\dagg,\bar{h}))$
\vspace{0.05in}
\STATE{\bf Greedy backward deletion to estimate intervention set}: initialize $\hat{\cc{I}} = \cc{I}$ and
\begin{enumerate}
\item[(a)] let $\{\hat{\Omega}_e\}_{e=1}^m$ be noise variance encoded in $\hat{\Theta}(\hat{\dagg}_\text{opt},\bar{h})$
\item[(b)] estimate intervention strengths for each $j\in[p]$: $\xi_j := \frac{1}{m}\sum_{e}([\hat{\Omega}_e]_{j,j}-\frac{1}{m}\sum_{e}[\hat{\Omega}_e]_{j,j})^2$
\item[(c)] remove weakest intervention: $\hat{\cc{I}}_\mathrm{opt} \leftarrow \hat{\cc{I}}_\mathrm{opt} \setminus \{\argmin_{j}\xi_j: j \in\hat{\cc{I}}_\mathrm{opt}\}$
\item[(d)] supply $(\hat{\dagg}_\text{opt},\hat{\cc{I}}_\mathrm{opt})$ to Algorithm~\ref{algo:dag_score} and obtain $\texttt{score}_{\lambda,\gamma}(\hat{\dagg}_\mathrm{opt},\hat{\Theta}(\hat{\dagg}_\mathrm{opt},\bar{h}))$
\item[(e)] repeat (c,d) until the likelihood score does not improve
\end{enumerate}
\vspace{0.05in}
\STATE{\bf Output:} equivalence class $\optimec(\hat{\dagg}_\text{opt})$ using Definition~\ref{defn:equiv_class}
\end{algorithmic} \label{algo:equiv_class}
\end{algorithm}
\noindent{\emph Remark 4:} The guarantees of Corollary~\ref{cor:incoherent} can be extended to Algorithm~\ref{algo:equiv_class}. Specifically, suppose the conditions of this corollary hold and the candidate set of DAGs $\dagg_{\text{cand}}$ contains a member of the population interventional equivalence class, i.e. $\dagg_{\text{cand}} \cap \simec({\dagg}^\star) \neq \emptyset$. Furthermore, suppose that the alternating minimization technique in Algorithm~\ref{algo:dag_score} obtains a globally optimal solution. Then, we show in Appendix~\ref{sec:consistency_algorithm_1} that in the infinite data limit, the output of Algorithm~\ref{algo:equiv_class} is consistent, i.e. $\hat{\cc{I}}_\mathrm{opt} = \cc{I}^\star$ and $\optimec(\hat{\dagg}_\text{opt}) = \simec({\dagg}^\star)$ with probability tending to one.

\noindent\textbf{Selecting $(\lambda,\gamma,\bar{h})$ via cross-validation:} In \suppmat{} Section~\ref{sec:supp_mat_cross_validation}, we propose an approach to exhaustively search over the equivalence classes indexed by $(\lambda,\gamma,\bar{h})$, and choose an optimal one based on validation with test data. The complexity of our validation approach is $\approx \tilde{t}((\bar{h}_\text{max}+1)|{\dagg}_{\text{cand}}| + p)$, where $\bar{h}_\text{max}$ is the maximum number of latent variables allowed in the model and $\tilde{t}$ represents the time it takes to score a DAG using Algorithm~\ref{algo:dag_score}. The value for $\tilde{t}$ depends on the DAG and the data generating mechanism; in our numerical experiments, this is typically on the order of $5$ seconds for $p = 20$ node graphs.

\subsection{Using \methName{} to remove spurious edges from `starting point' DAGs}
\label{sec:alg_starting_point}
We next consider settings where the DAGs in a candidate set are viewed as `starting points' and may contain spurious dependencies due to potential latent confounding. Such scenarios naturally arise in practice. For example, a domain expert may be unsure about some of the edges in a DAG and may include them in the analysis to be conservative. In other contexts, the user may have deployed their favorite structure learning algorithm(s) to obtain a set of DAGs. Since many of the computationally efficient structural learning approaches (e.g. GES) do not account for the presence of latent variables, the fitted graph may be more dense than the population DAGs. 

Our objective, in contexts where DAGs are viewed as starting points, is to use \methName{} to remove spurious dependencies and return a refined set of equally scoring DAGs. Our approach is based on the following simple observation: scoring starting point DAGs using Algorithm~\ref{algo:dag_score} may yield connectivity matrices that are more dense than the population connectivity matrix, although the magnitude of the spurious edges will be small. To remove these spurious edges, for each DAG in the candidate set, we greedily delete the weakest edge and compute a regularized likelihood score using Algorithm~\ref{algo:dag_score}. We repeat this process until the score can no longer be improved. After pruning, we obtain a refined collection of candidate DAGs, which are then supplied to Algorithm~\ref{algo:equiv_class} to identify an optimally scoring set of DAGs. A summary of the entire procedure is presented in Algorithm~\ref{algo:structure_learning}. Similar to Algorithm~\ref{algo:equiv_class}, in all our numerical experiments, we choose the regularization parameters $(\lambda,\gamma,\bar{h})$ via cross-validation; see \suppmat{} Section~\ref{sec:supp_mat_cross_validation}. The complexity of our exhaustive validation approach is $\approx \tilde{t}((\bar{h}_\text{max}+1)[\sum_{\dagg \in \tilde{\dagg}_{\text{cand}}} \|\dagg\|_{\ell_0}]+ p)$.

\begin{algorithm}
\caption{{{}Improving `starting point' DAGs via \methName{}}}
\begin{algorithmic}[1]
\vspace{0.1in}
\STATE {\bf Input}: data $\hat{\Sigma}_e,\hat{\pi}_e$ for $e \in [m]$; candidate DAG(s) $\tilde{\dagg}_\text{cand}$; regularization params. $\lambda,\gamma \geq 0$; $\#$ of latent vars. $\bar{h}$
\vspace{.05in}
\STATE{\bf Backward deletion to remove spurious edges}: initialize $\dagg_\text{cand} = \emptyset$; for each ${\dagg} \in \tilde{\dagg}_\text{cand}$:
\begin{enumerate}
    \item[(a)] supply data and $(\dagg,{\cc{I}} = [p])$ to Algorithm~\ref{algo:dag_score} to find score $\texttt{score}_{\lambda,\gamma}({\dagg},\hat{\Theta}(\dagg,\bar{h}))$
    \item[(b)] let ${\dagg}$ be the DAG after deleting the smallest edge in magnitude in ${\dagg}$
    \item[(c)] repeat (a-b) until the likelihood score does not improve;  add $\dagg$ to $\dagg_\text{cand}$
\end{enumerate}
\vspace{0.05in}
\STATE{\bf Output}: supply $\dagg_\text{cand}$ to Algorithm~\ref{algo:equiv_class} to find best scoring DAGs 
\end{algorithmic} \label{algo:structure_learning}
\end{algorithm}
\noindent{\emph Remark 5:} As with Remark 4, the guarantees of Corollary~\ref{cor:incoherent} can be extended to Algorithm~\ref{algo:structure_learning}. In particular, if the candidate set of DAGs $\tilde{\dagg}_{\text{cand}}$ contains a DAG that is a supergraph of a DAG in the population interventional equivalence class, then, we show in Appendix \ref{sec:consistency_algorithm_3} that in the infinite data limit, the output of Algorithm~\ref{algo:structure_learning} is consistent, i.e.  $\hat{\cc{I}}_\mathrm{opt} = \cc{I}^\star$ and $\optimec(\hat{\dagg}_\text{opt}) = \simec({\dagg}^\star)$ with probability tending to one.

%% file: sections/experiments.tex
\section{Synthetic and real experiments}
Code to reproduce all the experiments can be found here: \url{https://github.com/juangamella/ut-lvce-paper}.
\label{sec:experiments}
\subsection{Synthetic experiments: recovering the interventional equivalence class}
\label{sec:experiments_recovering}
{\bf{Setup}:} We consider a collection of $p = 20$ observed variables influenced by $h=2$ latent variables. The entries of the latent coefficient matrix $\Gamma^\star \in \mathbb{R}^{p \times h}$ are generated IID from the distribution $\mathcal{N}(0,1/\sqrt{p})$. The noise term $\epsilon_i$ for each coordinate $i$ is distributed according to a zero mean Gaussian with variance chosen uniformly and independently from the interval $[0.5,0.6]$. We suppose there are $m = 5$ environments, an observation environment $e = 1$, and four interventional environments $e \in \{2,3,4,5\}$. For the observational environment, $\delta^e$ is all zeros and for the interventional environments and every $i \in \cc{I}^\star$, $\delta^{e}_i$ is a zero mean Gaussian distribution whose variance will be specified later. Similarly, the distribution of each latent variable is taken to be $\mathcal{H}^e_i \sim \mathcal{N}(0,\text{Unif}[0.2,0.3]+\zeta^e_i)$ for every $i \in [h]$, where $\zeta^{e=1}_i = 0$ and is otherwise chosen uniformly and independently from the interval $[0.2,1]$. The population connectivity matrix $B^\star$, the choice of intervention targets $\cc{I}^\star$, and the amount of data in every environment are specified later. In \suppmat{} Section~\ref{sec:additional_synthetic_robustness}, we provide additional experiments for the following settings: weaker interventions on observed variables and stronger latent effects. Finally, in \suppmat{} Section~\ref{sec:additional_synthetic_robustness_number_denseness}, we illustrate the performance of our method with a varying number of latent variables ($h =3,4,10$).

{\bf{Metrics to assess the quality of an estimated equivalence class}:} To quantify the `closeness' of an estimated interventional equivalence class $\himec(\hat{\dagg})$ to the population interventional equivalence class $\simec({\dagg}^\star)$, we use the following two metrics:
\begin{equation}
    \begin{aligned}
        \mathrm{FDP}&:  \max_{\dagg_2 \in \himec(\hat{\dagg})} \min_{\dagg_1 \in \simec({\dagg}^\star)}[\# \text{ edges in }\dagg_2 \text{ not in }\dagg_1]/[\#\text{ edges in }\dagg_2], \\
        \mathrm{TDP}&:  \min_{\dagg_1 \in \simec({\dagg}^\star)} \max_{\dagg_2 \in \himec(\hat{\dagg})}[\# \text{ edges in }\dagg_1 \text{ and in }\dagg_2]/[\#\text{ edges in }\dagg_1].
    \end{aligned}
    \label{eqn:metrics}
\end{equation}  
The metric $\mathrm{FDP}$ is akin to false discovery proportion and measures the ratio of spurious edges contained in the DAGs of the estimated interventional equivalence class. The metric $\mathrm{TDP}$ is akin to true discovery proportion and measures the proportion of true edges (in the DAGs of the population interventional equivalence class) that are also DAGs in the estimated interventional equivalence class. It is straightforward to check that $\himec(\hat{\dagg}) = \simec({\dagg}^\star)$ if and only if $\mathrm{FDP} = 0$ and $\mathrm{TDP} = 1$. 

\subsubsection{\methName{} with specified input DAGs}

\begin{figure}
    \centering
       \begin{subfigure}[b]{0.5\textwidth}
    \centering
        \includegraphics[scale = 0.58]{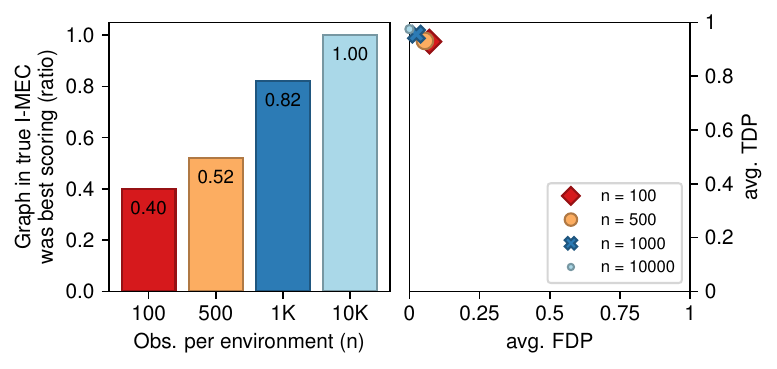}
        \caption{Equivalence class of best scoring DAGs\\ using Algorithm~\ref{algo:equiv_class}}
        \end{subfigure}
        \begin{subfigure}[b]{0.4\textwidth}
            \centering
        \includegraphics[scale = 0.58]{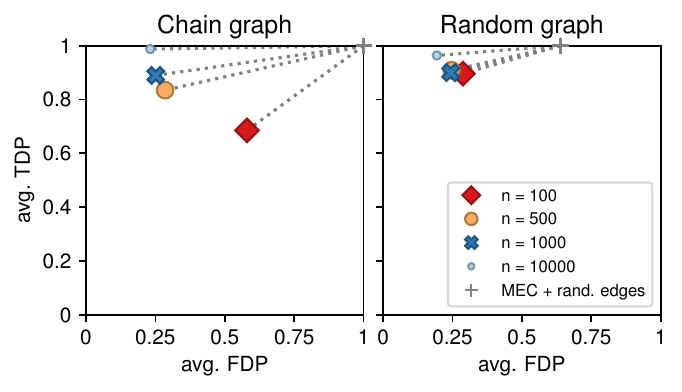}
        \caption{Removing spurious edges \\using  Algorithm~\ref{algo:structure_learning}}
        \end{subfigure}
    \caption{\small{a) proportion of instances, among $50$ independent trials, that the best scoring DAG in Step 3 of Algorithm~\ref{algo:equiv_class} is in $\simec({\dagg}^\star)$ and average $\mathrm{FDP}$ and $\mathrm{TDP}$ of the estimated $\optimec(\hat{\dagg}_\text{opt})$; b) performance of Algorithm~\ref{algo:structure_learning} with the `starting point' DAGs $\mathrm{MEC}(\dagg^\star) + \text{20 random edges}$.}} 
        \label{fig:equivalence_class_recovery}
\end{figure}

{\bf{\methName{} with a candidate set of DAGs}:} We generate the population DAG as follows: we first generate an Erd\"os-Renyi graph with edge probability $0.11$ and then we orient the edges according to a random total ordering of the variables. The edge strengths are drawn uniformly at random from the interval $[0.5,0.7]$. Let $\cc{I}^\star$ be ten indices chosen uniformly at random. The variance of the perturbations on the observed variables $\delta_i^e$ is taken uniformly and independently from the interval $[6,12]$. The candidate set of DAGs is taken to be the Markov equivalence class of $\dagg^\star$, which by definition is a superset of the interventional equivalence class $\simec({\dagg}^\star)$. We generate $n$ observations for each environment, where $n$ is chosen from the set $\{100,500,1000,10000\}$. To illustrate the effectiveness of the scoring function, we supply the data and each candidate DAG, with $\cc{I} = [p]$ to Algorithm~\ref{algo:dag_score}. The left plot in Figure~\ref{fig:equivalence_class_recovery}a displays the proportion of instances, across $50$ independent trials, that the best scoring DAG is inside $\simec({\dagg}^\star)$. We observe that as the sample size increases, the scoring function becomes more accurate and correctly outputs a member of the interventional equivalence class. In order to obtain an equivalence class of best scoring DAGs, we apply Algorithm~\ref{algo:equiv_class}. As shown in the right plot in Figure~\ref{fig:equivalence_class_recovery}a, the estimated equivalence class is close to the true equivalence class, even when $n = 100$. 

{\bf{\methName{} with `starting point' DAGs}:} We consider the setting described above and generate two different population DAGs $\dagg^\star$: a chain graph and an Erd{\"o}s-Renyi graph with edge probability $0.11$, with the edge strengths of each graph drawn uniformly at random from the interval $[0.5,0.7]$. In each case, the `starting point' DAGs are taken to be ones in $\mathrm{MEC}(\dagg^\star)$ with $20$ edges added at random to each graph (picked uniformly, without replacement, from all valid edge additions). We supply the data and the starting point DAGs to Algorithm~\ref{algo:structure_learning}. Ideally, Algorithm~\ref{algo:structure_learning} removes spurious edges and improves upon the original set of DAGs to identify a class of best-scoring DAGs that is close to the population interventional equivalence class. Figure~\ref{fig:equivalence_class_recovery}b confirms this to be the case. In particular, we observe that the estimated interventional equivalence class, averaged across $50$ independent trials, has a small average $\mathrm{FDP}$ and a large average $\mathrm{TDP}$. Furthermore, we see that as compared to the `starting point' DAGs, Algorithm~\ref{algo:structure_learning} produces an estimate with substantially smaller false discoveries without much loss in power.

\subsubsection{\methName{} as a structure learning procedure and comparisons to other methods}
\label{sec:experiments_comparison}
A set of input DAGs may not be available {a priori} and must be learned from data. Thus, we use GES to obtain a collection of DAGs, although, in principle, any structural learning algorithm may be deployed. Since GES does not account for latent confounding, its output DAGs are typically dense and contain many spurious edges. Thus, as prescribed in Section~\ref{sec:alg_starting_point}, we apply Algorithm~\ref{algo:structure_learning} to prune GES DAGs and return an interventional equivalence class. We compare the performance of our algorithm to three causal learning methods that account for latent effects: causal Dantzig \citep{Dantzig}, backShift \citep{backshift}, and LRpS-GES \citep{marloes}. We note here that the first two methods exploit interventional data while LRpS-GES only operates with observational data. Furthermore, causal Dantzig performs local structural learning around a target variable of interest while the other two methods yield a causal model over the entire graph. Throughout,  we consider the synthetic setup at the beginning of this section and generate $50$ Erd{\"o}s-Renyi DAGs with edge probability $0.11$ and edge strengths drawn uniformly at random from the interval $[0.5,0.7]$; we illustrate the robustness of our method to varying graph sparsity and varying number of latent variables in \suppmat{} Section~\ref{sec:additional_synthetic_robustness_number_denseness}. Furthermore, the magnitude of the perturbations $\xi_i^e$ are taken uniformly and independently from the interval $[3,6]$. Finally, we generate $n$ observations for each environment where $n$ is chosen from the set $\{100,500,1000\}$.

\begin{figure}
    \centering
        \includegraphics[scale = 0.53]{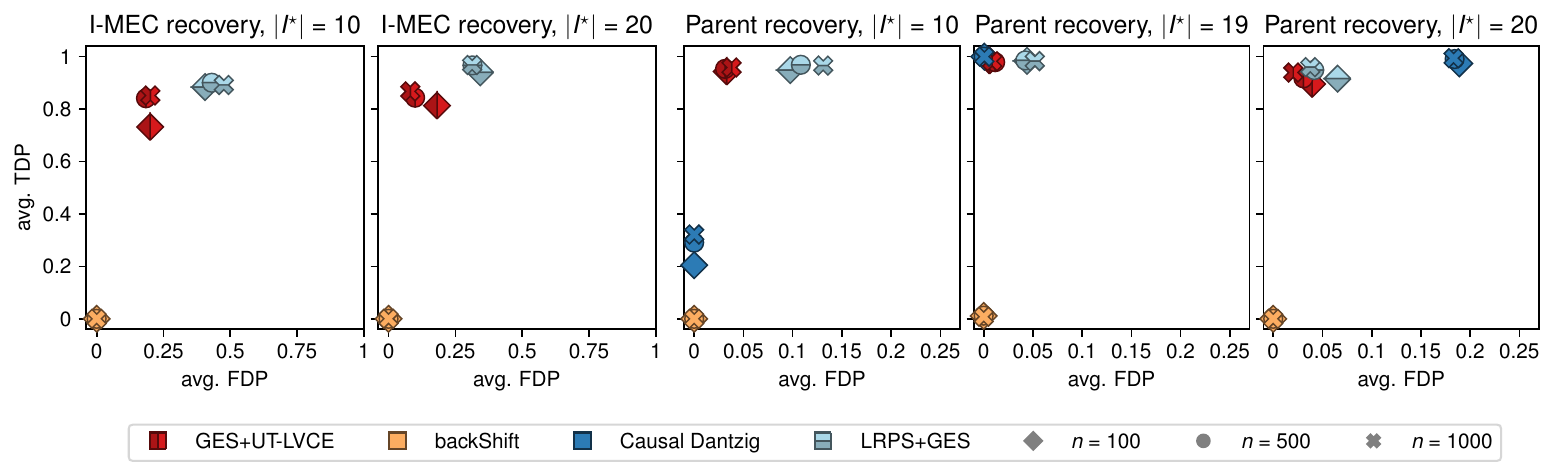}
    \caption{\small{Performance of Algorithm~\ref{algo:structure_learning} with GES starting DAGs and comparisons with other methods for different size of intervention targets $|\cc{I}^\star|$; the left two plots are performances over the entire graph and the right three graphs are performances locally for a target variable. Here, for assessing the quality of local structure recovery, we use the metrics in \eqref{eqn:metrics} restricted to the parental set(s) of the target variable.}} 
        \label{fig:comparisons}
\end{figure}

{\bf{Evaluating performance over the entire graph}:} We consider two settings: $|\cc{I}^\star| \in \{10,20\}$.  Figure~\ref{fig:comparisons} shows the average $\mathrm{FDP}$ and $\mathrm{TDP}$, averaged across all the $50$ DAGs and $10$ runs for each DAG, for the outputs of \methName{}, backShift and LRpS-GES. As observed in Figure~\ref{fig:comparisons}, \methName{} yields an estimated equivalence class with small average $\mathrm{FDP}$ and a large $\mathrm{TDP}$, 
and performs more favorably compared to the other methods, especially in the setting with partial interventions. We also observe that LRpS-GES produces substantially larger false discoveries as it does not exploit interventional data for improved identifiability, and that backShift yields poor estimates since there are interventions on the latent variables. We note that the performance of \methName{} is naturally affected by the `goodness' of the input GES DAGs. In particular, we show in \suppmat{} Section \ref{sec:additional_synthetic_GES_analysis} that if any of the GES DAGs is a supergraph of a DAG in $\simec({\dagg}^\star)$, the performance of \methName{} substantially improves. Finally, as announced earlier, our procedure \methName{} can take as input DAGs produced by any structural learning algorithm. As an example, the user may take the DAGs obtained by GES as well as those from backShift and LRpS-GES as input to Algorithm~\ref{algo:structure_learning}, although we do not explore this in our experiments. 

{\bf{Evaluating local structure recovery}:} We consider a similar setting as above and let $X_p$ be the target variable of interest. When generating our Erd{\"o}s-Renyi DAGs, we discard DAGs where the target variable has less than two parents and obtain a total of $50$ DAGs. We additionally consider the setting with $|\cc{I}^\star| = 19$ where all but the target variable has received an intervention. We observe that for $|\cc{I}^\star| = 10$, Causal Dantzig has very low power, which can be attributed to yielding a single estimate even if the parental set is unidentifiable and requiring interventions on all but the target variable for consistency. When $|\cc{I}^\star| = 19$, Causal Dantzig obtains accurate recovery of the parental set even though there are perturbations on the latent variables. In \suppmat{} Section \ref{sec:additional_synthetic_causal_dantzig}, we observe that when the magnitude of the latent perturbations are made to be stronger, Causal Dantzig performs poorly as compared to \methName{}. Finally, when $|\cc{I}^\star| = 20$, Causal Dantzig yields inaccurate estimates since there are perturbations on the target variable of interest.

\subsection{Analysis on real data}
\label{sec:real}
\subsubsection{Protein expressions}
 We next analyze the protein mass spectroscopy dataset \citep{sachs}. This dataset (downloaded from \url{https://www.bnlearn.com/research/sachs05/index.html}) contains a large number of measurements of the abundance of $11$ phosphoproteins and phospholipids recorded under different experimental conditions in primary human immune system cells. The different experimental conditions are characterized by associated reagents that inhibit or activate signaling nodes, corresponding to interventions at different points of the protein-signaling network. Following the previous works \citep{mooij,nicolai_pnas}, we take 8 environments consisting of an observational environment and 7 interventional environments.  
 
Multiple papers have applied their structural learning algorithm to identify the causal relationships among the $11$ proteins \citep{sachs,Eaton,mooij,nicolai_pnas}. Each proposed method returns a DAG (potentially multiple due to non-identifiability) with some commonalities among the output structures, but also many differences. Naturally, the following questions arise: i) how well does each DAG fit the data, and which one is most representative of the data? ii) are there other DAGs in the equivalence class of the best scoring DAG that fit the data equally well? and iii) can any spurious edges in the DAGs be removed to obtain a better fit to data? As outlined next, our procedure \methName{} is useful for addressing these questions.

\begin{figure}
    \centering
    \includegraphics[scale = 0.64]{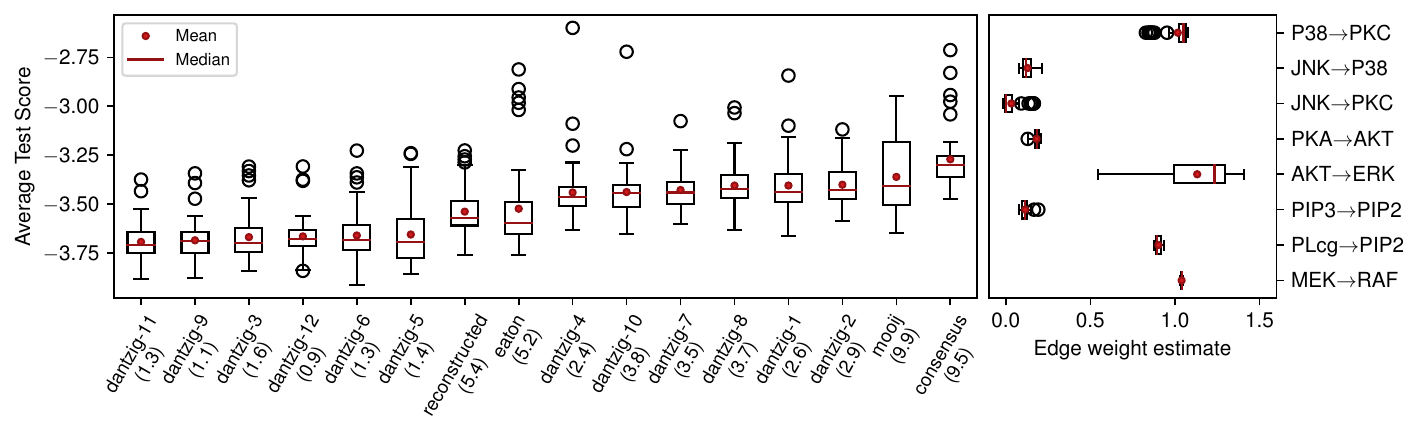}
    \caption{\small{Performance of Algorithms~\ref{algo:equiv_class} and \ref{algo:structure_learning} on the protein expressions dataset with $50$ random splits of the data: a) test score of each DAG produced in the literature using Algorithm~\ref{algo:equiv_class} and the average edges pruned (shown below for each DAG) using Algorithm~\ref{algo:structure_learning}; b) edge strengths of the best scoring DAG ``Causal Dantzig 11".}}
        \label{fig:sachs}
\end{figure}

{\bf Best scoring among reported DAGs in the literature:} We use Algorithm~\ref{algo:equiv_class} to score the DAGs obtained by previous methods. We keep $0.05\%$ of the data for computing test performance. Of the remaining $0.95\%$ of the data, we take $70\%$ for training and the remaining $30\%$ for validation. The number of latent variables $\bar{h}$ and the regularization parameters $\lambda,\gamma$ are selected via holdout validation. We obtain a causal model associated to each DAG and evaluate the corresponding negative log-likelihood score on the test set. For reproducibility, we repeat this experiment with $50$ different random splits of training/validation datasets. Figure~\ref{fig:sachs}a presents the box-plot of the test scores for each DAG. A number of remarks are in order. First, the top three best scoring DAGs (displayed in \suppmat{} Section~\ref{sec:sachs_best_scoring}) are produced by a method that accounts for latent variables \citep{nicolai_pnas}; the other structural learning procedures assume all relevant variables are observed. Related to the previous point, we find that there are strong latent effects on the protein network. As an example, for the best scoring DAG, our algorithm finds on average $1.16$ latent variables. Furthermore, for the top three scoring DAGs, we also find that many of the proteins have received a strong perturbation; this is likely due to off-target effects that were also reported in \cite{Eaton}. The presence of interventions on many of the variables implies that the equivalence class of all of these top three DAGs are singletons. Finally, in Figure~\ref{fig:sachs}b, we present a boxplot of the edge strengths for the top scoring DAG.

{\bf Removing spurious edges:} We next explore whether any spurious edges can be removed from these DAGs. To that end, we apply Algorithm~\ref{algo:structure_learning} to each DAG. We observe that the top scoring DAGs produced by \cite{nicolai_pnas} are rather stable as compared to the other DAGs, with $\approx 1$ edges removed on average by our procedure. This is consistent with the fact that \cite{nicolai_pnas} accounts for latent confounding and thus is likely to contain fewer spurious edges. We note that for the best scoring DAG, the edge that is removed most often is JNK $\rightarrow$ PKC; indeed, this edge has weak strength (see Figure~\ref{fig:sachs}b) and has not been reported in any other DAG in the literature.  

\subsubsection{California reservoirs}
The California reservoir network consists of $\approx{1530}$ reservoirs that act as buffers against severe drought conditions and are a major source of water for agricultural use, hydropower generation, and industrial use. Water managers of these reservoirs have to assess the likelihood of system-wide failure and the effectiveness of potential policies. Due to similarities in hydrological attributes (e.g. altitude, drainage area, spatial location), the reservoir network is highly interconnected. Thus, effective reservoir management requires an understanding of reservoir interdependencies. \cite{water} used historical data of volumes of the largest 55 reservoirs to obtain an undirected graphical model of the California reservoir network. This previous analysis, however, does not provide causal implications: namely, how change in the management of one reservoir (i.e. an intervention) affects the entire system. To that end, we explore the utility of \methName{} for learning causal relationships among the reservoirs. 

We consider the $10$ largest reservoirs (with respect to capacity) in California, where daily volume data (downloaded from \url{https://github.com/armeentaeb/WRR-Reservoir}) are available during the period of study (January 2003–December 2015). Following the preprocessing steps in \cite{water}, we average the data from daily down to 156 monthly observations. A seasonal adjustment step is performed to remove predictable seasonal patterns. The resulting data was demonstrated in \cite{water} to be well-approximated by a multivariate Gaussian distribution.

The reservoir data is not IID as its distribution varies depending on the severity of the drought. In particular, during a drought period, a reservoir manager may decide to reduce the outflow of water, and thus effectively decrease the variability in the reservoir volume; this is in contrast to a wet period where more outflow is allowed, as the reservoir is expected to be replenished. Based on the intuition described above, we organize our reservoir data into four `environments' or time-blocks based on the severity of the drought conditions: an environment during a normal period (2003-2006, 2010-2012) with no drought conditions, an environment associated to an abnormally dry period (2007, 2013), an environment associated to a moderate drought period (2008-2009), and an environment associated to a severe drought period (2014-2015). 

\begin{figure}
    \centering
    \begin{subfigure}[b]{0.4\textwidth}
        \includegraphics[scale = 0.47]{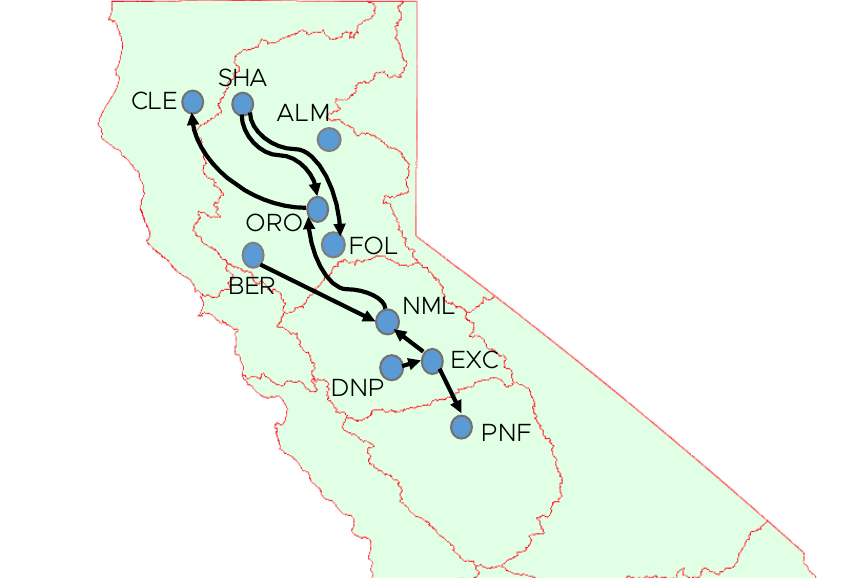}
    \caption{Output of Algorithm~\ref{algo:structure_learning}}
    \end{subfigure}
\begin{subfigure}[b]{0.5\textwidth}
\includegraphics[scale = 0.47]{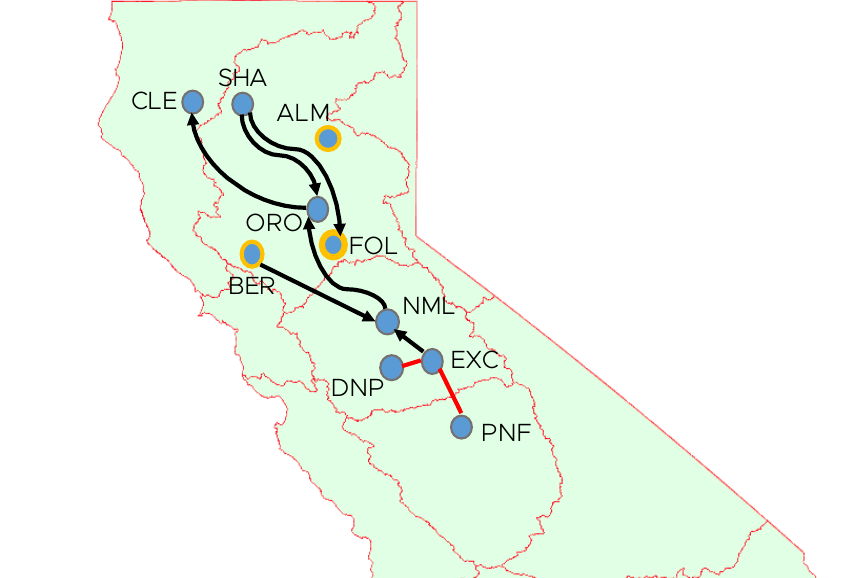}
\caption{`conservative' equivalence class{ (see text)}}
\end{subfigure}
    \caption{\small{Graphical structure of the 10 largest reservoirs in California (with respect to capacity): a) equivalence class consisting of a unique DAG obtained by Algorithm~\ref{algo:structure_learning};
    b) conservative equivalence class consisting of multiple DAGs (obtained from directing the red edges) after identifying strong interventions (shown in orange) from the output of Algorithm~\ref{algo:structure_learning}.}} 
        \label{fig:reservoirs}
\end{figure}

Unlike the protein expression dataset, no candidate DAGs are available a priori for the reservoir dataset. Thus, we employ GES on the first environment (normal period) to obtain a collection of `starting point' DAGs.  These DAGs are then supplied to Algorithm~\ref{algo:structure_learning}, where the number of latent variables $\bar{h}$ as well as the regularization parameters $\lambda,\gamma$ are selected via holdout validation with a $(70\%,30\%)$ training and validation set split for $10$ different random splits. For each split, we obtain a causal model and a corresponding equivalence class and then choose the model that obtains the best likelihood score on the overall data. The optimally scoring model consists of two latent variables ($\bar{h} = 2$) and an interventional equivalence class presented in Figure~\ref{fig:reservoirs}a. The connections in the learned DAG are between pairs of reservoirs with at least one of these commonalities: i) similar hydrological attributes (e.g. hydrological zone and elevation) and ii) coordinated management by a district or a state-wide project. For example, the reservoirs New Melones (NML), Don Pedro (NP), New Exchequer (EXC), and Pine Flat (PNF) are all in the San Joaquin district. Further, Shasta (SHA), Trinity (CLE), Oroville (ORO) and Folsom (FOL) are in the network of Central Valley and State Water projects and their reservoir operations are coordinated. 

We next analyze the estimated locations and magnitudes of the perturbations. Recall that the locations are encoded in the estimate $\hat{\cc{I}}_\text{opt}$ and the strength of the interventions are computed via the metric $\xi_j$ for every $j \in \hat{\cc{I}}_\text{opt}$ (see Section~\ref{sec:alg_candidate_dags}). Our model identifies perturbations on all reservoirs except Pine Flat. The strongest estimated intervention is on Lake Almanor, which is consistent with the fact that during the 2014-2015 drought period, there was little to no outflow of water in this reservoir. Finally, aside from the reservoirs \{`ALM', `BER', `FOL'\}, the intervention strengths on the remaining reservoirs are rather small (i.e. below the level $\xi_j \leq 0.01$). However, likely due to the small sample size, these reservoirs were included in the list of intervention targets after validation. Note that overestimating the list of intervention targets may lead to discarding plausible causal mechanisms, as more identifiability is claimed than present in the data. To remain `conservative', in Figure~\ref{fig:reservoirs}b, we present the interventional Markov equivalence class $\optimec({\hat{\dagg}}_\text{opt})$, where $\hat{\cc{I}}_\text{opt} = \text{\{`ALM', `BER', `FOL'\}}$ and $\hat{\dagg}_\text{opt}$ is the DAG in Figure~\ref{fig:reservoirs}a. The resulting structure highlights that certain edges (shown in red) may not be identifiable. 




%% file: sections/conclusion.tex
\section{Discussion and Future Work}
\label{section:discussion}
In this paper, we proposed a framework to model {unspecific} perturbation data among a collection of observed and latent variables. This framework allows for perturbations on all components of the system, including a response variable of interest or the latent variables. Further, we presented an algorithm \methName{} to fit DAGs to this model and obtain an equivalence class of DAGs that best explains the data. There are several interesting directions for further investigation that arise from our work. In Section~\ref{sec:alg_starting_point}, we discussed the setting where no DAGs are available a-priori and proposed using any structural learning algorithm to obtain a set of `starting point' DAGs; these are then subsequently pruned by Algorithm~\ref{algo:structure_learning} to arrive at an estimate for the interventional equivalence class. While the empirical results in Section~\ref{sec:experiments} support the utility of our heuristics, there is much room for more rigorous optimization techniques to search over the space of equivalent DAGs with respect to the scoring function \eqref{eqn:score} (e.g. provably consistent greedy methods). Further, the perturbation model \eqref{eq:SCM_env} assumes a linear relationship between the observed and latent variables. It would be of practical interest to explore extensions of our framework to non-linear settings, or alternatively, characterize the extent to which linear models capture the causal effects.

 \section*{Acknowledgements}
 AT received funding from the Royalty Research Fund at the University of Washington. JG and PB received funding from the European Research Council (ERC) under the
European Union’s Horizon 2020 research and innovation programme (grant agreement No.
786461).

%% file: sections/supp_mat.tex
\section*{Supplementary Material}
\section{Notations}
For a matrix $M \in \mathbb{R}^{d_1 \times d_2}$, we denote $\|M\|_2$ to be largest singular value. For a symmetric matrix $M \in \mathbb{S}^d$, we denote $\lambda_\text{max}(M)$ and $\lambda_\text{min}(M)$ to be the maximum and minimum eigenvalue of M, respectively. For a symmetric matrix $M\in\mathbb{S}^d$, we denote $\degree(M)$ to be the maximum number of nonzero elements in any row (or equivalently column) of the matrix. Recall that for an undirected graph $\mathcal{G}$, we denote $\degree[\mathcal{G}]$ to be the maximal degree of $\mathcal{G}$. 

\section{Proof of Theorem~\ref{thm:IMEC}}
\label{sec:supp_thm_equiv}
The proof of this theorem relies on some lemmas:
\begin{lemma}[A property of $(\id-\tilde{B})^{-1}(\id-B)^{-1}$ \citep{GamellaGnies2021}]Let $B,\tilde{B} \in \mathbb{R}^{p \times p}$ be two matrices that can be made to be lower-triangular with zeros on the diagonal after row and column permutations (or equivalently, the matrices correspond to two DAGs). If for $\tilde{\Omega},\Omega \in \mathbb{D}_{++}^p$, $(\id-B)^{-1}\Omega(\id-B)^{-T} = (\id-\tilde{B})^{-1}\tilde{\Omega}(\id-\tilde{B})^{-T}$, then the following statements are equivalent:
\begin{enumerate}
    \item[p1)] $\mathrm{support}(\tilde{B}_{i,:}) = \mathrm{support}(B_{i,:})$.
    \item[p2)] $[(\id-\tilde{B})(\id-B)^{-1}]_{i,:} = {\bf e}_i^T$.
    \item[p3)] $[(\id-\tilde{B})(\id-B)^{-1}]_{:,i} = {\bf e}_i$.
\end{enumerate}
\label{lemma:structure}
\end{lemma}
\begin{proof}
For completeness, we include the proof in \cite{GamellaGnies2021}. We break down the proof in the following steps:

\noindent\underline{$p1 \Leftrightarrow p2$}: The direction $p2 \Rightarrow p1$ follows immediately. For the direction $p1 \Rightarrow p2$, define the variables $\bar{X}$ via the following structural equation model
\begin{equation}
    \bar{X} = B\bar{X} + \epsilon~~,~~ \epsilon \sim \mathcal{N}(0,\Omega).
    \label{eqn:1}
\end{equation}
Since $(\id-B)^{-1}\Omega(\id-B)^{-T} = (\id-\tilde{B})^{-1}\tilde{\Omega}(\id-\tilde{B})^{-T}$, we have that equivalently:
\begin{equation}
    \bar{X} = \tilde{B}\bar{X}+\tilde{\epsilon}~~,~~\tilde{\epsilon}\sim \mathcal{N}(0,\tilde{\Omega}).
        \label{eqn:2}
\end{equation}
Let $S = \text{support}(B_{i,:})= \text{support}(\tilde{B}_{i,:})$. Then, equations \eqref{eqn:1} and \eqref{eqn:2} imply that $B_{i,:}$ and $\tilde{B}_{i,:}$ are both the regression coefficients from regressing $\bar{X}_S$ onto $\bar{X}_i$. Thus, $\tilde{B}_{i,:} = B_{i,:}$.

\noindent\underline{$p2 \Rightarrow p3$} The property $(\id-B)^{-1}\Omega(\id-B)^{-T} = (\id-\tilde{B})^{-1}\tilde{\Omega}(\id-\tilde{B})^{-T}$ means that $(\id-\tilde{B})(\id-B)^{-1}$ has orthogonal row vectors. Since ${\bf e}_i^T(\id-\tilde{B})(\id-B)^{-1} = {\bf e}_i^T$, it then follows that $[(\id-\tilde{B})(\id-B)^{-1}]_{:,i} = {\bf e}_i$. 

\noindent\underline{$p3 \Rightarrow p1$} Notice that:
\begin{equation*}
    (\id-\tilde{B})(\id-B)^{-1}\Omega(\id-B)^{-T}(\id-\tilde{B})^T = \tilde{\Omega},
\end{equation*}
which can be rewritten as:
\begin{equation}
    (\id-\tilde{B}) = \tilde{\Omega}(\id-\tilde{B})^{-T}(\id-B)^T\Omega^{-1}(\id-B).
    \label{eqn:3}
\end{equation}
Let $\Omega_{i,i}^{-1} = \alpha$ and $\tilde{\Omega}_{i,i} = \beta$, where $\alpha,\beta \neq 0$. Property $p2$ implies that  ${\bf e}_i^T(\id-\tilde{B})^{-T}(\id-B)^T = {\bf e}_i^T$. Thus, relation \eqref{eqn:3} lets us conclude that:
\begin{equation*}
     {\bf e}_i^T(\id-\tilde{B}) = \alpha\beta{{\bf e}}_i^T(\id-B),
\end{equation*}
and thus property $p1$. 
\end{proof}
\begin{lemma}[equivalent covariance models without latent variables] Consider an SCM \eqref{eq:SCM_env} with structure given by a DAG $\dagg$, with no latent confounders (i.e. $h = 0$), connectivity matrix $B$, noise variances $\{\Omega_e\}_{e=1}^m$ and intervention targets $\cc{I}$. For any DAG $\tilde{\dagg} \in \imec(\dagg)$, there exists a connectivity matrix $\tilde{B} \sim \tilde{\dagg}$ and noise variances $\{\tilde{\Omega}_e\}_{e=1}^m \subseteq \mathbb{D}_{++}^p$ such that $(\id-B)^{-1}\Omega_e(\id-B)^{-T} =(\id-\tilde{B})^{-1}\tilde{\Omega}_e(\id-\tilde{B})^{-T}$ for every $e = [m]$. Furthermore, $\mathbb{I}(\{\tilde{\Omega}_e\}_{e=1}^m) = \cc{I}$. 
\label{lemma:equiv_without_latent}
\end{lemma}
\begin{proof}
Decompose $\Omega_e$ as: $\Omega_e = \Omega_0 + \Delta\Omega_e$ where $\Omega_0 \in \mathbb{D}_{++}^p$ and $[\Delta\Omega_e]_{i,i} = 0$ for all $e = 1,2,\dots,m$ and $i \not\in \cc{I}$. Note that for any $\tilde{\dagg} \in \imec(\dagg)$, there exists a $\tilde{B}\sim\tilde{\dagg}$ and $\tilde{\Omega}_0 \in \mathbb{D}_{++}^p$ such that $(\id-B)^{-1}\Omega_0(\id-B)^{-T} = (\id-\tilde{B})^{-1} \tilde{\Omega}_0(\id-\tilde{B})^{-T}$. Since $\text{support}(\tilde{B}_{i,:}) = \text{support}({B}_{i,:})$ for all $i \in \cc{I}$, we have by Lemma~\ref{lemma:structure} that $[(\id-\tilde{B})(\id-B)^{-1}]_{:,i} = e_i$ and $[(\id-\tilde{B})(\id-B)^{-1}]_{i,:} = e_i^T$ for all $i\in\cc{I}$. Then, set $\Delta\tilde{\Omega}_e = \Delta{\Omega}_e$ and note that $(\id-\tilde{B})(\id-B)^{-1}\Delta{\Omega}_e(\id-B)^{-T}(\id-\tilde{B})^T = \Delta\tilde{\Omega}_e$. Defining $\tilde{\Omega}_e = \tilde{\Omega}_0 + \Delta\tilde{\Omega}_e$, we have that $(\id-{B})^{-1}\Omega_e(\id-B)^{-T} = (\id-\tilde{B})^{-1}\tilde{\Omega}_e(\id-\tilde{B})^{-T}$ for all $e = 1,2,\dots,m$ as desired. Furthermore, since $\tilde{\Omega}_e = (\id-\tilde{B})(\id-B)^{-1}\Omega_e(\id-B)^{-T}(\id-\tilde{B})^T$ and $\Omega_e$ is positive-definite matrix, we conclude that $\tilde{\Omega}_e$ is positive-definite. Finally, the property $\mathbb{I}(\{\tilde{\Omega}_e\}_{e=1}^m)=\cc{I}$ follows by construction. 
\end{proof}
\begin{lemma}[Sufficient condition for equally scoring parameters sets]Let ${\Theta}= ({B},{\Gamma},\cc{I},\{{\Omega}_e,{\Psi}_e\}_{e=1}^m)$ and 
$\tilde{\Theta}= (\tilde{B},\tilde{\Gamma},\allowbreak\tilde{\cc{I}},\{\tilde{\Omega}_e,\tilde{\Psi}_e\}_{e=1}^m)$ be a set of parameters that satisfy for every $e = 1,2,\dots,m$:
\begin{equation*}
    (\id-B)^{-1}(\Omega_e + \Gamma\Psi_e\Gamma^T)(\id-B)^{-T} = (\id-\tilde{B})^{-1}(\tilde{\Omega}_e + \tilde{\Gamma}\tilde{\Psi}_e\tilde{\Gamma}^T)(\id-\tilde{B})^{-T},
\end{equation*}
with $\cc{I} = \mathbb{I}(\{\Omega_e\}_{e=1}^m)$ and $\cc{\tilde{I}} = \mathbb{I}(\{\tilde{\Omega}_e\}_{e=1}^m)$. Furthermore, suppose that $|\tilde{\cc{I}}| = |\cc{I}|$, $\|\dagg\|_{\ell_0} = \|\tilde{\dagg}\|_{\ell_0}$, $\text{moral}(\dagg) = \text{moral}(\tilde{\dagg})$ where $B \sim {\dagg}$ and $\tilde{B}\sim\tilde{\dagg}$. Then, $\text{score}_{\lambda,\gamma}(\dagg,\Theta) = \text{score}_{\lambda,\gamma}(\tilde{\dagg},\tilde{\Theta})$ for every $\lambda,\gamma \geq 0$. 
\label{lemma:score_equiv}
\end{lemma}
\begin{proof}
The parameters ${\Theta}$ and $\tilde{\Theta}$ specify the precision matrices for every $e = 1,2,\dots,m$:
\begin{equation*}
    \begin{aligned}
    K_e &=  (\id-B)^T(\Omega_e + \Gamma\Psi_e\Gamma^T)^{-1}(\id-B), \\
    \tilde{K}_e &= (\id-\tilde{B})^{T}(\tilde{\Omega}_e + \tilde{\Gamma}\tilde{\Psi}_e\tilde{\Gamma}^T)^{-1}(\id-\tilde{B}).
    \end{aligned}
\end{equation*}
Furthermore, the regularized likelihood score for each model is given by:
\begin{equation*}
\begin{aligned}
    \text{score}_{\lambda,\gamma}(\dagg,\Theta) &= \sum_{e=1}^m \hat{\pi}_e(-\log\det(K_e)+\mathrm{trace}(K_e\hat{\Sigma}_e))+\lambda\mathcal{R}_\gamma(\dagg,\cc{I}), \\
    \text{score}_{\lambda,\gamma}(\tilde{\dagg},\tilde{\Theta}) &= \sum_{e=1}^m \hat{\pi}_e(-\log\det(\tilde{K}_e)+\mathrm{trace}(\tilde{K}_e\hat{\Sigma}_e))+\lambda\mathcal{R}_\gamma(\tilde{\dagg},\tilde{\cc{I}}).
    \end{aligned}
\end{equation*}
By the assumptions of the lemma, $K_e = \tilde{K}_e$ for every $e = 1,2,\dots,m$, and $\mathcal{R}_\gamma(\tilde{\dagg},\tilde{\cc{I}}) = \mathcal{R}_\gamma({\dagg},\cc{I})$. Thus, $\text{score}_{\lambda,\gamma}(\dagg,\Theta) = \text{score}_{\lambda,\gamma}(\tilde{\dagg},\tilde{\Theta})$. 
\end{proof}
We are now ready to prove Theorem~\ref{thm:IMEC}.
\begin{proof}[Proof of Theorem~\ref{thm:IMEC}]Take any $\tilde{\dagg} \in \imec(\dagg)$. By Lemma~\ref{lemma:equiv_without_latent}, we have that that there exists a connectivity matrix $\tilde{B} \sim \tilde{\dagg}$ and $\{\tilde{\Omega}_e\}_{e=1}^m \subseteq \mathbb{D}_{++}^p$ with $\cc{I} = \mathbb{I}(\{\tilde{\Omega}_e\}_{e=1}^m)$ such that $(\id-B)^{-1}\Omega_e(\id-B)^{-T} =(\id-\tilde{B})^{-1}\tilde{\Omega}_e(\id-\tilde{B})^{-T}$ for every $e = 1,2,\dots,m$. Furthermore, let $\tilde{\Gamma} = (\id-\tilde{B})(\id-B)^{-1}\Gamma$ and $\tilde{\Psi}_e = \Psi_e$ for every $e = 1,2,\dots,m$. It is straightforward to show that the parameters $\tilde{\Theta} = (\tilde{B},\tilde{\Gamma},\cc{I},\{\tilde{\Omega}_e,\tilde{\Psi}_e\}_{e=1}^m)$ satisfy the condition of Lemma~\ref{lemma:score_equiv} and we can conclude that $\text{score}_{\lambda,\gamma}(\dagg,\Theta) = \text{score}_{\lambda,\gamma}(\tilde{\dagg},\tilde{\Theta})$ for every $\lambda,\gamma \geq 0$.  
\end{proof}

\section{Characterization of optimally scoring DAGs via  \methName{} in the infinite sample regime}
\label{appendix:infinite_characterization}
Throughout, we consider the asymptotic regime where $p/n_e \to 0$ as $n_e \to \infty$ for all $e \in [m]$. For every $\dagg \in \simec(\dagg^\star)$, let $(B,\Gamma,\{\Omega_{e},\Psi_e\}_{e=1}^m)$ be any set of parameters with $B \sim \dagg$ that specify the population covariance matrix $\Sigma_e^\star$ in environment $e$, i.e. $\Sigma_{e}^\star = (\id-B)^{-1}(\Omega_e + \Gamma\Psi_e\Gamma)^{-1}(\id-B)^{-T}$. In our analysis, the parameter space $(B,\Gamma,\{\Omega_e,\Psi_e\}_{e =1}^m)$ {is assumed} to be compact. For notational ease, we denote the constraint set for the parameters $(B,\Gamma,\{\Omega_e,\Psi_e\}_{e =1}^m)$ as $\mathcal{F}$. {Assuming such a compactness constraint}
enables uniform convergence of M-estimators \citep{van2000empirical}. As an example of a compactness constraint, let $\tau_1,\tau_2,\tau_3> 0$ be scalars where for every such $(B,\Gamma,\{\Omega_{e},\Psi_e\}_{e=1}^m)$, $\|(\id-B)\|_{F} \leq \tau_1$, $\min_{e}\sigma_\text{min}(\Omega_e+\Gamma\Psi_e\Gamma^T) \geq \tau_2$, and $\max_{e}\sigma_\text{max}(\Omega_e+\Gamma\Psi_e\Gamma^T) \leq \tau_3$. Here, $\sigma_\text{max}(\cdot)$ and $\sigma_\text{min}(\cdot)$ denote maximum and minimum singular values of an input matrix, respectively. Then, the space of connectivity matrices and noise variances {has} the additional constraints $\|(\id-B)\|_{F} \leq \tau_1$, $\min_{e}\sigma_\text{min}(\Omega_e+\Gamma\Psi_e\Gamma^T) \geq \tau_2$, and $\max_{e}\sigma_\text{max}(\Omega_e+\Gamma\Psi_e\Gamma^T) \leq \tau_3$ in the score function \eqref{eqn:estimator_theory}. As before, we denote the score of a DAG $\dagg$ with a pre-specified number of latent variables $\bar{h}$ when constrained to the compact parameter space $\mathcal{F}$ as $\texttt{score}_{\lambda,\gamma}(\dagg,\bar{h})$. 

With this setup, we have the following characterization of the optimally scoring models, scored according to \eqref{eqn:estimator_theory}.
\begin{proposition}[Characterization of optimally scoring DAGs] Suppose that the set of parameters are constrained to be in the compact space $\mathcal{F}$. Suppose that $n_e \to \infty$, $\gamma$ is set to be any bounded scalar, and that $\lambda \to 0$ while satisfying
\begin{equation*}
\begin{aligned}
\lambda &\gg \left| \max_{(B,\Gamma,\{\Omega_e,\Psi_e\}_{e=1}^m)\in\mathcal{F}}
    \sum_{e=1}^m \mathrm{tr}((\id-B)^T(\Omega_e+\Gamma\Psi_e\Gamma^T)^{-1}(\id-B)(\hat{\pi}_e\hat{\Sigma}_e-{\pi}_{e}^\star\Sigma_{e}^\star))\right|,
   \\&+ \left|\max_{(B,\Gamma,\{\Omega_e,\Psi_e\}_{e=1}^m)\in\mathcal{F}}
   \sum_{e=1}^m(\hat{\pi}_e-\pi_{e}^\star)\log\det(\Omega_e+\Gamma\Psi_e\Gamma^T)\right|,
\end{aligned}
\end{equation*}
where $\Sigma^{\star}_e$ represents the true covariance in environment $e$ and $\pi^{\star}_e = {\lim_{\mbox{all}\ n_e \to \infty}} \frac{n_e}{\sum_{e=1}^mn_e}$. Then, with probability tending to one, the set of optimally scoring models (according to \eqref{eqn:estimator_theory}) are solution to the following optimization problem:
\begin{equation}
\begin{aligned}
    \argmin_{\dagg,\bar{h},B,\Gamma,\cc{I},\{\Omega_e,\Psi_e\}_{e=1}^m}&~~\mathcal{R}_\gamma(\dagg,\cc{I})\\\text{subject-to}:&~~ B \sim \dagg, \mathcal{I} = \mathbb{I}(\{\Omega_e\}_{e=1}^m),\text{ and}\\
    &~~\Sigma_e^\star = (\id-B)^{-1}(\Omega_e+\Gamma\Psi_e\Gamma^T)^{-1}(\id-B)^{-T} ~\text{for every }e\in[m].
    \end{aligned}
    \label{eq:optimal_params}
\end{equation}
\label{prop:asymptotic_property}
\end{proposition}
The proof of Proposition~\ref{prop:asymptotic_property} relies on the following lemma:
\begin{lemma} The minimizers of the following optimization
\begin{eqnarray*}
\begin{aligned}
 \argmin_{\dagg,\bar{h},\cc{I}}\argmin_{(B,\Gamma,\{\Omega_e,\Psi_e\}_{e=1}^m)\in\mathcal{F}}&\sum_{e=1}^m {\pi}^\star_e\Big(\log\det(\Omega_e+\Gamma\Psi_e\Gamma^T) \\&+ \mathrm{tr}((\id-B)^T(\Omega_e+\Gamma\Psi_e\Gamma^T)^{-1}(\id-B){\Sigma}^\star_e)\Big)\\
    \text{subject-to: }& B \sim \dagg~~;~~ \mathbb{I}(\{\Omega_e\}_{e=1}^m) \subseteq \cc{I}
    \end{aligned}
\end{eqnarray*}
are given by parameters $\dagg,\bar{h},B,\Gamma,\cc{I},\{\Omega_e,\Psi_e\}_{e=1}^m$ that satisfy:
\begin{equation*}
\begin{aligned}
 B \sim \dagg&, \mathcal{I} \subseteq\mathbb{I}(\{\Omega_e\}_{e=1}^m),\text{ and}\\
    &\Sigma_e^\star = (\id-B)^{-1}(\Omega_e+\Gamma\Psi_e\Gamma^T)^{-1}(\id-B)^{-T} ~\text{for every }e\in[m].
\end{aligned}
\end{equation*}
\label{lemma:intermed_model}
\end{lemma}

\begin{proof}[Proof of Lemma~\ref{lemma:intermed_model}]
Let $\mathcal{M}(B,\Gamma,\cc{I},\Omega_e,\Psi_e)$ denote a model associated with each equation in the SCM \eqref{eq:SCM_env} (main paper). For notational convenience, we use the short-hand notation $\mathcal{M}_e$ for this model. We let $\Sigma(\mathcal{M}_e) = (\id-{B})^{-1}(\Omega_e+\Gamma\Psi_e\Gamma^T)(\id-{B})^{-T}$ be the associated covariance model parameterized by the parameters $(B,\Gamma,\cc{I},\Omega_e,\Psi_e)$. The optimal solutions of the optimization problem in Lemma~\ref{lemma:intermed_model} can then be equivalently reformulated as:
\begin{equation}
    \begin{aligned}
    \argmin_{\{\mathcal{M}_e\}_{e = 1}^{m}} &  \sum_{e = 1}^{m}\pi^{\star}_e \texttt{KL}(\Sigma^{\star}_e,\Sigma(\mathcal{M}_e)),
    \end{aligned}
    \label{eqn:KL_formul}
\end{equation}
where $\texttt{KL}(\cdot,\cdot)$ represents the Gaussian KL-divergence. Notice that for the decision variables
{$\mathcal{M}_e^{\star} = (B^\star,\Gamma^\star,\cc{I}^\star,\Omega_e^\star,\Psi_e^\star)$}
for each $e = 1,2,\dots,m$, \eqref{eqn:KL_formul} achieves zero loss. Hence, any other optimal solution of \eqref{eqn:KL_formul} must yield zero loss, or equivalently, $\Sigma(\mathcal{M}_e) = \Sigma^\star_{e}$ for {any} optimal collection $\{\mathcal{M}_e\}_{e = 1}^{m}$.  
\end{proof}

\begin{proof}[Proof of Proposition~\ref{prop:asymptotic_property}]
For a fixed $\dagg$, number of latent variables $\bar{h}$, and set of intervention targets $\cc{I}$, we let $S_{\lambda,\gamma}(\dagg,\bar{h},\cc{I})$ be the following score:
\begin{equation*}
\begin{aligned}
    S_{\lambda,\gamma}(\dagg,\bar{h},\cc{I}) := \min_{(B,\Gamma,\{\Omega_e,\Psi_e\}_{e=1}^m)\in\mathcal{F}}&\sum_{e=1}^m \hat{\pi}_e\Big(\log\det(\Omega_e+\Gamma\Psi_e\Gamma^T) \\&+ \mathrm{tr}((\id-B)^T(\Omega_e+\Gamma\Psi_e\Gamma^T)^{-1}(\id-B)\hat{\Sigma}_e)\Big) + \lambda\mathcal{R}_\gamma(\dagg,\cc{I}).\\
    \text{subject-to: }& B\sim\dagg~;~\mathbb{I}(\{\Omega_e\}_{e=1}^m) \subseteq \cc{I}
    \end{aligned}
\end{equation*}
Notice the score $S_{\lambda,\gamma}(\dagg,\bar{h},\cc{I})$ is related to the the score $\texttt{score}_{\lambda,\gamma}(\dagg,\hat{\Theta}(\dagg,\bar{h}))$ in \eqref{eqn:score} via the following simple relation: $\texttt{score}_{\lambda,\gamma}(\dagg,\hat{\Theta}(\dagg,\bar{h})) = \min_{\cc{I}}S_{\lambda,\gamma}(\dagg,\bar{h},\cc{I})$, and thus:
\begin{equation} \min_{\dagg,\bar{h}} \texttt{score}_{\lambda,\gamma}(\dagg,\hat{\Theta}(\dagg,\bar{h})) = \min_{\dagg,\bar{h},\cc{I}} S_{\lambda,\gamma}(\dagg,\bar{h},\cc{I}).
\label{eqn:new_score_relation}
\end{equation}
 We define the population analogue of the score $S_{\lambda,\gamma}(\dagg,\bar{h},\cc{I})$ below:
\begin{equation*}
\begin{aligned}
   S^\star(\dagg,\bar{h},\cc{I}) := \min_{(B,\Gamma,\{\Omega_e,\Psi_e\}_{e=1}^m)\in\mathcal{F}}&\sum_{e=1}^m {\pi}^\star_e\Big(\log\det(\Omega_e+\Gamma\Psi_e\Gamma^T) \\&+ \mathrm{tr}((\id-B)^T(\Omega_e+\Gamma\Psi_e\Gamma^T)^{-1}(\id-B){\Sigma}^\star_e)\Big).\\
    \text{subject-to: }& B \sim \dagg~~;~~ \mathbb{I}(\{\Omega_e\}_{e=1}^m) \subseteq \cc{I},
    \end{aligned}
\end{equation*}
where $\Sigma^{\star}_e$ represents the true covariance in environment $e$ and $\pi^{\star}_e = {\lim_{\mbox{all}\ n_e \to \infty}} \frac{n_e}{\sum_{e=1}^mn_e}$. 

First, we show that for every $\dagg$, $\bar{h}$, $\cc{I}$, the score function $S_{\lambda,\gamma}(\dagg,\bar{h},\cc{I}) \overset{p}{\to} S^\star(\dagg,\bar{h},\cc{I})$ as $n_e \to \infty$ for every $e \in [m]$. This follows by the compactness of the parameter space leading to
uniform convergence of M-estimators,
and that $\lambda \to 0$ and bounded $\gamma$ \citep{van2000empirical}. For more details, note that:
\begin{equation*}
\begin{aligned}
S_{\lambda,\gamma}(\dagg,\bar{h},\cc{I})-\lambda\mathcal{R}_{\gamma}(\dagg,\cc{I}) &= \min_{(B,\Gamma,\{\Omega_e,\Psi_e\}_{e=1}^m)\in\mathcal{F}}  \sum_{e=1}^m {\pi}^\star_e\Big(\log\det(\Omega_e+\Gamma\Psi_e\Gamma^T) \\&+ \mathrm{tr}((\id-B)^T(\Omega_e+\Gamma\Psi_e\Gamma^T)^{-1}(\id-B){\Sigma}^\star_e)\Big)\\&+\mathrm{tr}((\id-B)^T(\Omega_e+\Gamma\Psi_e\Gamma^T)^{-1}(\id-B)(\hat{\pi}_e\hat{\Sigma}_e-{\pi}_{e}^\star\Sigma_{e}^\star))\\&+ (\hat{\pi}_e-\pi_{e}^\star)\log\det(\Omega_e+\Gamma\Psi_e\Gamma^T).
    \end{aligned}
\end{equation*}
From there, we have
\begin{equation*}
\begin{aligned}
S_{\lambda,\gamma}(\dagg,\bar{h},\cc{I}) &\geq  S^\star(\dagg,\bar{h},\cc{I}) \\&-\min_{(B,\Gamma,\{\Omega_e,\Psi_e\}_{e=1}^m)\in\mathcal{F}}
    \sum_{e=1}^m \mathrm{tr}((\id-B)^T(\Omega_e+\Gamma\Psi_e\Gamma^T)^{-1}(\id-B)(\hat{\pi}_e\hat{\Sigma}_e-{\pi}_{e}^\star\Sigma_{e}^\star))\\
    &-\min_{(B,\Gamma,\{\Omega_e,\Psi_e\}_{e=1}^m)\in\mathcal{F}}
   (\hat{\pi}_e-\pi_{e}^\star)\log\det(\Omega_e+\Gamma\Psi_e\Gamma^T)+\lambda\mathcal{R}_\gamma(\dagg,\cc{I}),\\
S_{\lambda,\gamma}(\dagg,\bar{h},\cc{I}) &\leq S^\star(\dagg,\bar{h},\cc{I}) \\&+ \max_{(B,\Gamma,\{\Omega_e,\Psi_e\}_{e=1}^m)\in\mathcal{F}}
    \sum_{e=1}^m \mathrm{tr}((\id-B)^T(\Omega_e+\Gamma\Psi_e\Gamma^T)^{-1}(\id-B)(\hat{\pi}_e\hat{\Sigma}_e-{\pi}_{e}^\star\Sigma_{e}^\star))
    \\&+\max_{(B,\Gamma,\{\Omega_e,\Psi_e\}_{e=1}^m)\in\mathcal{F}}
   (\hat{\pi}_e-\pi_{e}^\star)\log\det(\Omega_e+\Gamma\Psi_e\Gamma^T) + \lambda\mathcal{R}_\gamma(\dagg,\cc{I}).
\end{aligned}
\end{equation*}
By the compactness constraint, 
\begin{gather*}
\min_{(B,\Gamma,\{\Omega_e,\Psi_e\}_{e=1}^m)\in\mathcal{F}}
    \sum_{e=1}^m \mathrm{tr}((\id-B)^T(\Omega_e+\Gamma\Psi_e\Gamma^T)^{-1}(\id-B)(\hat{\pi}_e\hat{\Sigma}_e-{\pi}_{e}^\star\Sigma_{e}^\star)) \to 0, \\
    \min_{(B,\Gamma,\{\Omega_e,\Psi_e\}_{e=1}^m)\in\mathcal{F}}
   (\hat{\pi}_e-\pi_{e}^\star)\log\det(\Omega_e+\Gamma\Psi_e\Gamma^T) \to 0,
\end{gather*}    
    as $n_e \to \infty$. Furthermore, since $\lambda \to 0$ as $n_e \to \infty$ for every $e \in [m]$, and $\mathcal{R}_{\gamma}(\dagg,\cc{I})$ is bounded for a finite $\gamma$, we have that $\lambda\mathcal{R}_\gamma(\dagg,\cc{I})\to 0$. We can thus conclude  $S_{\lambda,\gamma}(\dagg,\bar{h},\cc{I}) \to  S^\star(\dagg,\bar{h},\cc{I}) $ in the infinite data limit for every environment. 
    

With the choice of $\lambda$ in Proposition~\ref{prop:asymptotic_property}, we have that it is above fluctuations due to sampling error; it follows that for two population score equivalent models $(\dagg_1,\bar{h}_1,\cc{I}_1)$ and $(\dagg_2,\bar{h}_2,\cc{I}_2)$ with $S^\star(\dagg_1,\bar{h}_1,\cc{I}_1) = S^\star(\dagg_2,\bar{h}_2,\cc{I}_2)$, if $\mathcal{R}_\gamma(\dagg_1,\cc{I}_1) < \mathcal{R}_\gamma(\dagg_2,\cc{I}_2)$, then, there exists $N$ such that for $n_e \geq N$ for every $e$, $S_{\lambda,\gamma}(\dagg_1,\bar{h}_1,\cc{I}_1) < S_{\lambda,\gamma}(\dagg_2,\bar{h}_2,\cc{I}_2)$. This allows us to conclude that:
\begin{equation*}
\mathbb{P}\left(\argmin S_{\lambda,\gamma}(\dagg,\bar{h},\cc{I}) = \argmin \mathcal{R}_{\gamma}(\dagg,\cc{I}) \text{  subject-to  } \dagg,\bar{h},\cc{I} \in \argmin S^\star(\dagg,\bar{h},\cc{I})\right) \to 1,
\end{equation*}
as $n_e \to \infty$ for every $e \in [m]$. In other words, we can conclude that in the infinite data regime, minimizers of \eqref{eqn:estimator_theory} converge (with probability tending to one) to
\begin{equation}
\begin{aligned}
    \argmin_{\dagg,\bar{h},B,\Gamma,\cc{I},\{\Omega_e,\Psi_e\}_{e=1}^m \in \Theta_\text{opt}}&~~\mathcal{R}_\gamma(\dagg,\cc{I}),
    \end{aligned}
    \label{eq:optimal_params}
\end{equation}
where 
\begin{eqnarray*}
\begin{aligned}
    \Theta_\text{opt}:= \argmin_{\dagg,\bar{h},\cc{I}}\argmin_{(B,\Gamma,\{\Omega_e,\Psi_e\}_{e=1}^m)\in\mathcal{F}}&\sum_{e=1}^m {\pi}^\star_e\Big(\log\det(\Omega_e+\Gamma\Psi_e\Gamma^T) \\&+ \mathrm{tr}((\id-B)^T(\Omega_e+\Gamma\Psi_e\Gamma^T)^{-1}(\id-B){\Sigma}^\star_e)\Big)\\
    \text{subject-to: }& B \sim \dagg~~;~~ \mathbb{I}(\{\Omega_e\}_{e=1}^m) \subseteq \cc{I}.
    \end{aligned}
\end{eqnarray*}
By Lemma~\ref{lemma:intermed_model}, we then conclude that with probability tending to one, the minimizers of \eqref{eqn:estimator_theory} in the infinite data limit are:

\begin{equation*}
\begin{aligned}
    \argmin_{\dagg,\bar{h},B,\Gamma,\cc{I},\{\Omega_e,\Psi_e\}_{e=1}^m}&~~\mathcal{R}_\gamma(\dagg,\cc{I})\\\text{subject-to}:&~~ B \sim \dagg, \mathbb{I}(\{\Omega_e\}_{e=1}^m) \subseteq \cc{I},\text{ and}\\
    &~~\Sigma_e^\star = (\id-B)^{-1}(\Omega_e+\Gamma\Psi_e\Gamma^T)^{-1}(\id-B)^{-T} ~\text{for every }e\in[m].
    \end{aligned}
    \label{eq:optimal_params_between}
\end{equation*}
Finally, note that $\mathcal{R}_\gamma(\dagg,\cc{I})$ is monotonic in the size of $\cc{I}$, we can replace the constraint $\mathbb{I}(\{\Omega_e\}_{e=1}^m) \allowbreak\subseteq \cc{I}$ with $\mathbb{I}(\{\Omega^e\}_{e=1}^m) =\cc{I}$ and attain the desired result. 
\end{proof}

\section{Equivalent causal models}
\subsection{Proof of Proposition~\ref{prop:all_dags_scm}}
\label{proof_prop_all_dags_scm}
Consider a causal model $(B,\Gamma,\{\Omega_e,\Psi_e\}_{e=1}^m)$ specifying the SCM \eqref{eq:SCM_env} for the data among observed variables (these parameters can for example be the population parameters), so that:
$$\Sigma_e^\star = (\id-{B})^{-1}({\Omega}_e+{\Gamma}{\Psi}_e{\Gamma}^T)^{-1}(\id-{B})^{-T} ~\text{for every }e\in[m].$$
We will construct an equivalent SCM with parameters $(\tilde{B},\tilde{\Gamma},\{\tilde{\Omega}_e,\tilde{\Psi}_e\}_{e=1}^m)$ where the connectivity matrix $\tilde{B}$ can be arbitrary and compatible with any DAG, and the coefficient matrix $\tilde{\Gamma}$ is any arbitrary $p \times p$ and invertible matrix. 

Specifically, Let $\tilde{\dagg}$ be any DAG and $\tilde{B}$ be any connectivity matrix associated with $\tilde{\dagg}$. Let $\tilde{\Gamma} \in \mathbb{R}^{p \times p}$ be an arbitrary invertible matrix. For every $e \in [m]$, choose a diagonal positive matrix $\tilde{\Omega_e} \in \mathbb{D}^p_{++}$ such that:
\begin{equation*}
    (\id-B)^{-1}(\Omega_e+\Gamma\Psi_e{\Gamma}^T)(\id-B)^{-T} \succ (\id-\tilde{B})^{-1}\tilde{\Omega}_e(\id-\tilde{B})^{-T}.
\end{equation*}
Notice that such a matrix exists since $(\id-\tilde{B})(\id-B)^{-1}(\Omega_e+\Gamma\Psi_e{\Gamma}^T)(\id-B)^{-T}(\id-\tilde{B})^T$ is a positive definite matrix. Define then for every $e \in [m]$:
$$\tilde{\Psi}_e = \tilde{\Gamma}^{-1}\left[(\id-B)^{-1}(\Omega_e+\Gamma\Psi_e{\Gamma}^T)(\id-B)^{-T}-  (\id-\tilde{B})^{-1}\tilde{\Omega}_e(\id-\tilde{B})^{-T}\right]\tilde{\Gamma}^{-T}.$$
By construction, $\tilde{\Psi}_e \succ 0$ and $(\id-B)^{-1}(\Omega_e+\Gamma\Psi_e{\Gamma}^T)(\id-B)^{-T} = (\id-\tilde{B})^{-1}(\tilde{\Omega}_e+\tilde{\Gamma}\tilde{\Psi}_e\tilde{\Gamma}^T)(\id-\tilde{B})^{-T}$ for every $e \in [m]$. We have thus shown that:
$$\Sigma_e^\star = (\id-\tilde{B})^{-1}(\tilde{\Omega}_e+\tilde{\Gamma}\tilde{\Psi}_e\tilde{\Gamma}^T)^{-1}(\id-\tilde{B})^{-T} ~\text{for every }e\in[m].$$
That is, the model specified by the parameters $(\tilde{B},\tilde{\Gamma},\{\tilde{\Omega}_e,\tilde{\Psi}_e\}_{e=1}^m)$ specifies an SCM that is compatible with the underlying data distributions. 

Figure \ref{fig:illustration_faithfulness_main} displays the two equivalent models. Figure \ref{fig:illustration_faithfulness_main}(a) can be taken for example to be the population model with a single latent variable. Figure \ref{fig:illustration_faithfulness_main}(b) is a model with three latent variables that is an equally good representation of the data among the observed variables.

\subsection{Connection to violation of faithfulness via an illustration}
\label{sec:illutration_faithfulness}
For simplicity, we first consider the scenario without any perturbations, e.g. there are no nodes $\mathcal{E}$ in Figure~\ref{fig:illustration_faithfulness_main}. Suppose that the graphical models in Figure~\ref{fig:illustration_faithfulness_main}(a) and  Figure~\ref{fig:illustration_faithfulness_main}(b) specify the distribution among the observed variables. Notice that Figure~\ref{fig:illustration_faithfulness_main}(a) implies that $X_1\perp X_3$, while the same conclusion cannot be made in Figure~\ref{fig:illustration_faithfulness_main}(b). In other words, if the model in Figure~\ref{fig:illustration_faithfulness_main}(b) is the population DAG, the conditional independence relationships among the observed variables in Figure~\ref{fig:illustration_faithfulness_main}(b) are not encoded in the data distribution. Thus, the faithfulness assumption is not satisfied.  

Now we consider the scenario where there are perturbations $\mathcal{E}$. In our modeling assumption, we assume that the interventions are independent among the observed variables, that is the matrix $\Omega_e$ encoding noise variances among the observed variables is diagonal. Then, again, Figure~\ref{fig:illustration_faithfulness_main}(a) concludes that $X_1 \perp X_3$, while Figure~\ref{fig:illustration_faithfulness_main}(b) does not.

\subsection{Proof of Corollary~\ref{prop:all_dags}}
\label{proof_prop_all_dags}
Let $(B,\Gamma,\{\Omega_e,\Psi_e\}_{e=1}^m)$ be the population parameters. Consider the construction in the proof of Proposition~\ref{prop:all_dags_scm}. Let $\tilde{\dagg}$ be the empty graph and $\tilde{B} = 0$. Let $\tilde{\Omega}_e = \alpha\id$, where $\alpha$ is chosen such that:
\begin{equation*}
    (\id-B)^{-1}(\Omega_e+\Gamma\Psi_e{\Gamma}^T)(\id-B)^{-T} \succ \alpha(\id-\tilde{B})^{-1}(\id-\tilde{B})^{-T}.
\end{equation*}
Setting $\tilde{\Gamma}$ and $\tilde{\Psi}_e$ as in the proof of Proposition~\ref{prop:all_dags_scm}, we have that $(\tilde{B},\tilde{\Gamma},\{\tilde{\Omega}_e,\tilde{\Psi}_e\}_{e=1}^m)$ is an equivalent model. We have found a model satisfying the constraint in the optimization \eqref{eq:optimal_params}. By the construction of $\tilde{\Omega}_e$, we have that the set of intervention targets encoded by this equivalent model is empty, i.e. $\tilde{\cc{I}} = \mathbb{I}(\{\tilde{\Omega}_e\}) = \emptyset$. Furthermore, $\mathcal{R}_\gamma(\tilde{\dagg},\cc{\tilde{I}}) = 0$, which is the minimal value the regularization function $\mathcal{R}_\gamma(\cdot,\cdot)$ can attain for any value of $\gamma$. We thus conclude from Proposition~\ref{prop:asymptotic_property} that the constructed model with an empty graph is optimal. 

\section{Discussions of Assumption~\ref{assum:dense_incoherence}}
\label{sec:discussion_dense}
\subsection{When is the incoherence parameter $\inc^\star_e$ small?}
\label{sec:inc_small}
Recall that Assumption~\ref{assum:dense_incoherence} states that the product of an incoherence parameter $\inc^\star_e$ (capturing the denseness of the latent effects) and the degree of the moral graph is $\mathcal{O}(1)$. We provide a simple illustration of a model \eqref{eq:SCM_env} for which the incoherence parameter $\inc^\star_e$ is small in that it is close to its lower-bound $\sqrt{\frac{h}{p}}$. 

\emph{Illustration}: We consider a model \eqref{eq:SCM_env} where there is one latent variable (i.e. $h = 1$) and the latent coefficient matrix $\Gamma \in \mathbb{R}^{p \times 1}$ has identical entries so that the effect of the latent variable on the observed variables is \emph{equally spread out}. Let $k$ be the maximum number of parents for any node in the DAG $\dagg^\star$ among the observed variables. Let $\text{cond}(\Omega_e^\star) = \frac{\max_{i}[\Omega_e^\star]_{i,i}}{\min_i [\Omega_e^\star]_{i,i}}$ represent the condition number of the noise variance matrix $\Omega_e^\star$. Suppose that the edge weights of the DAG $\dagg^\star$, i.e. the nonzero entries of $B^\star$ are sufficiently small, i.e. are smaller in magnitude than $1/(k~\text{cond}(\Omega_e^\star))$. Then, some manipulations yield the following bound for $\inc^\star_e$:
\begin{equation*}
\begin{aligned}
\inc^\star_e &= \inc[\text{col-space}((\id-B^\star)^T{\Omega_e^\star}^{-1}\Gamma^\star] 
= \frac{\max_{i}|\left[(\id-B^\star)^T{\Omega_e^\star}^{-1}\Gamma^\star\right]_{i,1}|}{\|(\id-B^\star)^T{\Omega_e^\star}^{-1}\Gamma^\star\|_2}\\
&\leq \frac{\max_{i}|\left[(\id-B^\star)^T{\Omega_e^\star}^{-1}\Gamma^\star\right]_{i,1}|}{\sqrt{p}\min_{i}|\left[(\id-B^\star)^T{\Omega_e^\star}^{-1}\Gamma^\star\right]_{i,1}|}
\leq \sqrt{\frac{h}{p}} \frac{\text{cond}(\Omega^\star_e)(1+k\|B^\star\|_\infty)}{1-k\|B^\star\|_\infty\text{cond}(\Omega_e^\star)}.
\end{aligned}
\end{equation*}
Then, supposing that the noise variances on each observed variable are not too different, i.e. $\text{cond}(\Omega_e^\star) = \mathcal{O}(1)$, we have that $\inc^\star_e  = \mathcal{O}\left(\sqrt{\frac{h}{p}}\right)$. 

We have shown via the above illustration that when the effect of the latent variables is spread out among all the observed variables, the incoherence parameter is small with $\inc^\star_e = \mathcal{O}\left(\sqrt{\frac{h}{p}}\right)$.

\subsection{Examples of models \eqref{eq:SCM_env} that satisfy Assumption~\ref{assum:dense_incoherence}} 
Throughout, we consider models in which the low-rank matrix $L^\star_e$ is almost maximally incoherent (dense), that is $\inc[\text{col-space}(L^\star_e)] = \mathcal{O}\left(\sqrt{\frac{h}{p}}\right)$ so the effect of marginalization over the latent variables is diffuse across all the observed variables (see Section \ref{sec:inc_small}). We will suppress the constants involved in Assumption~\ref{assum:dense_incoherence} and focus on the trade-off between $\inc[\text{col-space}(L^\star_e)]$ and maximal degree of the moral graph of $\dagg^\star$ represented by the quantity $\degree[\text{moral}(\dagg^\star)]$. So we study models \eqref{eq:SCM_env} in which:
\begin{equation} \inc[\text{col-space}(L^\star_e)]^2\degree[\text{moral}(\dagg^\star)]= \mathcal{O}\left(\frac{h}{p}\degree[\text{moral}(\dagg^\star)]\right) = \mathcal{O}(1).
\label{eqn:inc_condition_mine}
\end{equation}
As we describe next, there are nontrivial classes of models in which the above condition holds. 

\emph{Polynomial degree}: The next class of models we consider are those in which the degree of the moral graph of $\dagg^\star$ grows polynomially with $p$:
$$ \degree[\text{moral}(\dagg^\star)] = \mathcal{O}(p^q)~~;~~h = \mathcal{O}\left(\frac{p}{p^q}\right),$$
where $q \in (0,1)$. Here, according to the theorems in Section~\ref{sec:dense_sparse}, consistent estimation of the underlying equivalence class of DAGs is possible with the number of latent variables growing with $p$. 

\emph{Bounded degree}: the first class of models that we consider are the moral graph of the DAG $\dagg^\star$ among the observed variables has constant degree:
$$ \degree[\text{moral}(\dagg^\star)] = \mathcal{O}(1)~~;~~h = \mathcal{O}(p).$$
Here again, consistent estimation of the underlying equivalence class of DAGs is possible even when the number of latent variables is on the same order as the number of observed variables. 

\subsection{Comparison to assumptions in \cite{Chandrasekaran2011RankSparsityIF,Chand2012,marloes}}
Building on the methodology and results of \cite{Chandrasekaran2011RankSparsityIF,Chand2012}, \cite{marloes} impose a similar but more stringent condition (than Assumption~\ref{assum:dense_incoherence}) on the denseness of the latent effects for equivalence class recovery in observational settings. In particular, \cite{marloes} provide guarantees for models in which:
\begin{equation}
\inc[\text{col-space}(L^\star)]\degree[\text{moral}(\dagg^\star)] = \mathcal{O}\left(\sqrt{\frac{h}{p}}\degree[\text{moral}(\dagg^\star)]\right) = \mathcal{O}(1),
\label{eqn:inc_condition_frot}
\end{equation}
where $L^\star$ is the matrix $L^\star_e$ for an observational environment $e$. Comparing \eqref{eqn:inc_condition_mine} with \eqref{eqn:inc_condition_frot}, we see that our guarantees apply to a broader class of models \footnote{The method in \cite{marloes} uses a nuclear norm penalty to induce low-rank structure; due to the facial structure of the nuclear norm ball, the incoherence condition that they impose is more stringent than the one in our paper.}. Furthermore, as described in the main text, the method in \cite{marloes} is only appropriate for observational settings, whereas our method also exploits interventional data for additional identifiability. 

\section{Proof of Proposition \ref{thm:sparse_dense_1}}
\label{proof_prop_dense_without_interventions}
The proof relies on a few lemmas. We first state a known result (see \cite{Chandrasekaran2011RankSparsityIF}) on the spectral norm of a sparse matrix. For completeness, we include a proof.  
\begin{lemma}[spectral norm of a low-degree matrix] Let $N \in \mathbb{S}^p$ and $\degree(N)$ be the maximum number of non-zeros in any column of $N$. Then, $\|N\|_2 \leq \degree(N)\|N\|_\infty$. 
\label{lemma:degree_2}
\end{lemma}
\begin{proof}
Let $v \in \mathbb{R}^{p}$ with $\|v\|_2 = 1$. Notice that for any standard basis element $e_i$:
\begin{eqnarray*}
|e_i^TNv|_\infty &\leq& \|e_i^T{N}\|_2\sqrt{\sum_{j:N_{i,j}\neq 0} v_j^2} \\
&\leq& \sqrt{\degree(N)}\|N\|_\infty\sqrt{\sum_{j:N_{i,j}\neq 0} v_j^2}.
\end{eqnarray*}
Combining this with the inequality $\sum_{i=1}^p\sum_{j:N_{i,j}\neq 0} v_j^2 \leq \degree(N)$, $\|Nv\|_2^2$ is bounded as follows:
\begin{eqnarray*}
\|Nv\|_2^2 &\leq& \sum_{i = 1}^p \degree(N)\|N\|_\infty^2\sum_{j:N_{i,j}\neq 0} v_j^2\\
&=& \degree(N)\|N\|_\infty^2\left[\sum_{i=1}^p\sum_{j:N_{i,j}\neq 0} v_j^2\right]\\
&\leq& \degree(N)^2\|N\|_\infty^2.
\end{eqnarray*}
Since $v$ was arbitrary, we arrive at the desired result. 
\end{proof}
\begin{lemma}[sparse/low rank incoherence]
Let $K \in \mathbb{S}^{p}$ and $L \in \mathbb{S}^{p}$. If
$\degree(K)\inc[\allowbreak\text{col-space}(L)]^2 < 1$, then, $K = L$ if and only if $K = L = 0$. 
\label{lemma:sparse_low_rank}
\end{lemma}
\begin{proof} The proof follows very similarly to proof of Lemma 2 in \cite{Chandrasekaran2011RankSparsityIF}. Let $\Omega$ be the following subspace induced by $K$:
\begin{eqnarray*}
\Omega = \{M \in \mathbb{S}^{p}: M_{ij} = 0 \text{ if } K_{ij} = 0\}. 
\end{eqnarray*}
Let $T$ be the following subspace induced by $L$:
\begin{eqnarray*}
T = \{\mathcal{P}_{\text{col-space}(L)}M\mathcal{P}_{\text{col-space}(L)} \text{ for } M \text{ symmetric and non-singular}\}.
\end{eqnarray*}
It suffices to show that under the condition stated above $\Omega \cap T = \{0\}$. Note that:
\begin{equation*}
    \max_{\|N\|_2 = 1, N \in T} \|\Proj_{\Omega}(N)\|_2 <1 \Rightarrow \Omega \cap T = \{0\},
\end{equation*}
since if $N \in \Omega \cap T$ with $\|N\|_2=1$, $\|\Proj_{\Omega}(N)\|_2 = \|N\|_2 = 1$, leading to a contradiction. Furthermore, by Lemma~\ref{lemma:degree_2}, we have that: 
\begin{equation*}
    \begin{aligned}
    \max_{\|N\|_2 = 1, N \in T} \|\Proj_{\Omega}(N)\|_2 &\leq \degree(K)\max_{\|N\|_2 = 1, N \in T}\|\Proj_{\Omega}(N)\|_\infty \\
    &\leq \degree(K)\max_{\|N\|_2 = 1, N \in T}\|N\|_\infty \\
    &= \degree(K)\max_{\|N\|_2 = 1, N \in T}\max_{i,j}|e_i^T\Proj_{\text{col-space}(L)}N\Proj_{\text{col-space}(L)}e_j| \\
    &\leq  \degree(K)\max_{i}\|\Proj_{\text{col-space}(L)}(e_i)\|_2^2\\
    &= \degree(K)\text{inc}[\text{col-space}(L)]^2.
    \end{aligned}
\end{equation*}

\end{proof}
\begin{lemma}[Sum of incoherent matrices] Let $L_1,L_2 \in \mathbb{S}$ with column spaces $\mathcal{C}_1$ and $\mathcal{C}_2$, respectively. Then:
\begin{equation*}
    \inc[\text{col-space}(L_1+L_2)] \leq 2\min\{\inc[\mathcal{C}_1],\inc[\mathcal{C}_2]\} + \max\{\inc[\mathcal{C}_1],\inc[\mathcal{C}_2]\}.
\end{equation*}
\label{lemma:incoherent_col_spaces}
\end{lemma}
\begin{proof}
Without loss of generality, let $\inc[\mathcal{C}_1] \leq \inc[\mathcal{C}_2]$. First, notice that for any $v \in \text{col-space}(L_1+L_2)$, we have that $v = \text{span}(u_1,u_2)$ where $u_1 \in \mathcal{C}_1$ and $u_2 \in \mathcal{C}_2$. Notice that:
\begin{equation*}
\begin{aligned}
\inc[\text{col-space}(L_1+L_2)] &= \max_i\max_{v \in \text{col-space}(L_1+L_2)} |v^Te_i|/\|v\|_2\\
&= \max_i\max_{\substack{u_1 \in \mathcal{C}_1,u_2\in\mathcal{C}_2,\|u_1\|_2=\|u_2\|_2 = 1\\v = c_1u_1+c_2u_2}}  |v^Te_i|/\|v\|_2 \\
&= \max_i\max_{\substack{u_1 \in \mathcal{C}_1,u_2\in\mathcal{C}_2,\|u_1\|_2=\|u_2\|_2 = 1\\ u_3 = u_2 - (u_2^Tu_1)u_1 \\ v = c_1u_1+c_2u_3}}  |v^Te_i|/\|v\|_2 \\
&\leq \max_i\max_{\substack{u_1 \in \mathcal{C}_1,u_2\in\mathcal{C}_2,\|u_1\|_2=\|u_2\|_2 = 1\\ u_3 = u_2 - (u_2^Tu_1)u_1 \\ v = c_1u_1+c_2u_3}}  \frac{|c_1|}{\sqrt{c_1^2+c_2^2}}|u_1^Te_i| + \frac{|c_2|}{\sqrt{c_1^2+c_2^2}}|u_3^Te_i| \\
&\leq \max_i \left[\max_{\substack{u_1 \in \mathcal{C}_1,\|u_1\|_2=1}} 2|u_1^Te_i|+\max_{\substack{u_2 \in \mathcal{C}_2,\|u_2\|_2=1}}|u_2^Te_i|\right]\\
&\leq 2\min\{\inc[\mathcal{C}_1],\inc[\mathcal{C}_2]\} + \max\{\inc[\mathcal{C}_1],\inc[\mathcal{C}_2]\}.
\end{aligned}
\end{equation*}

\end{proof}
\begin{proof}[Proof of Proposition~\ref{thm:sparse_dense_1}]
Consider the optimization problem
\begin{equation}
\begin{aligned}
    \argmin_{\dagg,\bar{h},B,\Gamma,\cc{I},\{\Omega_e,\Psi_e\}_{e=1}^m}&~~\mathcal{R}_\gamma(\dagg,\cc{I})\\\text{subject-to}:&~~ B \sim \dagg, \mathcal{I} = \mathbb{I}(\{\Omega_e\}_{e=1}^m),\text{ and}\\
    &~~\Sigma_e^\star = (\id-B)^{-1}(\Omega_e+\Gamma\Psi_e\Gamma^T)^{-1}(\id-B)^{-T} ~\text{for every }e\in[m]\\
    &~~\inc(\allowbreak\text{col-space}((\id-{B})^T{\Omega}_e^{-1}{\Gamma})) \leq 2\inc_e^\star \text{ for every }e\in[m],
    \end{aligned}
        \label{eq:optimal_params_inc_condition}
\end{equation}
Where compared to \eqref{eq:optimal_params}, we have added the incoherence constraint $\inc(\allowbreak\text{col-space}((\id-{B})^T{\Omega}_e^{-1}{\Gamma})) \leq 2\inc_e^\star$. Following exactly similar logic as proof of Proposition~\ref{prop:asymptotic_property}, one can show that with probability tending to one, the optimal solutions of \eqref{eqn:estimator_theory} with the incoherence constraint equal to the optimal solution of \eqref{eq:optimal_params_inc_condition}. Thus, we will analyze the estimates produced by \eqref{eq:optimal_params_inc_condition}. 

Let $$(\dagg,{B},{\Gamma},{\cc{I}},\allowbreak\{\Omega_e,\Psi_e\}_{e=1}^m)$$ be any optimal set of parameters in \eqref{eq:optimal_params_inc_condition}. Since the parameters $(\dagg^\star,{B}^\star,{\Gamma}^\star,{\cc{I}}^\star,\allowbreak\{\Omega_e^\star,\Psi_e^\star\}_{e=1}^m)$ are feasible in \eqref{eq:optimal_params_inc_condition}, we have that:
\begin{equation*}
    p~\degree[\text{moral}({\dagg})]+\|{\dagg}\|_{\ell_0} + \gamma|{\cc{I}}| \leq pd^\star+\|{\dagg}^\star\|_{\ell_0} + \gamma|\cc{I}^\star|.
\end{equation*}
Since $\|{\dagg}^\star\|_{\ell_0} \leq pd^\star$, $|\cc{I}^\star|\leq p$, and $\gamma \in (0,d^\star]$, we have the inequality $\degree\allowbreak[\text{moral}({\dagg})] \leq 3d^\star$. Again, by the feasibility of the parameters $(\dagg,{B},{\Gamma},{\cc{I}},\allowbreak\{\Omega_e,\Psi_e\}_{e=1}^m)$ in \eqref{eq:optimal_params_inc_condition}, we have that
\begin{equation*}
    (\id-B)^{-1}({\Omega}_e+{\Gamma}{\Psi}_e{\Gamma}^T)(\id-{B})^{-T} = (\id-{B}^\star)^{-1}({\Omega}^\star_e+{\Gamma}^\star{\Psi}^\star_e{\Gamma^\star}^T)(\id-{B}^\star)^{-T} \text{ for all }e\in[m].
\end{equation*}
Equivalently, taking the inverse of both sides in the previous equation, and using the Woodbury Inversion Lemma, we have for $e=1,2,\dots,m$
\begin{equation}
\begin{aligned}
    &(\id-{B})^{T}{\Omega}_e^{-1}(\id-{B})
    -(\id-{B})^{T}{\Omega}_e^{-1}{\Gamma}({\Psi}_e^{-1}+{\Gamma}^T{\Omega}_e^{-1}{\Gamma})^{-1}{\Gamma}^T{\Omega}_e^{-1}(\id-{B})\\
    &= (\id-{B}^\star)^{T}{{\Omega}_e^\star}^{-1}(\id-{B}^\star)
    -(\id-{B}^\star)^{T}{\Omega^\star_e}^{-1}{\Gamma^\star}({\Psi_e^\star}_e^{-1}+{\Gamma^\star}^T{\Omega^\star_e}^{-1}{\Gamma}^\star)^{-1}{\Gamma^\star}^T{{\Omega_e^\star}}^{-1}(\id-{B}^\star).
    \end{aligned}
    \label{eqn:Omega_results}
\end{equation}
Define for every $e=1,2,\dots,m$ the following quantities: \begin{equation}
    \begin{aligned}
        {K}_e &:= (\id-{B})^{T}{\Omega}_e^{-1}(\id-{B}),\\
    K_e^\star &= (\id-{B}^\star)^{T}{{\Omega}_e^\star}^{-1}(\id-{B}^\star),\\
    {L}_e&:= (\id-{B})^{T}{\Omega}_e^{-1}{\Gamma}({\Psi}_e^{-1}+{\Gamma}^T\hat{\Omega}_e^{-1}{\Gamma})^{-1}{\Gamma}^T{\Omega}_e^{-1}(\id-{B}),\\
    L_e^\star &:= (\id-{B}^\star)^{T}{\Omega^\star_e}^{-1}{\Gamma^\star}({\Psi_e^\star}^{-1}+{\Gamma^\star}^T{\Omega^\star_e}^{-1}{\Gamma}^\star)^{-1}{\Gamma^\star}^T{\Omega_e^\star}^{-1}(\id-{B}^\star).
    \end{aligned}
    \label{eqn:K_L_defn}
\end{equation}
Notice that $\degree({K}_e^\star) = d^\star$. Since $\degree[\text{moral}({\dagg})] \leq 3d^\star$, we have that $\degree({K}_e) \leq 3d^\star$. Furthermore, by the constraint on the optimization \eqref{eq:optimal_params_inc_condition}, $\inc[\text{col-space}({L}_e] \allowbreak\leq \allowbreak2\inc_e^\star$.

Notice that \eqref{eqn:Omega_results} can be equivalently written for every $e = 1,2,\dots,m$:
\begin{equation}
\begin{aligned}
    {K}_e - K_e^\star = {L}_e-L_e^\star.
    \end{aligned}
    \label{eqn:relation_L_K}
\end{equation}
Since $\degree(A+B) \leq \degree(A)+\degree(B)$ for matrices $A,B \in \mathbb{S}^{p}$, $\degree({K}_e-K_e^\star) \leq 4d^\star$ for every $e$. Furthermore, by Lemma~\ref{lemma:incoherent_col_spaces}, we have that $\inc[\text{col-space}({L}_e-L_e^\star)] \leq 4\inc_e^\star$ for every $e =1,2,\dots,m$. Thus, by Assumption~\ref{assum:dense_incoherence}, we have that: $\degree({K}_e-K_e^\star)\inc[\text{col-space}({L}_e-L_e^\star)]<1$ for all $e = 1,2,\dots,m$. Hence, by Lemma~\ref{lemma:sparse_low_rank}, ${K}_e-K_e^\star = 0$ or equivalently $(\id-{B})^{-1}{\Omega}_e(\id-{B})^{-T} = (\id-{B}^\star)^{-1}{{\Omega}_e^\star}(\id-{B}^\star)^{-T}$ for every $e = 1,2,\dots,m$. 

Let $\dagg_\text{all.opt},\cc{I}_\text{all.opt}$ be the collection of optimal DAGs and intervention targets in the optimization \eqref{eq:optimal_params_inc_condition}. Letting $\Sigma_{X^e|H^e}^\star$ be the covariance of $X^e|H^e$, the analysis above allows us to conclude that:
\begin{equation}
\begin{aligned}
 (\dagg_\text{all.opt},\cc{I}_\text{all.opt}) = \argmin_{\dagg,\cc{I}} \, &~~~\mathcal{R}_\gamma(\dagg,\cc{I}) \\\text{subject-to}\, &~~~ \text{ there exists }B \sim \dagg~:~\mathbb{I}(\{\Omega_e\}_{e=1}^m) = \cc{I}\\ \, &~~~~\Sigma_{X^e|H^e}^\star = (\id-{B})^{-1}{\Omega}_e(\id-{B})^{-T} \text{ for all }e=1,2,\dots,m.
    \end{aligned}
    \label{eqn:equivalent_opt}
\end{equation}

We first show that for any feasible $\cc{I}$ in \eqref{eqn:equivalent_opt}, $|\cc{I}| = |\cc{I}^\star|$. For simplicity, let $e = 1$ be the observational environment. Then, the relations $\Sigma_{X^e|H^e}^\star = (\id-{B})^{-1}{\Omega}_e(\id-{B}) \text{ for all }e=1,2,\dots,m$ imply for every $e = 2,3,\dots,m$
\begin{equation*}
(\id-B^\star)^{-1}\left[\Omega_e^\star-\Omega_1^\star\right](\id-B^\star)^{-T} = (\id-B)^{-1}\left[\Omega_e-\Omega_1\right](\id-B)^{-T}.
\end{equation*}
Since $\Omega_e^\star-\Omega_1^\star \succeq 0$ by Assumption~\ref{assump:observational}, the relation above lets us conclude that $\Omega_e - \Omega_1 \succeq 0$. Hence, $|\cc{I}^\star| = \text{rank}\left(\sum_{e=2}^m\Omega_e^\star-\Omega_1^\star\right)$ and $|\cc{I}| = \text{rank}\left(\sum_{e=2}^m\Omega_e-\Omega_1\right)$. Finally, again by the relation $\Sigma_{X^e|H^e}^\star = (\id-{B})^{-1}{\Omega}_e(\id-{B})^{-T} \text{ for all }e=1,2,\dots,m$, we have that:
\begin{equation*}
    (\id-B^\star)^{-1}\left[\sum_{e=2}^m\Omega_e^\star-\Omega_1^\star\right](\id-B^\star)^{-T} = (\id-B)^{-1}\left[\sum_{e=2}^m\Omega_e-\Omega_1\right](\id-B)^{-T},
\end{equation*}
which lets us conclude that $\text{rank}\left(\sum_{e=2}^m\Omega_e^\star-\Omega_1^\star\right) = \text{rank}\left(\sum_{e=2}^m\Omega_e-\Omega_1\right)$ and consequently $|\cc{I}| = |\cc{I}^\star|$. 

Next, we show that any optimal DAG in \eqref{eqn:equivalent_opt} must be in the Markov equivalence class of $\dagg^\star$. Consider the distribution $X^e|H^e$ which is specified by the covariance $\Sigma^\star_{X^e|H^e}$. For any DAG $\dagg$ compatible\footnote{By compatible, we mean that there exists $B$ compatible with $\dagg$ and $\Omega \in \mathbb{D}_{++}^p$ such that $\Sigma^\star_{X^e|H^e} = (\id-B)^{-1}\Omega(\id-B)^{-T}$.} with $X^e|H^e$, the following are satisfied:
\begin{equation}
    \begin{aligned}
      \{(i,j,S): X^e_i \independent X^e_j | X^e_S,H^e \text{ for some set }S\} \subseteq \{(i,j,S): X_S \text{ d-separates } X_i \text{ and }X_j \text{ in }\dagg\}, \\
      \{(i,j): X^e_i \independent X^e_j | X^e_{\backslash\{i,j\}},H^e\} \subseteq \{(i,j): X^e_i \text{ and }X^e_j \text{ are not connected in }\text{moral}(\dagg)\},
    \end{aligned}
    \label{eqn:DAG_rel}
\end{equation}
where set equality in the relations above hold if $X^e|H^e$ is faithful with respect to the DAG $\dagg$. By Assumption~\ref{assum:faithful} (i.e. faithfulness of $X^e|H^e$ for every environment with respect to the DAG $\dagg^\star$), and relation \eqref{eqn:DAG_rel}, we have for any DAG $\dagg$ consistent with $X^e|H^e$:
\begin{equation}
\begin{aligned}
    \|\dagg^\star\|_{\ell_0} \leq \|\dagg\|_{\ell_0}~~\,;~~
    \text{moral}(\dagg^\star) \subseteq \text{moral}(\dagg),
    \end{aligned}
    \label{eqn:dagg_comparison}
\end{equation}
where equality holds if and only if $\dagg \in \mathrm{MEC}(\dagg^\star)$. Combining this fact with $|\cc{I}| = |\cc{I}^\star|$ for any feasible $\cc{I}$ in \eqref{eqn:equivalent_opt}, we conclude that $\dagg_\text{all.opt}\subseteq \mathrm{MEC}(\dagg^\star)$.

\end{proof}

\section{Uninformative interventions and proof of Theorem~\ref{thm:sparse_dense}}
\subsection{Worst-case interventions}
\label{necessary_intervention}
In Section~\ref{proof_prop_dense_without_interventions}, we proved that in the setting of Proposition~\ref{thm:sparse_dense_1}, the optimal DAGs $\dagg_\mathrm{all.opt}$ in the optimization \eqref{eq:optimal_params_inc_condition} satisfy $\dagg_\mathrm{all.opt}\subseteq \mathrm{MEC}(\dagg^\star)$. We next show that there are worst-case intervention configurations such that $\dagg_\mathrm{all.opt} = \mathrm{MEC}(\dagg^\star)$. As an example, suppose for every $e,e' \in 1,2,\dots,m$, there exists $\alpha_{e,e'}>0$ such that
\begin{equation}
    \Omega_e^\star = \alpha_{e,e'}\Omega_{e'}^\star.
    \label{eqn:construction}
\end{equation}
By our construction \eqref{eqn:construction}, $\Sigma_{X^e|H^e} = \alpha_{e,e'}\Sigma_{X^{e'}|H^{e'}}$. Recall that the optimal DAGs satisfy the relation \eqref{eqn:equivalent_opt}. However, since $\Sigma_{X^e|H^e} = \alpha_{e,e'}\Sigma_{X^{e'}|H^{e'}}$, it is straightforward to see that when \eqref{eqn:construction} is satisfied, there is no additional information gained over just data in a single environment, i.e. \eqref{eqn:equivalent_opt} is simplified to:
\begin{equation}
\begin{aligned}
    \dagg_\text{all.opt} = \argmin_{D}\, &~~~ \mathcal{R}_{\gamma}(\dagg,\{1,2,\dots,p\}) \\\text{subject-to}\, &~~~ \text{ there exists }B \sim \dagg, \Omega  \in \mathbb{D}_{++}^p \text{ such that}\\ \, &~~~~\Sigma_{X^1|H^1}^\star = (\id-{B})^{-1}{\Omega}(\id-{B})^{-T}.
    \end{aligned}
    \label{eqn:equivalent_opt_reduced}
\end{equation}
Since $X^e|H^e$ is faithful with respect to $\dagg^\star$, following relation \eqref{eqn:dagg_comparison}, we conclude that $\dagg_\text{all.opt} = \mathrm{MEC}(\dagg^\star)$. 

\subsection{Proof of Theorem~\ref{thm:sparse_dense}}
\label{proof_thm_dense}
The proof of this theorem relies on two lemmas.
\begin{lemma}Let $B,\tilde{B} \in \mathbb{R}^{p \times p}$ be two matrices that can be made to be lower-triangular with zeros on the diagonal after row and column permutations (or equivalently, the matrices correspond to two DAGs). Suppose that there exists $\tilde{\Omega},\Omega \in \mathbb{D}_{++}^p$ such that $(\id-B)^{-1}\Omega(\id-B)^{-T} = (\id-\tilde{B})^{-1}\tilde{\Omega}(\id-\tilde{B})^{-T}$. Then, if for some $\alpha \neq 0$, $[(\id-\tilde{B})^{-1}(\id-B)]_{:,i} = \alpha\mathbf{e}_j$, then $[\id-B]_{i,:} \propto [\id-\tilde{B}]_{j,:}$.
\label{lemma:connecting_B_Btild}
\end{lemma}
\begin{proof}[Proof of Lemma~\ref{lemma:connecting_B_Btild}]
By the condition of the lemma, we have that: $\alpha\mathbf{e}_j^T(\id-\tilde{B})^{-T}(\id-B)^T = \mathbf{e}_i^T$ and that $(\id-\tilde{B})= \Omega(\id-\tilde{B})^{-T}(\id-B)^T\Omega^{-1}(\id-B)$. Combining these two, it follows that for some constant $\beta \neq 0$, $\mathbf{e}_j^T(\id-\tilde{B}) = \alpha\beta\mathbf{e}_i^T(\id-B)$.
\end{proof}
\begin{lemma}[Equivalent characterization of $\simec(\dagg^\star)$]
Under Assumptions~\ref{assump:heterogenous} and \ref{assump:truthful}, the following statements are equivalent for $\dagg \in \mathrm{MEC}(\dagg^\star)$:
\begin{enumerate}
    \item[p1)] There exists a connectivity matrix $B$ compatible with $\dagg$ and $\{\Omega_e\}_{e=1}^m \subseteq \mathbb{D}_{p}^{++}$ such that $(\id-B^\star)^{-1}\Omega_e^\star(\id-B^\star)^{-T} = (\id-B)^{-1}\Omega_e(\id-B)^{-T}$ for all $e = 1,2,\dots,m$
    \item[p2)] $\dagg \in \simec(\dagg^\star)$
\end{enumerate}
\label{lem:equivalence_IMEC}
\end{lemma}
\begin{proof}[Proof of Lemma~\ref{lem:equivalence_IMEC}]
The direction $p2 \Rightarrow p1$ follows from Lemma~\ref{lemma:equiv_without_latent}. We next prove $p1 \Rightarrow p2$. In particular, we must show that if for a $\dagg \in \mathrm{MEC}(\dagg^\star)$ with a compatible connectivity matrix $B$ satisfying $(\id-B^\star)^{-1}\Omega_e^\star(\id-B^\star)^{-T} = (\id-B)^{-1}\Omega_e(\id-B)^{-T}$ for all $e = 1,2,\dots,m$, then the parents of nodes indexed by $\cc{I}^\star$ in $\dagg$ are fixed to be the same as parents of $\dagg^\star$. More concretely, we will show that:
\begin{equation*}
    B_{i,:} = B^\star_{i,:}~~\,~~ \text{ for all }i\in \cc{I}^\star.
\end{equation*}
We define the following parameters:
\begin{equation*}
    M:= (\id-B)(\id-B^\star)^{-1}~~~;~~~N_e := (\id-B)(\id-B^\star)^{-1}\Omega_e^\star \,~,~ e=1,2,\dots,m.
\end{equation*}
Notice that the condition of $p1$ implies
\begin{equation}
    M_{k,:} \perp [N_e]_{l,:} \text{ for }k \neq l.
    \label{eqn:orthogonality_constraint}
\end{equation}
Since the rows of the matrix $M$ are linearly independent, the relation \eqref{eqn:orthogonality_constraint} implies that for every $k = 1,2,\dots,p$, the vectors $\{[N_e]_{k,:}\}_{e=1}^m$ live in a one-dimensional null space of the matrix formed by concatenating the vectors $\{M_{l,:}\}_{l \neq k}$, i.e.
\begin{equation}
   \text{dim}\left(\text{span}\left(\{[N_e]_{k,:}\}_{e=1}^m\right)\right)=1.
       \label{eqn:one_dim}
\end{equation}
We focus on a particular $i \in \cc{I}^\star$. Take an environment $e$ satisfying Assumption~\ref{assump:heterogenous} for $i \in \cc{I}^\star$. Then, \eqref{eqn:one_dim} implies that for every $k$, there exists a constant $c_k \neq 0$ (nonzero since the matrix $N_{e'}$ is non-singular for every $e'$) such that $[N_{1}]_{k,:} = c_k[N_{e}]_{k,:}$ or equivalently $M_{k,:}\Omega_1^\star = c_kM_{k,:}\Omega_{e}^\star$. Thus, combining this fact with Assumption 3, we conclude that:
\begin{equation}
\begin{aligned}
&\text{for every } k = 1,2,\dots,p, \\
& [M]_{k,i} = 0 ~~~~ \text{OR}~~~~[M]_{k,i} \text{ has one nonzero and }[M]_{k,\{1,2,\dots,p\}\backslash{i}} = 0.
    \end{aligned}
    \label{eqn:structure_M}
\end{equation}
Combining \eqref{eqn:structure_M} with the fact that the rows of $M$ are linearly independent, we conclude that:
\begin{equation}
    \begin{aligned}
    [M]_{:,i} = \alpha{\bf e}_j \text{ for some standard basis element }{\bf e}_j \text{ and }\alpha \neq 0.
    \end{aligned}
    \label{eqn:structure_M_new}
\end{equation}
Appealing to Lemma~\ref{lemma:connecting_B_Btild}, we conclude that \eqref{eqn:structure_M_new} can be equivalently written as: 
\begin{equation}
    \begin{aligned}
    [M]_{i,:} = \alpha{\bf e}_j^T \text{ for some standard basis element }{\bf e}_j \text{ and }\alpha \neq 0.
    \end{aligned}
    \label{eqn:structure_M_new_equivalent}
\end{equation}
We consider two scenarios. Scenario 1 is when $j \neq i$ in \eqref{eqn:structure_M_new} and Scenario 2 is when $j = i$. Our proof strategy is to show that under Assumption~\ref{assump:truthful},  Scenario 1 \emph{cannot} occur, implying that only Scenario 2 is possible. For Scenario 2, we conclude that $B_{i,:} = B^\star_{i,:}$. 

\noindent\underline{Scenario 1: $j \neq i$ in \eqref{eqn:structure_M_new_equivalent}} Since $B$ is a connectivity matrix associated with a DAG $\dagg \in \mathrm{MEC}(\dagg^\star)$, by Assumption 4, this case is not allowed.

\noindent\underline{Scenario 2: $j=i$ in \eqref{eqn:structure_M_new}} Here, we have that $[M]_{:,i} = \alpha{\bf e}_i$ for nonzero $\alpha$. Hence, $M_{i,i} \neq 0$. By \eqref{eqn:structure_M}, we then conclude that $M_{i,:}= \alpha{\bf e}_i^T$, or equivalently $[\id-B]_{i,:} = \alpha[\id-B^\star]_{i,:}$. Since $\id-B$ and $\id-B^\star$ have ones on the diagonal, $\alpha = 1$ and thus $B_{i,:} = B^\star_{i,:}$.\\
We repeat the above arguments for every $i \in \cc{I}^\star$ to conclude that $B_{i,:} = B^\star_{i,:}$.
\end{proof}
\begin{proof}[Proof of Theorem~\ref{thm:sparse_dense}]
Let $\dagg_\text{all.opt},B_\text{all.opt},\cc{I}_\text{all.opt}$ be the collection of optimal DAGs and intervention targets in the optimization \eqref{eq:optimal_params_inc_condition}. From \eqref{eqn:equivalent_opt} and Proposition~\ref{thm:sparse_dense_1}, we have that any $\dagg \in \dagg_\text{all.opt} \subseteq \mathrm{MEC}(\dagg^\star)$ with an associated connectivity matrix $B$ and noise variances $\Omega_e$ satisfies: 
\begin{equation*}
    (\id-B)(\id-B^\star)^{-1}\Omega_e^\star(\id-B^\star)^{1}(\id-B)^{T} = \Omega_e ~~~e=1,2,\dots,m,
\end{equation*}
where $\{\Omega_e\}_{e=1}^m$ are diagonal and $\cc{I} = \mathbb{I}(\{\Omega_e\}_{e=1}^m)$. By Lemma~\ref{lem:equivalence_IMEC}, we have that $\dagg \in \simec(\dagg^\star)$. Thus, the associated connectivity matrix $B$ satisfies for every $i \in \cc{I}^\star$: $\text{support}(B_{i,:}) = \text{support}(B^\star_{i,:})$. Then, appealing to Lemma~\ref{lemma:structure}, we conclude that $[\Omega^\star_e]_{i,i} = [\Omega_e]_{i,i}$ for every $i \in \cc{I}^\star$ and $e \in [m]$. Further, since the matrix $(\id-B)(\id-B^\star)^{-1}$ is invertible, we have that $\text{rank}(\Omega_e^\star-\Omega_f^\star) = \text{rank}(\Omega_e - \Omega_f)$. Combining the previous two facts, we conclude that $\cc{I} = \cc{I}^\star$ so that $\cc{I}_\mathrm{all.opt} = \{\cc{I}^\star\}$. Appealing to Lemma~\ref{lem:equivalence_IMEC}, we conclude that \eqref{eqn:equivalent_opt} can be equivalently expressed as:
\begin{equation*}
    \dagg_\text{all.opt} = \arg\min_{\text{DAG }\dagg} \mathcal{R}_\gamma(\dagg,\cc{I}^\star)~~\text{subject-to}~~\dagg \in \simec(\dagg^\star).
\end{equation*}
Since $\simec(\dagg^\star) \subseteq \mathrm{MEC}(\dagg^\star)$, the regularizer $\mathcal{R}_\gamma(\dagg,\cc{I}^\star) = \mathcal{R}_\gamma(\dagg^\star,\cc{I}^\star)$ for all $\dagg \in \simec(\dagg^\star)$. We thus conclude that $\dagg_\text{all.opt} = \simec(\dagg^\star)$. Finally, since the population parameters are feasible in the optimization \eqref{eq:optimal_params_inc_condition} and that $\dagg^\star$ and $\cc{I}^\star$ achieve the optimum loss $\mathcal{R}_\gamma(\cdot,\cdot)$, we conclude that $B^\star \in B_\text{all.opt}$.
\end{proof}

\subsection{Proof of Corollary~\ref{cor:incoherent}}
\label{proof_corr_sufficient}
Consider the optimization problem
\begin{equation}
\begin{aligned}
    \argmin_{\dagg,\bar{h},B,\Gamma,\cc{I},\{\Omega_e,\Psi_e\}_{e=1}^m}&~~\mathcal{R}_\gamma(\dagg,\cc{I})\\\text{subject-to}:&~~ B \sim \dagg, \mathcal{I} = \mathbb{I}(\{\Omega_e\}_{e=1}^m),\text{ and}\\
    &~~\Sigma_e^\star = (\id-B)^{-1}(\Omega_e+\Gamma\Psi_e\Gamma^T)^{-1}(\id-B)^{-T} ~\text{for every }e\in[m]\\
    &~~\bar{h} \leq h_\text{max},
    \end{aligned}
        \label{eq:optimal_params_latent_bound}
\end{equation}
Where compared to \eqref{eq:optimal_params}, we have added a constraint $\bar{h} \leq h_\text{max}$ on the number of latent variables included in the model. Following exactly similar logic as proof of Proposition~\ref{prop:asymptotic_property}, one can show that with probability tending to one, in the infinite data regime, the optimal solutions of \eqref{eqn:estimator_theory} with the constraint on the number of latent variables equal to the optimal solution of \eqref{eq:optimal_params_inc_condition}. Thus, we will analyze the estimates produced by \eqref{eq:optimal_params_latent_bound}. 

We follow a very similar proof technique as proof of Theorem~\ref{thm:sparse_dense}. Specifically, let $(\dagg,{B},{\Gamma},{\cc{I}},\allowbreak\{\Omega_e,\Psi_e\}_{e=1}^m)$ be any optimal set of parameters in \eqref{eq:optimal_params_latent_bound}. Then, we can arrive at the equality \eqref{eqn:relation_L_K} where the matrices $K_e,K^\star_e, L_e,L^\star_e$ are defined in \eqref{eqn:K_L_defn}. Note that from Lemma~\ref{lemma:sparse_low_rank} that for all $e = 1,2,\dots,m$:
\begin{equation}
    \degree({K}_e-K_e^\star)\inc[\text{col-space}({L}_e-L_e^\star)]^2 < 1 \Rightarrow {K}_e = K_e^\star.
    \label{eqn:desired}
\end{equation}
In this analysis, we show that under the conditions described in Corollary~\ref{cor:incoherent}, $\degree({K}_e-K_e^\star)\inc[\text{col-space}({L}_e-L_e^\star)]^2 < 1$, allowing us to conclude that ${K}_e = K_e^\star$. Following then an exact line of reasoning as the last paragraph of proof of Theorem~\ref{thm:sparse_dense}, we conclude that $\dagg_\text{all.opt} = \simec(\dagg^\star)$, $B^\star \in B_\text{all.opt}$ and $\cc{I}_\text{all.opt} = \{\cc{I}^\star\}$, where $\dagg_\text{all.opt},B_\text{all.opt},\cc{I}_\text{all.opt}$ is the collection of optimal DAGs, connectivity matrices, and intervention sets, respectively, in the optimization \eqref{eq:optimal_params_latent_bound}. 

Notice that $\degree({K}_e-K_e^\star) \leq \degree[\text{moral}({\dagg})]+d^\star$. Further, by Lemma~\ref{lemma:incoherent_col_spaces}, $\inc[\text{col-space}({L}_e-L_e^\star)] \leq 2(\inc[\text{col-space}({L}_e)]+\inc^\star_e)$. Thus, it suffices to show for all $e\in[m]$ that:
\begin{equation}
    4(\degree[\text{moral}({\dagg})]+d^\star) (\inc[\text{col-space}({L}_e)]^2+(\inc^\star_e)^2)<1.
    \label{eqn:sufficient_cond_moral}
\end{equation}
Since the population parameters satisfy the constraint of the optimization problem \eqref{eq:optimal_params_latent_bound}, we have that $\mathcal{R}_\gamma({\dagg},{\cc{I}}) \leq \mathcal{R}_\gamma({\dagg}^\star,\cc{I}^\star)$. Thus, we can conclude with the choice of $\gamma$ that $\degree[\text{moral}({\dagg})] \leq 3d^\star$. Therefore, the following conditions are satisfied for every $e=1,2,\dots,m$ due to Assumption 1, the conditions of Corollary~\ref{cor:incoherent}, and the bound $d^\star \leq \nu^\star$:
\begin{equation*}
    \begin{aligned}
    &\degree[\text{moral}({\dagg})]\inc[\text{col-space}({L}_e)]^2 \leq 3\nu^\star\inc[\text{col-space}({L}_e)]^2 <1/16,\\
    &\degree[\text{moral}({\dagg})]\inc[\text{col-space}({L}_e^\star)]^2 \leq 3\nu^\star\inc[\text{col-space}({L}_e^\star)]<1/16, \\
    &d^\star\inc[\text{col-space}({L}_e)]^2 \leq \nu^\star\inc[\text{col-space}({L}_e)]^2 <1/16, \\
    &d^\star(\inc^\star_e)^2 < 1/16.
    \end{aligned}
\end{equation*}
Combining these relations, we arrive at the inequality in \eqref{eqn:sufficient_cond_moral}.

\section{Proof of Theorem~\ref{thm:perturbed_latent_known} with known latent perturbations}
\label{sec:known_latent_proof}
Consider the optimization problem
\begin{equation}
\begin{aligned}
    \argmin_{\dagg,\bar{h},B,\Gamma,\cc{I},\{\Omega_e,\Psi_e\}_{e=1}^m}&~~\mathcal{R}_\gamma(\dagg,\cc{I})\\\text{subject-to}:&~~ B \sim \dagg, \mathcal{I} = \mathbb{I}(\{\Omega_e\}_{e=1}^m),\text{ and}\\
    &~~\Sigma_e^\star = (\id-B)^{-1}(\Omega_e+\Gamma\Psi_e\Gamma^T)^{-1}(\id-B)^{-T} ~\text{for every }e\in[m]\\
    &~~\Psi_e = \psi_e^\star\id \text{ for every }e \in [m],
    \end{aligned}
    \label{eq:optimal_params_known_latent_perturb}
\end{equation}
Where compared to \eqref{eq:optimal_params}, we have added the known latent perturbations constraint $\Psi_e = \psi_e^\star\id$. Following exactly similar logic as proof of Proposition~\ref{prop:asymptotic_property}, one can show that with probability tending to one, the optimal solutions of \eqref{eqn:estimator_theory} with the known latent perturbations constraint equal to the optimal solution of \eqref{eq:optimal_params_known_latent_perturb}. Thus, we will analyze the estimates produced by \eqref{eq:optimal_params_known_latent_perturb}. We will let $\dagg_\text{all.opt},B_\text{all.opt},\cc{I}_\text{all.opt}$ be optimal DAGs, connectivity matrices, and intervention sets according to the optimization \eqref{eq:optimal_params_known_latent_perturb}.

Consider any optimal set of parameters $(\dagg,B,\Gamma,\{\Omega_e,\Psi_e)$ of \eqref{eq:optimal_params_known_latent_perturb}. Then for every $e = 1,2,\dots,m$
\begin{equation}
\begin{aligned}
&(\id-B^\star)^{-1}(\Omega_e^\star + {\psi_e^\star}\Gamma^\star{\Gamma^\star}^T)(\id-B^\star)^{-T} = (\id-B)^{-1}(\Omega_e + {\psi_e^\star}\Gamma{\Gamma}^T)(\id-B)^{-T}.  \end{aligned}
\label{eqn:known_psi}
\end{equation}
The relations \eqref{eqn:known_psi} imply:
\begin{equation}
\begin{aligned}
    (\id-B^\star)^{-1}\left(\Omega_2^\star-{\psi_2^\star}/{\psi_1^\star}\Omega_1^\star\right)(\id-B^\star)^{-T} &=(\id-B)^{-1}\left(\Omega_2-{\psi_2^\star}/{\psi_1^\star}\Omega_1\right)(\id-B)^{-T}, \\
     (\id-B^\star)^{-1}\left(\Omega_e^\star-{\psi_e^\star}/{\psi_1^\star}\Omega_1^\star\right)(\id-B^\star)^{-T} &=(\id-B)^{-1}\left(\Omega_e-{\psi_e^\star}/{\psi_1^\star}\Omega_1\right)(\id-B)^{-T}.
    \end{aligned}
    \label{eqn:known_psi_1}
\end{equation}
Appealing to Assumption 3', the first relation in   \eqref{eqn:known_psi_1} yields:
\begin{equation}
    (\id-B^\star)^{-1}\Omega_1^\star(\id-B^\star)^{-T} = \frac{1}{(1-\psi_2^\star/\psi_1^\star)}(\id-B)^{-1}\left(\Omega_2-{\psi_2^\star}/{\psi_1^\star}\Omega_1\right)(\id-B)^{-T}.
        \label{eqn:known_psi_2}
\end{equation}
Relation \eqref{eqn:known_psi_2} states that $B$ is an equivalent connectivity with respect to distribution $X^1|H^1$ in the sense $\Sigma_{X^1|H^1}^\star = (\id-B)^{-1}\Omega(\id-B)^{-T}$ for some $\Omega \in \mathbb{D}_{++}^p$. Due to Assumption 2 (faithfulness condition), we appeal to relation \eqref{eqn:dagg_comparison}, and conclude that: $\|\dagg^\star\|_{\ell_0} \leq \|\dagg\|_{\ell_0}$ and $\text{moral}(\dagg^\star) \subseteq \text{moral}(\dagg)$, where equality holds if and only if $\dagg \in \mathrm{MEC}(\dagg^\star)$. Thus, since $\gamma \leq \frac{1}{|\cc{I}^\star|}$, the following holds:
\begin{equation*}
\begin{aligned}
    &\text{for any DAG }\dagg \text{ with a connectivity matrix }B \text{ satisfying } \eqref{eqn:known_psi_2} \text{ where } \dagg \not\in \mathrm{MEC}(\dagg^\star), \\
    & \Rightarrow, \\
    &\mathcal{R}_{\gamma}(\dagg,\cc{I}) > \mathcal{R}_{\gamma}(\dagg^\star,\cc{I}^\star) \text{ for all }\cc{I} \subseteq [p].
\end{aligned}
\end{equation*}
The relation above allows us to conclude that $\dagg \in \mathrm{MEC}(\dagg^\star)$.

We will next show that $B_{i,:} = B^\star_{i,:}$ for all $i \in \cc{I}^\star$. Consider a particular $i \in \cc{I}^\star$. Let $e$ be the environment satisfying Assumption 4'. Let $M := (\id-B)^{-1}(\id-B^\star)$ and $N_1 :=M(\Omega_2^\star-{\psi_2^\star}/{\psi_1^\star}\Omega_1^\star) = (1-{\psi_2^\star}/{\psi_1^\star})\Omega_1^\star$, $N_2 :=M(\Omega_e^\star-{\psi_e^\star}/{\psi_1^\star}\Omega_1^\star)$. From Assumption 4', we have that $N_1, N_2$ are non-singular. Further, from \eqref{eqn:known_psi_1}:
\begin{equation*}
    M_{k,:} \perp [N_1]_{l,:} \text{ and }M_{k,:} \perp [N_2]_{l,:}~~\text{ for }k\neq{l}.    
\end{equation*}
Since the rows of the matrix $M$ are linearly independent, we have for every $k = 1,2,\dots,p$:  $[N_1]_{k,:}$ and $[N_2]_{k,:}$ are linearly independent. Thus, there exists constant $c\neq{0}$ such that $c(1-{\psi_2^\star}/{\psi_1^\star})M_{k,:}\Omega_1^\star = M_{k,:}(\Omega_e^\star-{\psi_e^\star}/{\psi_1^\star}\Omega_1^\star)$. Suppose $M_{k,i} \neq 0$. We will argue that $M_{k,j} = 0$ for all $j\neq i$. Specifically, suppose $M_{k,j} \neq 0$ for $j\neq i$. We would have that $[\Omega_e^\star-\psi_e^\star/\psi_1^\star\Omega_1^\star]_{i,i}[\Omega_1^\star]_{j,j}^{-1} =[\Omega_e^\star-\psi_e^\star/\psi_1^\star\Omega_1^\star]_{j,j}[\Omega_1^\star]_{j,j}^{-1}$. However, we arrive at a contradiction by Assumption 4'. We have thus argued that the matrix $M$ has the structure described in \eqref{eqn:structure_M_new}. Furthermore, going through the scenarios described in the proof of Lemma~\ref{lem:equivalence_IMEC} and appealing to the intervention truthfulness condition in Assumption \ref{assump:truthful} (using relation \eqref{eqn:known_psi_2}), we conclude that $B_{i,:} = B^\star_{i,:}$ for all $i \in \cc{I}^\star$. In other words, we have now shown that $\dagg_\text{opt}\subseteq \simec(\dagg^\star)$. 

We must now show that for all $\dagg \in \simec(\dagg^\star)$, $\dagg \in \dagg_\text{opt}$. Let $(\tilde{\dagg},\tilde{B},\tilde{\Gamma},\tilde{\cc{I}},\{\tilde{\Omega}_e,\tilde{\Psi}_e\}_{e=1}^m)$ be any optimal set of parameters of \eqref{eq:optimal_params_known_latent_perturb} where we have shown that $\tilde{\dagg} \in \simec(\dagg^\star)$. By Lemma~\ref{lemma:equiv_without_latent}, there exists a connectivity matrix $B \sim \dagg$, and noise variances $\Omega_e$ such that for all $e=1,2,\dots,m$
\begin{eqnarray}
\id-\tilde{B})^{-1}\tilde{\Omega}_e(\id-\tilde{B})^{-T} = (\id-{B})^{-1}{\Omega}_e(\id-{B})^{-T}.
\label{eqn:temp1}
\end{eqnarray}
Furthermore, let $\Gamma = \tilde{\Gamma}$ and $\Psi_e = \psi_e^\star\id$. We then have that the model $({\dagg},{B},{\Gamma},{\cc{I}},\{{\Omega}_e,{\Psi}_e\}_{e=1}^m)$ is feasible in \eqref{eq:optimal_params_known_latent_perturb}. It remains to check that $\mathcal{R}_\gamma(\dagg,\cc{I}) = \mathcal{R}_\gamma(\tilde{\dagg},\tilde{\cc{I}})$. Since $\dagg$ and $\tilde{\dagg}$ are in the same Markov equivalence class, it suffices to show that $\cc{I} = \tilde{\cc{I}}$. Since $\dagg$ and $\tilde{\dagg}$ are both in the set $\simec(\dagg^\star)$, we have that $\tilde{B}_{i,:} = B_{i,:}$ for every $i \in \tilde{\cc{I}}$. From Lemma~\ref{lemma:structure}, we have that: $[\Omega_e]_{i,i} = [\tilde{\Omega}_e]_{i,i}$ for every $i \in \tilde{\cc{I}}$. Let $\cc{I} = \mathbb{I}(\{\Omega_e\}_{e=1}^m)$. We have shown that $\tilde{\cc{I}} \subseteq \cc{I}$. Suppose that there exists $j \in \cc{I} \setminus \tilde{\cc{I}}$. Then, there must exist $e,f$ such that $[\Omega_e]_{j,j} = [\Omega_f]_{j,j}$. From \eqref{eqn:temp1}, we have that:
\begin{eqnarray}
\id-\tilde{B})^{-1}(\tilde{\Omega}_e-\tilde{\Omega}_f)(\id-\tilde{B})^{-T} = (\id-{B})^{-1}({\Omega}_e-\Omega_f)(\id-{B})^{-T}. 
\label{eqn:temp2}
\end{eqnarray}
We have that $[\tilde{\Omega}_{e}]_{i,i}-[\tilde{\Omega}_{f}]_{i,i} = 0$ for every $i \not\in \tilde{\cc{I}}$. We also have that $[\tilde{\Omega}_{e}]_{i,i}-[\tilde{\Omega}_{f}]_{i,i} = [{\Omega}_{e}]_{i,i}-[{\Omega}_{f}]_{i,i}$ for every $i \in \cc{I}$. Since $[{\Omega}_{e}]_{j,j}-[{\Omega}_{f}]_{j,j} \neq 0$, we have that $\text{rank}({\Omega}_e-\Omega_f) > \text{rank}(\tilde{\Omega}_e-\tilde{\Omega}_f)$ which is a contradiction given \eqref{eqn:temp2}. Therefore, we conclude that $\cc{I} = \tilde{\cc{I}}$.

\section{Three environments are required if the number of latent variables is not constrained}
\label{sec:two_environments}
Without imposing a condition on the number of latent variables, two environments can only offer identifiability up the Markov equivalence class, even if all the variables are perturbed. To offer some intuition, we sketch a quick argument below. Consider the setting in Section~\ref{sec:known_latent_perturb} where the number of latent variables is allowed to be arbitrary. For simplicity, we assume there are two environments with no perturbations on the latent variables and parameters $(B^\star,\Gamma^\star,\{\Omega_e^\star\}_{e = 1}^2)$ where $B^\star$ represents the connectivity matrix, $\Gamma^\star$ encodes the effect of latent variables, and $\Omega_e^\star$ is a diagonal matrix encoding the noise terms on the observed variables for environments $e = 1,2$. Here, for example, $\Omega_1^\star$ is the noise variance associated with an observational environment and $\Omega_2^\star$ is the noise term associated with an interventional environment with $\Omega_2^\star \succ \Omega_1^\star$ (since there are interventions on all variables). Any compatible causal model, parameterized by $(B,\Gamma,\{\Omega_e\}_{e=1}^2)$ must entail the same covariance model, i.e. for $e = 1,2$:
\begin{equation}
    (\id-B^\star)^{-1}({\Gamma^\star}{\Gamma^\star}^T + \Omega_e^\star)^{-1}(\id-B^\star)^{-T} =  (\id-B)^{-1}({\Gamma}{\Gamma}^T + \Omega_e)^{-1}(\id-B)^{-T}.
    \label{eqn:relation}
\end{equation}
An equivalent reformulation of \eqref{eqn:relation} is:
\begin{equation}
\begin{aligned}
    (\id-B^\star)^{-1}(\Omega_2^\star-\Omega_1^\star)(\id-B^\star)^{-T} &=     (\id-B)^{-1}(\Omega_2-\Omega_1)(\id-B)^{-T},\\
    (\id-B^\star)^{-1}(\Omega_1^\star+\Gamma^\star{\Gamma^\star}^T)(\id-B^\star)^{-T} &= (\id-B)^{-1}({\Gamma}{\Gamma}^T + \Omega_1)^{-1}(\id-B)^{-T}. 
\end{aligned}
\label{eqn:reformulation}
\end{equation}
It is straightforward to check that for any DAG in the Markov equivalence class of the population DAG, there exists a connectivity matrix $B$ and diagonal matrix $D$ such that $(\id-B^\star)^{-1}(\Omega_2^\star-\Omega_1^\star)(\id-B^\star)^{-T} = (\id-B)^{-1}D(\id-B)^{-T}$. Let $\Omega_1$ be any positive definite diagonal matrix such that $(\id-B)(\id-B^\star)^{-1}(\Omega_2^\star-\Omega_1^\star)(\id-B^\star)^{-T}(\id-B)^T \succ \Omega_1$. Furthermore, let $\Gamma$ be any matrix square root of $(\id-B)(\id-B^\star)^{-1}(\Omega_2^\star-\Omega_1^\star)(\id-B^\star)^{-T}(\id-B)^T-\Omega_1$. Finally, let $\Omega_2 = \Omega_1+D$. Notice that by construction, the parameters $(B,\Gamma,\{\Omega_e\}_{e=1}^2)$ satisfy the relations in \eqref{eqn:reformulation}. In other words, we have shown that even though all observed variables have received an intervention, we are not able to attain any additional identifiability than the Markov equivalence class. A similar analysis can also be done in the case where there are perturbations on the latent variables.  
 \section{Approximately known latent perturbations}
 \label{sec:approximately_known}
 In this section, we relax the assumption of knowing the latent perturbations to having access to approximate values, where the level of approximation is given by $C_\psi$. In particular, we are given approximations $\tilde{\psi}_1,\tilde{\psi}_2,\dots,\tilde{\psi}_m$ where $\max\{\tilde{\psi}_e-\psi_e^\star,|1/\tilde{\psi}_e - 1/\psi_e^\star|\} \leq C_\psi$ for every $e = 1,2,\dots,m$. We then apply $\methName{}$ with the additional constraint that: $\max\{|1/\psi_e-1/\tilde{\psi}_e|,|\psi_e-\tilde{\psi}_e|\}\leq C_\psi$ for every $e = 1,2,\dots,m$. Having only an approximation to the latent perturbations comes at an expense of additional assumptions for partial identifiability. Specifically, we assume that the perturbations on the observed variables in $\cc{I}^\star$ are sufficiently strong as compared to the approximation error in the latent perturbations. Furthermore, we assume that the latent effects induce some confounding dependencies among the observed variables, although this assumption is generally far weaker than the incoherence condition in Assumption 1 (see Section~\ref{sec:dense_sparse}).

We impose the following assumptions where as introduced in the main paper, $d^\star = \degree\allowbreak[\text{moral}(\dagg^\star)]$:
\begin{customassump}
{1'}latent effects induce some confounding dependencies: $\text{for every }i \in \cc{I}^\star \text{ and }$ $|\kappa| \leq \frac{4C_\psi}{\max_{e,e'}\psi_{e'}^\star/\psi_{e}^\star-1}, \kappa \neq 0, |\delta| \leq C_\psi$, $\text{there exists } e \in [m]$ such that the following conditions hold: i) $[\Omega_e^\star]_{i,i} > [\Omega_1^\star]_{i,i}$, ii) $\degree\left((\id-B^\star)^T(\Omega_e^\star- \Omega_1^\star(\psi_e^\star+\delta)+\kappa\Gamma^\star{\Gamma^\star}^T)^{-1}(\id-B^\star)\right) \geq 3d^\star$ and iii) $\degree\left((\id-B^\star)^T(\Omega_1^\star+\kappa\Gamma^\star{\Gamma^\star}^T)^{-1}(\id-B^\star)\right) \geq 3d^\star$.
\label{assump:new_confound}
\end{customassump}
\begin{customassump}{4''}\text{heterogeneous interventions on observed and latent variables}: i) for  every $e \neq e' \in [m],  \max\left\{\frac{\psi_{e}^\star}{\psi_{e'}^\star},\frac{\psi_{e'}^\star}{\psi_{e}^\star}\right\} > 1+\frac{C_\psi(\lambda_\text{max}(\Omega_1^\star)+\|\Gamma^\star\|_2^2)}{\lambda_\text{min}(\Omega_1^\star)}$, ii) for every $i,j\in \cc{I}^\star, i \neq j, \text{ there exists }e \in [m] \text{ such that }[\Omega^\star_e]_{i,i} > [\Omega^\star_{1}]_{i,i}$ and\\ $\left[\left(\Omega_{e}^\star-{\psi_{e}^\star}\Omega_{1}^\star\right)\left(\Omega_{1}^\star\right)^{-1}\right]_{i,i}\neq\left[\left(\Omega_{e}^\star-{\psi_{e}^\star}\Omega_{1}^\star\right)\left(\Omega_{1}^\star\right)^{-1}\right]_{j,j}$. 
\end{customassump}

\begin{customassump}{6}$\text{sufficiently strong perturbations on the observed variables in }\cc{I}^\star$: $\text{ for every}\allowbreak{i} \in \cc{I}^\star, \text{ there exists } e\in[m] \text{ such that } [\Omega_e^\star]_{i,i} > (2\psi_e^\star+1)[\Omega_1^\star]_{i,i}$.
\end{customassump}

Assumption 1' ensures that the latent effects induce some confounding dependencies. This condition is far weaker than an incoherence-type assumption such as Assumption 1; for a comprehensive discussion, see Section~\ref{sec:comparison_to_incoherence}. Assumption 4'' (analogous to Assumption 4) ensures that the interventions on the latent variables and observed variables are informative for additional identifiability. One can show that if the parameters $\Omega^{\star}_e,\Omega^{\star}_1$ and $\psi^\star_1,\psi^{\star}_2,\psi^{\star}_e$ are drawn from continuous distributions, Assumption 4'' is satisfied almost surely. Finally, Assumption 6 requires that the perturbations on the observed variables in $\cc{I}^\star$ are sufficiently strong.

\begin{theorem}[Equivalence class characterization under approximately known latent perturbations] Consider the estimator \eqref{eqn:estimates_theory} with $\Psi_e = \psi_e\id$ and the constraints $\max\{|\tilde{\psi}_e - \psi_e|,|{1}/{\tilde{\psi}_e}-{1}/{{\psi}_e}|\} \leq C_\psi$ where $C_\psi$ is chosen conservatively so that $\max\{|\tilde{\psi}_e - \psi_e^\star|,|{1}/{\tilde{\psi}_e}-{1}/{{\psi}_e^\star}|\} \leq C_\psi$. Suppose Assumptions 1',2,3', 4'', and 6 are satisfied. Letting $\frac{1}{|\cc{I}^\star} > \gamma >0$, then $\hat{\dagg}_\mathrm{all.opt} = \simec(\dagg^\star)$ with probability tending to one.
\label{thm:perturbed_latent_grouped}
\end{theorem}

To prove Theorem~\ref{thm:perturbed_latent_grouped}, we consider the optimization problem
\begin{equation}
\begin{aligned}
    \argmin_{\dagg,\bar{h},B,\Gamma,\cc{I},\{\Omega_e,\psi_e\id\}_{e=1}^m}&~~\mathcal{R}_\gamma(\dagg,\cc{I})\\\text{subject-to}:&~~ B \sim \dagg, \mathcal{I} = \mathbb{I}(\{\Omega_e\}_{e=1}^m),\text{ and}\\
    &~~\Sigma_e^\star = (\id-B)^{-1}(\Omega_e+\psi_e\Gamma\Gamma^T)^{-1}(\id-B)^{-T} ~\text{for every }e\in[m]\\
    &~~\max\{|\tilde{\psi}_e - \psi_e|,|{1}/{\tilde{\psi}_e}-{1}/{{\psi}_e}|\} \leq C_\psi
    \end{aligned}
        \label{eq:optimal_params_latent_grouped}
\end{equation}
Where compared to \eqref{eq:optimal_params}, we have added the constraint $\Psi_e = \psi_e\id$ (the latent variables are iid), and the constraint $\max\{|\tilde{\psi}_e - \psi_e|,|{1}/{\tilde{\psi}_e}-{1}/{{\psi}_e}|\} \leq C_\psi$ that controls the deviation to which we know the interventions on the latent variables. Following exactly similar logic as proof of Proposition~\ref{prop:asymptotic_property}, one can show that with probability tending to one, the optimal solutions of \eqref{eqn:estimator_theory} with the constraint on the number of latent variables equal to the optimal solution of \eqref{eq:optimal_params_latent_grouped}. Thus, we will analyze the estimates produced by \eqref{eq:optimal_params_latent_grouped}. 

The proof of Theorem~\ref{thm:perturbed_latent_grouped} relies on the following lemmas:
\begin{lemma} Let $\dagg$ and $B$ be any DAG and associated connectivity matrix that is optimal with respect to \eqref{eq:optimal_params_latent_grouped}. Suppose there exists a non-singular diagonal matrix $\Omega \in \mathbb{D}^p$ satisfying the following relation for any $e \in [m]$, $|\kappa| \leq 4C_\psi/\max_{e,e'}(\psi_{e'}^\star/\psi_{e}^\star-1)$, $|\delta| \leq C_\psi$:
\begin{equation}
    (\id-B^\star)^{-1}(\Omega_e^\star -\Omega_1^\star(\psi_e^\star+\delta)-\kappa\Gamma^\star{\Gamma^\star}^T)(\id-B^\star)^{-T} = (\id-B)^{-1}\Omega(\id-B)^{-T}.
\end{equation}
Then, under Assumption 1', $\kappa = 0$.
\label{lemma:intermediate_1}
\end{lemma}
\begin{lemma}[Sufficient conditions for estimated and true latent perturbations to be equal]Under Assumptions 1',2,3',4'' and 6, we have that for any optimal solution $(B,\Gamma,\cc{I},\{\Omega_e,\Psi_e\}_{e=1}^m)$ of \eqref{eq:optimal_params_latent_grouped}, $\psi_1 = \psi_1^\star$, $\psi_2 = \psi_2^\star$. Furthermore, for any $i \in \cc{I}^\star$, letting $e \in [m]$ be the environment satisfying Assumption 4'', we have that $\psi_e = \psi_e^\star$.  
\label{lemma:equal_perturb}
\end{lemma}
\begin{proof}[Proof of Theorem~\ref{thm:perturbed_latent_grouped}]
We appeal to Lemma~\ref{lemma:equal_perturb} to conclude that for any optimal solution $(B,\Gamma,I,\{\Omega_e,\Psi_e\}_{e=1}^m)$, $\psi_1 = \psi_1^\star$, $\psi = \psi_2^\star$ and for every $e$ satisfying Assumption 6 on the intervention on the observed variables, $\psi_e^\star = \psi_e$. We then follow a similar strategy to prove Theorem~\ref{thm:perturbed_latent_known} to conclude the desired result. 
\end{proof}

\begin{proof}[Proof of Lemma~\ref{lemma:intermediate_1}]
Taking matrix inverses of both sides of the equation in the lemma, we have that:
\begin{equation}
    (\id-B^\star)^{T}(\Omega_e^\star-\Omega_1^\star(\psi_e^\star+\delta)-\kappa\Gamma^\star{\Gamma^\star}^T)^{-1}(\id-B^\star) = (\id-B)^{T}\Omega^{-1}(\id-B).
    \label{relation_lemma_intermediate}
\end{equation}
By Assumption \ref{assump:new_confound}
 and the relation~\eqref{relation_lemma_intermediate}, we have that $\degree[\text{moral}(\dagg)] \geq 3d^\star$. We arrive at a contradiction with $\dagg$ being optimal however since for any $\cc{I}$, 
\begin{equation*}
    \mathcal{R}_\gamma(\dagg,I) > \degree[\text{moral}(\dagg)] \geq 3d^\star \geq \mathcal{R}_\gamma(\dagg^\star,\cc{I}^\star).
\end{equation*}

\end{proof}

\begin{proof}[Proof of Lemma~\ref{lemma:equal_perturb}] Let $(B,\Gamma,\cc{I},\{\Omega_e,\psi_e\id\}_{e=1}^m)$ be any optimal parameters of \eqref{eq:optimal_params_latent_grouped}. Then, by feasibility, we have for every $e \in [m]$
\begin{equation}
\begin{aligned}
&(\id-B^\star)^{-1}(\Omega_e^\star + {\psi_e^\star}\Gamma^\star{\Gamma^\star}^T)(\id-B^\star)^{-T} = (\id-B)^{-1}(\Omega_e + {\psi_e}\Gamma{\Gamma}^T)(\id-B)^{-T}.  \end{aligned}
\label{eqn:known_psip}
\end{equation}
Notice that without loss of generality, $\psi_{1}=1$ as $\Gamma$ can be appropriately scaled. We will first show that the assumptions imply that $\psi_{2}^\star = \psi_2$. We prove $\psi_{2} = \psi_{2}^\star$ both in the settings where $\psi_{2}^\star > \psi_{1}^\star$ and $\psi_{2}^\star < \psi_{1}^\star$.

\noindent\underline{Scenario $\psi_{2}^\star > \psi_{1}^\star$}: The relation \eqref{eqn:known_psip} implies that:
\begin{equation}
\begin{aligned}
    &(\id-B^\star)^{-1}\left((1-\psi_{2}^\star)\Omega_{1}^\star+(\psi_{2}^\star-\psi_{2})({\Gamma^\star}{\Gamma^\star}^T+\Omega_{1}^\star)\right)(\id-B^\star)^{-T}\\ &=(\id-B)^{-1}\left(\Omega_{2}-{\psi_{2}}\Omega_{1}\right)(\id-B)^{-T}. 
    \end{aligned}
    \label{eqn:known_psi_1p}
\end{equation}
By Assumption 4'', we have that the matrix  $(1-\psi_{2}^\star)\Omega_{1}^\star+(\psi_{2}^\star-\psi_{2})({\Gamma^\star}{\Gamma^\star}^T+\Omega_{1}^\star)$ is non-singular and $\psi_{2} \neq 1$. Rearranging the left-hand side of \eqref{eqn:known_psi_1p}, we have that:
\begin{equation}
\begin{aligned}
    &(\id-B^\star)^{T}\left(\Omega_{1}^\star-{(\psi_{2}^\star-\psi_{2})}/(\psi_{2}-1){\Gamma^\star}{\Gamma^\star}^T\right)^{-1}(\id-B^\star)\\ &=({\psi_{2}-1})(\id-B)^{T}\left(\Omega_{2}-{\psi_{2}}\Omega_{1}\right)^{-1}(\id-B).
    \end{aligned}
\end{equation}
Let $\kappa := (\psi_{2}^\star-\psi_{2})/(\psi_{2}-1)$. By the constraint on how close $\psi_e$ is to $\psi_e^\star$ in \eqref{eq:optimal_params_latent_grouped}, we have that $|\kappa| \leq {2C_\psi}/{\psi_{2}^\star-1}$. Appealing to Lemma~\ref{lemma:intermediate_1}, we have that $\kappa = 0$ or equivalently $\psi_{2}^\star=\psi_{2}$. 

\noindent\underline{Scenario $\psi_{2}^\star < \psi_{1}^\star$}: The relation \eqref{eqn:known_psip} implies that:
\begin{equation}
\begin{aligned}
    &(\id-B^\star)^{-1}\left(({1}/{\psi_{2}^\star}-1)\Omega_{2}^\star+({1}/{\psi_{2}^\star}-{1}/{\psi_{2}})({\Gamma^\star}{\Gamma^\star}^T-\Omega_{2}^\star)\right)(\id-B^\star)^{-T}\\ &=(\id-B)^{-1}\left(\Omega_{1}-{1}/{\psi_{2}}\Omega_{2}\right)(\id-B)^{-T}.
    \end{aligned}
    \label{eqn:known_psi_2p}
\end{equation}
By Assumption 4'', we have that that the matrix  $(1/\psi_{2}^\star-1)\Omega_{2}^\star+(1/\psi_{2}^\star-1/\psi_{2})({\Gamma^\star}{\Gamma^\star}^T-\Omega_{2}^\star)$ is non-singular and $\psi_{2} \neq 1$. Rearranging the left-hand side of \eqref{eqn:known_psi_2p}, we have that:
\begin{equation}
\begin{aligned}
    &(\id-B^\star)^{-1}\left(\Omega_{1}^\star-{(1/\psi_{2}^\star-1/\psi_{2})}/(1/\psi_{2}-1){\Gamma^\star}{\Gamma^\star}^T\right)(\id-B^\star)^{-T}\\ &=\frac{1}{1/\psi_{2}-1}(\id-B)^{-1}\left(\Omega_{e_1}-{1/\psi_{2}}\Omega_{2}\right)(\id-B)^{-T}.
    \end{aligned}
\end{equation}
Let $\kappa := (1/\psi_{2}^\star-1/\psi_{2})/(1/\psi_{2}-1)$. By the constraint on how close $\psi_e$ is to $\psi_e^\star$ in \eqref{eq:optimal_params_latent_grouped}, we have that $|\kappa| \leq {2C_\psi}/({1/\psi_{2}^\star-1})$. Appealing to Lemma~\ref{lemma:intermediate_1}, we have that $\kappa = 0$ or equivalently $\psi_{2}^\star=\psi_{2}$. 

Consider any $i \in \cc{I}^\star$ and let $e$ be an environment satisfying Assumption 4'', i.e. the observed variable $X_i$ receives strong heterogeneous interventions at environment $e$. We will show that $\psi_e^\star = \psi_e$. Again, we consider two settings: $\psi_e^\star > \psi_1^\star$ and $\psi_e^\star < \psi_1^\star$ (notice that $\psi_1^\star \neq \psi_e^\star$ by Assumption 4''):

\noindent\underline{$\psi_e^\star > \psi_1^\star$:} The relation \eqref{eqn:known_psip} implies that:
\begin{equation}
\begin{aligned}
    &(\id-B^\star)^{-1}\left(\Omega_{e}^\star-\psi_e^\star\Omega_1^\star-(\psi_{e}^\star-\psi_{e})({\Gamma^\star}{\Gamma^\star}^T+\Omega_{1}^\star)\right)(\id-B^\star)^{-T}\\ &=(\id-B)^{-1}\left(\Omega_{e}-{\psi_{e}}\Omega_{1}\right)(\id-B)^{-T}.
    \end{aligned}
    \label{eqn:known_psi_1pp}
\end{equation}
Letting $\sigma_\text{min}(\cdot)$ be the minimum singular value of an input matrix, we have by Assumption 6 that:
\begin{equation}
    \sigma_\text{min}(\Omega_e^\star-\psi_e^\star\Omega_1^\star) \geq \min\{|1-\psi_e^\star|\sigma_\text{min}(\Omega_1^\star),(1+\psi_e^\star)\sigma_\text{min}(\Omega_1^\star)\} \geq |1-\psi_e^\star|\sigma_\text{min}(\Omega_1^\star).
\end{equation}
By Assumption 4'', we have that: $(\Omega_{e}^\star-\psi_e^\star\Omega_1^\star-(\psi_{e}^\star-\psi_{e})({\Gamma^\star}{\Gamma^\star}^T+\Omega_1^\star)$ is non-singular. Therefore, re-arranging \eqref{eqn:known_psi_1pp}, we have that:
\begin{equation}
    (\id-B^\star)^{T}\left(\Omega_{e}^\star-\Omega_1^\star\psi_e-{(\psi_{e}^\star-\psi_{e})}{\Gamma^\star}{\Gamma^\star}^T\right)^{-1}(\id-B^\star) =(\id-B)^{T}\left(\Omega_{e}-{\psi_{e}}\Omega_{1}\right)^{-1}(\id-B).
\end{equation}
Let $\kappa := (\psi_{e}^\star-\psi_{e})$. By the constraint on how close $\psi_e$ is to $\psi_e^\star$ in \eqref{eq:optimal_params_latent_grouped}, we have that $|\kappa| \leq {2C_\psi}/{\psi_{e}^\star-1}$. Appealing to Lemma~\ref{lemma:intermediate_1}, we have that $\kappa = 0$ or equivalently $\psi_{e}^\star=\psi_{e}$.

\noindent\underline{$\psi_e^\star < \psi_1^\star$:} The relation \eqref{eqn:known_psip} implies that:
\begin{equation}
\begin{aligned}
    &(\id-B^\star)^{-1}\left(\Omega_e^\star/{\psi_e^\star}-\Omega_1^\star+({1}/{\psi_{e}^\star}-{1}/{\psi_{e}})({\Gamma^\star}{\Gamma^\star}^T-\Omega_{e}^\star)\right)(\id-B^\star)^{-T}\\ &=(\id-B)^{-1}\left(\Omega_{1}-{1}/{\psi_{e}}\Omega_{e}\right)(\id-B)^{-T}. 
    \end{aligned}
    \label{eqn:known_psi_2ppp}
\end{equation}
We have by Assumption 6 that:
\begin{equation}
    \sigma_\text{min}(1/{\psi_e^\star}\Omega_e^\star-\Omega_1^\star) \geq \min\{|1/\psi_e^\star-1|\sigma_\text{min}(\Omega_1^\star),(1/\psi_e^\star+1)\sigma_\text{min}(\Omega_1^\star)\} \geq |1/\psi_e^\star-1|\sigma_\text{min}(\Omega_1^\star)
\end{equation}
By Assumption 4'', we have that that the matrix $\left(\Omega_e^\star/{\psi_e^\star}-\Omega_1^\star+({1}/{\psi_{e}^\star}-{1}/{\psi_{e}})({\Gamma^\star}{\Gamma^\star}^T-\Omega_{e}^\star)\right)$ is non-singular and $\psi_e \neq 0$. Rearranging the left-hand side of \eqref{eqn:known_psi_2p}, we have that:
\begin{equation}
\begin{aligned}
    &(\id-B^\star)^{-1}\left(\Omega_{e}^\star-\Omega_1^\star\psi_e-\psi_e(1/\psi_{e}^\star-1/\psi_{e}){\Gamma^\star}{\Gamma^\star}^T\right)(\id-B^\star)^{-T}\\ &=\psi_e(\id-B)^{-1}\left(\Omega_{e}/\psi_e-\Omega_{1}\right)(\id-B)^{-T}. 
\end{aligned}
\end{equation}
Let $\kappa := (1/\psi_{e}^\star-1/\psi_{e})/(1/\psi_{e})$. By the constraint on how close $\psi_e$ is to $\psi_e^\star$ in \eqref{eq:optimal_params_latent_grouped}, we have that $|\kappa| \leq {2C_\psi}/({1/\psi_{e}^\star-1})$. Appealing to Lemma~\ref{lemma:intermediate_1}, we have that $\kappa = 0$ or equivalently $\psi_{e}^\star=\psi_{e}$. 
\end{proof}

\subsection{Assumptions~\ref{assump:new_confound} is weaker than an incoherence-type condition on the latent effects}
\label{sec:comparison_to_incoherence} 
We first show that Assumption~\ref{assump:new_confound} is generally satisfied even when the number of latent variables is large, whereas the incoherence-type assumption requires that the number of latent variables is far smaller than the ambient dimension. Specifically, using the Woodbury inversion lemma, consider the following decomposition of $(\id-B^\star)^T(\Omega_e^\star- \Omega_1^\star(\psi_e^\star+\delta)+\kappa\Gamma^\star{\Gamma^\star}^T)^{-1}(\id-B^\star)$ (which appears in Assumption~\ref{assump:new_confound} ) when $\kappa \neq 0$
\begin{equation*}
    (\id-B^\star)^T(\Omega_e^\star- \Omega_1^\star(\psi_e^\star+\delta)+\kappa\Gamma^\star{\Gamma^\star}^T)^{-1}(\id-B^\star) = S_e - L_e,
\end{equation*}
where
\begin{equation*}
\begin{aligned}
    S_e &= (\id-B^\star)^T(\Omega_e^\star- \Omega_1^\star(\psi_e^\star+\delta))^{-1}(\id-B^\star), \\
    L_e &= (\id-B^\star)^T(\Omega_e^\star- \Omega_1^\star(\psi_e^\star+\delta))^{-1}\Gamma^\star(\kappa\id + {\Gamma^\star}^T\\&(\Omega_e^\star- \Omega_1^\star(\psi_e^\star+\delta))^{-1}\Gamma^\star)^{-1}{\Gamma^\star}^T(\Omega_e^\star-\Omega_1^\star(\psi_e^\star+\delta))^{-1}(\id-B^\star).
\end{aligned}
\end{equation*}
Assumption~\ref{assump:new_confound} imposes a lower-bound on the degree of the matrix sum $S_e - L_e$, which we show is not very stringent and holds even when the number of latent variables is large. Specifically, notice that $\degree(S_e) \leq \degree[\text{moral}(\dagg^\star)]$. Furthermore, generally, even when the number of latent variables is large, $\degree(L_e)$ is large relative to the ambient dimension. Due to the basic inequality $\degree(S_e+L_e) \geq \degree(L_e) - \degree(S_e)$, it is then straightforward to see that the condition in Assumption~\ref{assump:new_confound}  is generally satisfied.  

Furthermore, even in settings where the number of latent variables is far smaller than the ambient dimension, Assumption~\ref{assump:new_confound}  can be far weaker than the incoherence condition in Assumption~\ref{assump:new_confound}. For illustration, we consider a simple setting when the number of latent variables is equal to one and show that an incoherence-type condition implies the condition $\degree(S_e+L_e) \geq 3\degree[\text{moral}(\dagg^\star)]$ in Assumption~\ref{assump:new_confound}. Specifically, some linear algebraic manipulations yield that for any rank-1 symmetric matrix $M$, ${\degree[M]} \geq 1/\inc[\text{col-space}(M)]^2$. Employing this inequality in conjunction with the bound $\degree(S_e+L_e) \geq \degree(L_e) - \degree(S_e)$, we find that 
\begin{equation*}
    \begin{aligned}
    \degree(S_e+L_e) &\geq 1/\inc[\text{col-space}(L_e)]^2 - \degree[\text{moral}(\dagg^\star)] \\
    &= \degree[\text{moral}(\dagg^\star)](1/(\inc[\text{col-space}(L_e)]^2\degree[\text{moral}(\dagg^\star)])-1).
    \end{aligned}
\end{equation*}
The above inequality suggests than the incoherence-type condition $\inc[\text{col-space}(L_e)]^2\degree\allowbreak[\text{moral}(\dagg^\star)] < 4$ would imply the condition in Assumption~\ref{assump:new_confound} .

\section{Illustration with unperturbed latent variables}
\label{sec:no_latent_perturb}
We consider the following illustration to show that if no constraints are imposed on the number of latent variables, the equivalence class of optimally scoring DAGs when $|\cc{I}^\star| < p$ could be very different than $\simec(\dagg^\star)$. In this section, we will construct a simple example where the equivalence class of optimally scoring DAGs is the empty graph, even when the population graph has multiple edges. Specifically, consider the following structural equation model (specialization of \eqref{eq:SCM_env}) among DAG of three nodes and a single normally distributed latent variable for all environments $e\in\mathcal{E}$:
\begin{equation}
    \begin{aligned}
       X_1^e &= c_1H + \epsilon_1, \\
       X_2^e &= c_2H + \epsilon_2, \\
       X_3^e &= c_3H + \alpha{X}_1^e + \epsilon_3 + \delta^e.
    \end{aligned}
    \label{eqn:SCM_example}
\end{equation}
Here, $\epsilon_1,\epsilon_2,\epsilon_3 \sim \mathcal{N}(0,1)$ and $\delta^e$ is identically zero for environment $e = 1$ (observational) and $\delta^e \sim \mathcal{N}(0,w^e)$ for $e > 1$. Note that by construction, $\cc{I}^\star = \{3\}$. The SCM \eqref{eqn:SCM_example} is then parameterized by the following quantities:
\begin{equation}
\begin{aligned}
    B^\star &= \begin{pmatrix}1&0& 0 \\ \beta &1 &0\\ \alpha &0& 1\end{pmatrix}~~;~~\Omega_1^\star= \id~~;~~\Omega_e^\star = \Omega_1^\star+\text{diag}\begin{pmatrix}0&0&w_e \end{pmatrix} \text{ for }e>1\\\Gamma^\star &= \begin{pmatrix}c_1&c_2& c_3\end{pmatrix}^T.
\end{aligned}    
\end{equation}
We then have the following theorem statement:
\begin{proposition}[Equivalence class with unperturbed latent variables]Consider the SCM \eqref{eqn:SCM_example}. In population, for any $\gamma>0$, $\dagg_\text{regul.opt}^\gamma$ is the empty graph. 
\end{proposition}
\begin{proof}
We will construct $\Gamma,\Omega_e$ such that together with the connectivity matrix associated to an empty DAG, they entail the same covariance model as the population. Specifically, consider the following set of parameters:
\begin{equation}
\begin{aligned}
    B &= 0~~;~~
    \Omega_1 = \lambda_\text{min}(\Sigma_e^\star)/2\id~;~\Omega_e = \Omega_1+\text{diag}\begin{pmatrix} 0 & 0 &w_e\end{pmatrix} \text{ for }e > 1~;~
    \\\Gamma&= \text{matrix square root of }\Sigma_{e}^\star -\lambda_\text{min}(\Sigma_e^\star)/2\id,
    \end{aligned}
    \label{eqn:est_model}
\end{equation}
where $\Sigma_e^\star$ is the population covariance. Thus, the intervention set encoded by the model \eqref{eqn:est_model} is $\cc{I} = \{3\}$. It is straightforward to see that the parameters $(B,\Gamma,\Omega_e)$ entail the same covariance as the population model, that is:
\begin{equation*}
    (\id-B)^{-1}(\Omega_e+\Gamma\Gamma^T)(\id-B)^{-T} = \Sigma_e^\star ~~\text{ for all }e\in\mathcal{E}.
\end{equation*}
Note that the graph encoded by $B$ is the empty graph. Furthermore, note that any model that yields the same covariance as the population must contain at least a single intervention target since $\Sigma_e^\star-\Sigma_1^\star$ has rank-1. We have concluded the result.
\end{proof}

\section{Consistency guarantees of Algorithm 1 and Algorithm 2}
Throughout, we assume that for a given DAG $\dagg$, number of latent variables $\bar{h}$ and intervention targets $\cc{I}$, Algorithm 0 obtains the global optimum solution. As required in Corollary~\ref{cor:incoherent}, we will assume that the input number of latent variables $\bar{h}$ is greater than the true number of latent variables, i.e. $\bar{h} \geq \text{dim}(H)$. 
\subsection{Consistency guarantees of Algorithm 1}
\label{sec:consistency_algorithm_1}
We will denote the output $\hat{\Theta}(\hat{\dagg}_\text{opt},\bar{h})$ of Algorithm 1 by $(\hat{\dagg}_\text{opt},\hat{B}_\text{opt},\hat{\Gamma}_\text{opt},\{\hat{\Omega}_{\text{opt},e},\hat{\Psi}_{\text{opt},e}\}_{e=1}^m)$. We also take $\lambda \to 0$ with the rate given in Proposition~\ref{prop:asymptotic_property}, and assume that the parameters are in a compact space $\mathcal{F}$ for technical reasons; see Section~\ref{appendix:infinite_characterization}. We will prove the following formal statement, where recall that $d^\star := \degree[\text{moral}(\dagg^\star)]$.
\begin{theorem}Consider a set of candidate DAGs $\dagg_\text{cand}$ and suppose $\exists \dagg \in \dagg_\text{cand}$ with the following two properties:
\begin{enumerate}
\item there are parameters $(B,\Gamma,\{\Psi_e,\Omega_e\}_{e=1}^m)$ with $B \sim \dagg$ and $\Sigma_e^\star = (\id-B)^{-1}(\Omega_e+\Gamma\Psi_e\Gamma^T)^{-1}(\id-B)^{-T}$ for every $e \in [m]$.
\item $\degree[\text{moral}(\dagg)] + \|\dagg\|_{\ell_0} \leq \degree[\text{moral}(\dagg^\star)] + \|\dagg^\star\|_{\ell_0}$
\end{enumerate}
Suppose that Assumptions \ref{assum:dense_incoherence}-\ref{assum:faithful} and \ref{assump:heterogenous}-\ref{assump:truthful} are satisfied. Let $d^\star\geq \gamma > 0$.  Let $\nu^\star$ be a positive integer with $\nu^\star \geq d^\star$. Suppose that $48\nu^\star\inc[\mathrm{col}\text{-}\mathrm{space}((\id-\hat{B}_\text{opt})^T\hat{\Omega}_{\text{opt},e}^{-1}\hat{\Gamma}_\text{opt})]<1$ for all $e\in[m]$. {Then, $\hat{\cc{I}}_\text{opt} = \cc{I}^\star$ and  $\optimec(\hat{\dagg}_\text{opt}) = \simec({\dagg}^\star)$} with probability tending to one.
\label{thm:consistency_algorithm_1}
\end{theorem}
We immediately have the following corollary noting that DAGs in the same Markov equivalence class have the same moral graph and the same number of edges.
\begin{corollary} Suppose that $\dagg_\text{cand} \cap \simec(\dagg^\star) \neq \emptyset$. {Then, $\hat{\cc{I}}_\text{opt} = \cc{I}^\star$ and  $\optimec(\hat{\dagg}_\text{opt}) = \simec({\dagg}^\star)$} with probability tending to one. 
\end{corollary}

\begin{proof}[Proof of Theorem~\ref{thm:consistency_algorithm_1}]
The proof will rely on the following fact about the candidate set of DAGs $\dagg_\text{cand}$: there exists a DAG $\dagg \in \dagg_\text{cand}$ and associated parameters $(B,\Gamma,\{\Omega_e,\Psi_e\}_{e=1}^m)$ with $B \sim \dagg$ that is compatible with the data distribution, i.e. $\Sigma_e^\star = (\id-B)^{-1}(\Omega_e+\Gamma\Psi_e\Gamma^T)^{-1}(\id-B)^{-T}$ for every $e \in [m]$. Indeed, By the assumption of the theorem, we have that $\dagg_\text{cand}$ contains at least one of the DAGs in $\simec(\dagg^\star)$. We will denote this DAG by $\tilde{\dagg}^\star$. By Theorem~\ref{thm:IMEC}, we have that there exists a model associated with the DAG $\tilde{\dagg}^\star$ that is compatible with the data distributions. 

We will analyze different components of Algorithm 1. \\

\noindent\underline{Steps 2-3 of Algorithm 1}: As described in these steps in the main text, we take $\cc{I} = [p]$. Step 2 of Algorithm 1 scores different DAGs in the candidate set, with the score:
\begin{equation*}
\begin{aligned}
    S_{\lambda,\gamma}(\dagg,\bar{h},\cc{I}) := \min_{(B,\Gamma,\{\Omega_e,\Psi_e\}_{e=1}^m)\in\mathcal{F}}&\sum_{e=1}^m \hat{\pi}_e\Big(\log\det(\Omega_e+\Gamma\Psi_e\Gamma^T) \\&+ \mathrm{tr}((\id-B)^T(\Omega_e+\Gamma\Psi_e\Gamma^T)^{-1}(\id-B)\hat{\Sigma}_e)\Big) + \lambda\mathcal{R}_\gamma(\dagg,\cc{I}).\\
    \text{subject-to: }& B\sim\dagg, \mathbb{I}(\{\Omega_e\}_{e=1}^m) \subseteq \cc{I}
    \end{aligned}
\end{equation*}
Notice the score $S_{\lambda,\gamma}(\dagg,\bar{h},\cc{I})$ is the same as the score $\texttt{score}_{\lambda,\gamma}(\dagg,\hat{\Theta}(\dagg,\bar{h}))$ in \eqref{eqn:score}. We define the population analogue of the score $S_{\lambda,\gamma}(\dagg,\bar{h},\cc{I})$ below:
\begin{equation*}
\begin{aligned}
   S^\star(\dagg,\bar{h},\cc{I}) := \min_{(B,\Gamma,\{\Omega_e,\Psi_e\}_{e=1}^m)\in\mathcal{F}}&\sum_{e=1}^m {\pi}^\star_e\Big(\log\det(\Omega_e+\Gamma\Psi_e\Gamma^T) \\&+ \mathrm{tr}((\id-B)^T(\Omega_e+\Gamma\Psi_e\Gamma^T)^{-1}(\id-B){\Sigma}^\star_e)\Big).\\
    \text{subject-to: }& B \sim \dagg,\mathbb{I}(\{\Omega_e\}_{e=1}^m) \subseteq \cc{I}
    \end{aligned}
\end{equation*}
Under the rate of regularization $\lambda$ described in Proposition~\ref{prop:asymptotic_property}, following very similar proof strategy to that of Proposition~\ref{prop:asymptotic_property}, we conclude that $S_{\lambda,\gamma}(\dagg,\bar{h},\cc{I}) \to S^\star(\dagg,\bar{h},\cc{I})$ in the infinite data limit. As described earlier, we have that there exists a model associated with the DAG $\tilde{\dagg}^\star$ that is compatible with the data distributions. By Lemma~\ref{lemma:intermed_model}, we have that that the minimizers of $\argmin_{\dagg} S^\star(\dagg,\bar{h},\cc{I})$ are precisely models that are compatible with the data distributions; that is:
$$(B,\Gamma,\{\Omega_e,\Psi_e\}_{e=1}^m) \text{ optimal } \Leftrightarrow \Sigma_e^\star = (\id-B)^{-1}(\Omega_e+\Gamma\Psi_e\Gamma^T)^{-1}(\id-B)^{-T} ~\text{for every }e\in[m],$$
where optimal here means that they result in the smallest score according to $S^\star(\cdot,\cdot,\cdot)$. Since the DAG $\tilde{\dagg}^\star$ attains optimal score, any candidate DAG that is selected in this step must also attain this optimal score.  

With the choice of $\lambda$ in Proposition~\ref{prop:asymptotic_property}, we have that it is above fluctuations due to sampling error; it follows that for two population score equivalent models $(\dagg_1,\bar{h},\cc{I})$ and $(\dagg_2,\bar{h},\cc{I})$ with $S^\star(\dagg_1,\bar{h},\cc{I}) = S^\star(\dagg_2,\bar{h},\cc{I})$, if $\mathcal{R}_\gamma(\dagg_1,\cc{I}) < \mathcal{R}_\gamma(\dagg_2,\cc{I})$, then, there exists $N$ such that for $n_e \geq N$, $S_{\lambda,\gamma}(\dagg_1,\bar{h},\cc{I}) < S_{\lambda,\gamma}(\dagg_2,\bar{h},\cc{I})$. Thus, in the infinite data limit, with probability tending to one, the output of steps 3 of Algorithm 1 is with probability tending to one the minimizer of the following optimization problem with $\cc{I} = [p]$
\begin{equation}
\begin{aligned}
    \argmin_{\dagg \in \dagg_\text{cand},B,\Gamma,\{\Omega_e,\Psi_e\}_{e=1}^m}&~~\mathcal{R}_\gamma(\dagg,\cc{I})\\\text{subject-to}:&~~ B \sim \dagg, \mathbb{I}(\{\Omega_e\}_{e=1}^m) \subseteq \cc{I}\text{ and}\\
    &~~\Sigma_e^\star = (\id-B)^{-1}(\Omega_e+\Gamma\Psi_e\Gamma^T)^{-1}(\id-B)^{-T} ~\text{for every }e\in[m].
    \end{aligned}
    \label{eq:optimal_params_step3}
\end{equation}
Let $\hat{\dagg}_\mathrm{opt}$ be the output of Step 3 of Algorithm 2. From the analysis above, we have that in the infinite data limit and with probability tending to one, $\hat{\dagg}_\mathrm{opt}$ and its associated parameters of minimizers of \eqref{eq:optimal_params_step3}. Furthermore, since $\tilde{\dagg}^\star$ and its associated parameters are feasible in \eqref{eq:optimal_params_step3}, we can make two conclusions with probability tending to one:
\begin{equation}
\begin{aligned}
1.~& p~\degree[\mathrm{moral}(\hat{\dagg}_\mathrm{opt})] + \|\hat{\dagg}_\mathrm{opt}\|_{\ell_0} \leq pd^\star + \|{\dagg}^\star\|_{\ell_0},\\
2.~& S^\star(\hat{\dagg}_\text{opt},\bar{h},[p]) = \min_{\dagg,\cc{I}}S^\star(\dagg,\bar{h},\cc{I}).
\end{aligned}
\label{prop:optimal}
\end{equation}
Here, the first fact is based on every member of the Markov equivalence class having the same moral graph and the same number of edges. The second fact is based on noting that the population score $S^\star(\cdot,\cdot,\cdot)$ is minimized by models that are compatible with the data distribution, and for any DAG $\dagg$, $S^\star(\dagg,\bar{h},\cc{I}_1) \leq S^\star(\dagg,\bar{h},\cc{I}_2)$ with $\cc{I}_1 \supseteq \cc{I}_2$.  

\noindent\underline{Step 4 of Algorithm 1}: Recall that Step 4 of Algorithm 1 takes the best scoring model (among the candidate sets). With this best model, it greedily removes intervention targets until the score of the resulting model does not improve any further. From the analysis of Steps 2-3, we have with probability tending to one that the best scoring DAG $\hat{\dagg}_\mathrm{opt}$ is a minimizer of the optimization \eqref{eq:optimal_params_step3}. As we described in property 2 of \eqref{prop:optimal}, this model minimizes the population score $\argmin_{\dagg,\cc{I}} S^\star(\dagg,\bar{h},\cc{I})$. Furthermore, recall that for any $\cc{I}$, $S_{\lambda,\gamma}(\hat{\dagg}_\mathrm{opt},\bar{h},\cc{I}) \to S^\star(\hat{\dagg}_\mathrm{opt},\bar{h},\cc{I})$. Therefore, at every step of the greedy algorithm, an intervention target can only be removed if the resulting intervention targets $\cc{I}$ satisfies: $S^\star(\dagg_\mathrm{opt},\bar{h},\cc{I}) = S^\star(\dagg_\mathrm{opt},\bar{h},[p])$. Since the model obtained after step 3 is already optimally scoring (i.e. minimizing $S^\star(\cdot,\cdot,\cdot)$ with probability tending to one), we have that every iteration of step 4, the associated model $(B,\Gamma,\{\Omega_e,\Psi_e\}_{e=1}^m)$ satisfies the condition $\Sigma_e^\star = (\id-B)^{-1}(\Omega_e+\Gamma\Psi_e\Gamma^T)^{-1}(\id-B)^{-T} ~\text{for every }e\in[m],$
with $B \sim \hat{\dagg}_\mathrm{opt}$.\\

\noindent\underline{Putting things together}: For notational ease, let $(B,\Gamma,\{\Omega_e,\Psi_e\}_{e=1}^m)$ be the output of Algorithm 1. From the analysis above we have that:
\begin{equation*}
    (\id-B)^{-1}({\Omega}_e+{\Gamma}{\Psi}_e{\Gamma}^T)(\id-{B})^{-T} = (\id-{B}^\star)^{-1}({\Omega}^\star_e+{\Gamma}^\star{\Psi}^\star_e{\Gamma^\star}^T)(\id-{B}^\star)^{-T} \text{ for all }e\in[m].
\end{equation*}
Equivalently, taking the inverse of both sides in the previous equation, and using the Woodbury Inversion Lemma, we have for $e=1,2,\dots,m$
\begin{equation*}
\begin{aligned}
    &(\id-{B})^{T}{\Omega}_e^{-1}(\id-{B})
    -(\id-{B})^{T}{\Omega}_e^{-1}{\Gamma}({\Psi}_e^{-1}+{\Gamma}^T{\Omega}_e^{-1}{\Gamma})^{-1}{\Gamma}^T{\Omega}_e^{-1}(\id-{B})\\
    &= (\id-{B}^\star)^{T}{{\Omega}_e^\star}^{-1}(\id-{B}^\star)
    -(\id-{B}^\star)^{T}{\Omega^\star_e}^{-1}{\Gamma^\star}({\Psi_e^\star}^{-1}+{\Gamma^\star}^T{\Omega^\star_e}^{-1}{\Gamma}^\star)^{-1}{\Gamma^\star}^T{{\Omega_e^\star}}^{-1}(\id-{B}^\star).
    \end{aligned}
    \label{eqn:Omega_results1}
\end{equation*}
Define for every $e=1,2,\dots,m$ the following quantities: \begin{equation*}
    \begin{aligned}
        {K}_e &:= (\id-{B})^{T}{\Omega}_e^{-1}(\id-{B}),\\
    K_e^\star &= (\id-{B}^\star)^{T}{{\Omega}_e^\star}^{-1}(\id-{B}^\star),\\
    {L}_e&:= (\id-{B})^{T}{\Omega}_e^{-1}{\Gamma}({\Psi}_e^{-1}+{\Gamma}^T\hat{\Omega}_e^{-1}{\Gamma})^{-1}{\Gamma}^T{\Omega}_e^{-1}(\id-{B}),\\
    L_e^\star &:= (\id-{B}^\star)^{T}{\Omega^\star_e}^{-1}{\Gamma^\star}({\Psi_e^\star}^{-1}+{\Gamma^\star}^T{\Omega^\star_e}^{-1}{\Gamma}^\star)^{-1}{\Gamma^\star}^T{\Omega_e^\star}^{-1}(\id-{B}^\star).
    \end{aligned}
    \label{eqn:K_L_defn1}
\end{equation*}
Note that from Lemma~\ref{lemma:sparse_low_rank}, we have the following implication for any $e \in [m]$
\begin{equation}
    \degree({K}_e-K_e^\star)\inc[\text{col-space}({L}_e-L_e^\star)]^2 < 1 \Rightarrow {K}_e = K_e^\star.
    \label{eqn:desired}
\end{equation}
In this analysis, we show that under the conditions described in Theorem~\ref{thm:consistency_algorithm_1}, $\degree({K}_e-K_e^\star)\inc[\text{col-space}({L}_e-L_e^\star)]^2 < 1$, allowing us to conclude that ${K}_e = K_e^\star$. Notice that $\degree({K}_e-K_e^\star) \leq \degree[\text{moral}({\dagg})]+d^\star$. Further, by Lemma~\ref{lemma:incoherent_col_spaces}, $\inc[\text{col-space}({L}_e-L_e^\star)] \leq 2(\inc[\text{col-space}({L}_e)]+\inc[\text{col-space}({L}_e^\star)])$. Thus, it suffices to show for all $e\in[m]$ that:
\begin{equation}
    4(\degree[\text{moral}({\dagg})]+d^\star) (\inc[\text{col-space}({L}_e)]^2+\inc[\text{col-space}({L}_e^\star)]^2)<1.
    \label{eqn:sufficient_cond_moral1}
\end{equation}
From the property 1 in \eqref{prop:optimal}, we have that
$\degree[\text{moral}({\dagg})] \leq 2d^\star$. Therefore, the following conditions are satisfied for every $e\in[m]$ due to Assumption 1, the conditions of Corollary~\ref{cor:incoherent}, and the bound $d^\star \leq \nu^\star$:
\begin{equation*}
    \begin{aligned}
    &\degree[\text{moral}({\dagg})]\inc[\text{col-space}({L}_e)]^2 \leq 3\nu^\star\inc[\text{col-space}({L}_e)]^2 <1/16,\\
    &\degree[\text{moral}({\dagg})]\inc[\text{col-space}({L}_e^\star)]^2 \leq 3\nu^\star\inc[\text{col-space}({L}_e^\star)]<1/16, \\
    &d^\star\inc[\text{col-space}({L}_e)]^2 \leq \nu^\star\inc[\text{col-space}({L}_e)]^2 <1/16, \\
    &d^\star\inc[\text{col-space}({L}_e^\star)]^2 < 1/16.
    \end{aligned}
\end{equation*}
Combining these relations, we arrive at the inequality in \eqref{eqn:sufficient_cond_moral1}. We have thus concluded
\begin{equation*}
    (\id-B)(\id-B^\star)^{-1}\Omega_e^\star(\id-B^\star)^{1}(\id-B)^{T} = \Omega_e ~~~e=1,2,\dots,m,
\end{equation*}
Following the same steps as proof of Theorem~\ref{thm:sparse_dense}, we conclude that $\mathbb{I}(\{\Omega_e\}_{e=1}^m) = \cc{I}^\star$ and $\hat{\dagg}_\mathrm{opt} \in \simec(\dagg^\star)$. Finally, it remains to check that the output of the intervention targets $\hat{\cc{I}}_\mathrm{opt}$ is equal to $\cc{I}^\star$. Since, $\hat{\cc{I}}_\mathrm{opt} \supseteq \mathbb{I}(\{\Omega_e\}_{e=1}^m)$, we clearly have that $\hat{\cc{I}}_\mathrm{opt}\supseteq \cc{I}^\star$. Suppose that there exists a $j \in \hat{\cc{I}}_\mathrm{opt}\setminus \cc{I}^\star$. Notice that $S^\star(\hat{\dagg}_\mathrm{opt},\bar{h},\hat{\cc{I}}_\mathrm{opt}\setminus \{j\}) = \argmin_{\dagg,\cc{I}} S^\star(\dagg,\bar{h},\cc{I}).$ With the choice of $\lambda$ in Proposition~\ref{prop:asymptotic_property}, we have that it is above fluctuations due to sampling error; this, if follows that for two population score equivalent models $(\hat{\dagg}_\mathrm{opt},\bar{h},\cc{I}_1)$ and $(\hat{\dagg}_\mathrm{opt},\bar{h},\cc{I}_2)$ with $S^\star(\hat{\dagg}_\mathrm{opt},\bar{h},\cc{I}_1) = S^\star(\hat{\dagg}_\mathrm{opt},\bar{h},\cc{I}_2)$, if $\mathcal{R}_\gamma(\hat{\dagg}_\mathrm{opt},\cc{I}_1) < \mathcal{R}_\gamma(\hat{\dagg}_\mathrm{opt},\cc{I}_2)$, then, there exists $N$ such that for $n_e \geq N$, $S_{\lambda,\gamma}(\hat{\dagg}_\mathrm{opt},\bar{h},\cc{I}_1) < S_{\lambda,\gamma}(\hat{\dagg}_\mathrm{opt},\bar{h},\cc{I}_2)$. Letting $\cc{I}_1 = \hat{\cc{I}}_\mathrm{opt}\setminus \{j\}$, $\cc{I}_2 = \hat{\cc{I}}_\mathrm{opt}$, we can conclude that the target $j$ could have been removed to improve the score. Repeating this argument, we can conclude that $\hat{\cc{I}}_\mathrm{opt} = \cc{I}^\star$.
\end{proof}

\subsection{Consistency guarantees of Algorithm 2}
\label{sec:consistency_algorithm_3}
We will denote the output $\hat{\Theta}(\hat{\dagg}_\text{opt},\bar{h})$ of Algorithm 2 by $(\hat{\dagg}_\text{opt},\hat{B}_\text{opt},\hat{\Gamma}_\text{opt},\{\hat{\Omega}_{\text{opt},e},\hat{\Psi}_{\text{opt},e}\}_{e=1}^m)$. We also take $\lambda \to 0$ with the rate given in Proposition~\ref{prop:asymptotic_property}, and assume that the parameters are in a compact space $\mathcal{F}$ for technical reasons; see Section~\ref{appendix:infinite_characterization}. We will prove the following formal statement, where recall that $d^\star := \degree[\text{moral}(\dagg^\star)]$.
\begin{theorem}Consider a set of candidate DAGs $\tilde{\dagg}_\text{cand}$ and suppose $\exists \bar{\dagg} \in \tilde{\dagg}_\text{cand}, \bar{\dagg}^\star \in \simec(\dagg^\star)$  such that $\bar{\dagg} \supseteq \bar{\dagg}^\star$. Suppose that Assumptions \ref{assum:dense_incoherence}-\ref{assum:faithful} and \ref{assump:heterogenous}-\ref{assump:truthful} are satisfied. Let $d^\star\geq \gamma > 0$.  Let $\nu^\star$ be a positive integer with $\nu^\star \geq d^\star$. Suppose that $48\nu^\star\inc[\mathrm{col}\text{-}\mathrm{space}((\id-\hat{B}_\text{opt})^T\hat{\Omega}_{\text{opt},e}^{-1}\hat{\Gamma}_\text{opt})]<1$ for all $e\in[m]$. {Then, $\hat{\cc{I}}_\text{opt} = \cc{I}^\star$ and  $\optimec(\hat{\dagg}_\text{opt}) = \simec({\dagg}^\star)$} with probability tending to one.
\label{thm:consistency_algorithm_2}
\end{theorem}
\begin{proof}
Recall that Algorithm 2 starts with a candidate set of `starting point' DAGs $\tilde{\dagg}_\text{cand}$ and greedily deletes spurious edges in each DAG to obtain a modified set of candidate DAGs $\dagg_\text{cand}$. Our proof proceeds by showing that with probability tending to 1, the candidate set of DAGs $\dagg_\text{cand}$ contains a DAG $\dagg$ with associated parameters $(B,\Gamma,\{\Psi_e,\Omega_e\}_{e=1}^m)$ such that i) $\Sigma_e^\star = (\id-B)^{-1}(\Omega_e+\Gamma\Psi_e\Gamma^T)^{-1}(\id-B)^{-T} ~\text{for every }e\in[m]$, and ii) $\degree[\text{moral}(\dagg)] + \|\dagg\|_{\ell_0} \leq \degree[\text{moral}(\dagg^\star)] + \|\dagg^\star\|_{\ell_0}$. Then, we appeal to Theorem~\ref{thm:consistency_algorithm_1} to obtain the desired result. 

\noindent\underline{Step 2a of Algorithm 2:} Recall the scores $S_{\lambda,\gamma}(\dagg,\bar{h},\cc{I})$ and its population analogue $S^\star(\dagg,\bar{h},\cc{I})$ defined in the proof of Theorem~\ref{thm:consistency_algorithm_1}. Step 2a computes the score $S_{\lambda,\gamma}(\dagg,\bar{h},\cc{I})$. Under the rate of regularization $\lambda$ described in Proposition~\ref{prop:asymptotic_property}, following very similar proof strategy to that of Proposition~\ref{prop:asymptotic_property}, we can conclude that $S_{\lambda,\gamma}(\dagg,\bar{h},\cc{I}) \to S^\star(\dagg,\bar{h},\cc{I})$ in the infinite data limit for every $\dagg,\cc{I},\bar{h}$. Let $\bar{\dagg} \in \tilde{\dagg}_\text{cand}$ be a DAG in the original candidate set that is a supergraph of a DAG in $\simec(\dagg^\star)$. By Theorem~\ref{thm:IMEC}, we have that there exists a model associated with the DAG $\bar{\dagg}$ that is compatible with the data distributions. By Lemma~\ref{lemma:intermed_model}, we have that that the minimizers of $\argmin_{\dagg} S^\star(\dagg,\bar{h},\cc{I})$ are precisely models that are compatible with the data distributions; that is:
$$(B,\Gamma,\{\Omega_e,\Psi_e\}_{e=1}^m) \text{ optimal } \Leftrightarrow \Sigma_e^\star = (\id-B)^{-1}(\Omega_e+\Gamma\Psi_e\Gamma^T)^{-1}(\id-B)^{-T} ~\text{for every }e\in[m],$$
where optimal here means that they result in the smallest score according to $S^\star(\cdot,\cdot,\cdot)$. \\

\noindent\underline{Step 2b of Algorithm 2:} Since the DAG $\bar{\dagg}$ attains optimal score, any removal of the edges that is selected in this step must also attain this optimal score. Let $\hat{\dagg}$ be the DAG after removing spurious edges. Then, the associated model is also compatible with the data distribution.  

Suppose that $\hat{\dagg}$ is a strict supergraph of $\bar{\dagg}^\star$ (see theorem statement for definition). We will show that the score of $\hat{\dagg}$ will be larger than that of $\bar{\dagg}^\star$, so we conclude that the output of Step 2 must be a DAG $\hat{\dagg}$ that is a subgraph of $\bar{\dagg}^\star$ and thus $p~\degree[\text{moral}(\hat{\dagg})]+\|\hat{\dagg}\|_{\ell_0} < p~\degree[\text{moral}(\bar{\dagg})]+\|\bar{\dagg}\|_{\ell_0}$.

To show the above statement, let $(i,j)$ be a pair of edges that are connected in $\hat{\dagg}$ but are not connected in $\bar{\dagg}^\star$. Since the DAG $\hat{\dagg} \setminus (i\to j)$ that is obtained by removing the edge between the pair $(i,j)$ from $\hat{\dagg}$ is still a supergraph of $\bar{\dagg}^\star$, we have that $S^\star(\hat{\dagg}\setminus(i\to j),\bar{h},[p]) = \argmin_{\dagg,\cc{I}} S^\star(\dagg,\bar{h},\cc{I}).$ With the choice of $\lambda$ in Proposition~\ref{prop:asymptotic_property}, we have that it is above fluctuations due to sampling error; it follows that for two population score equivalent models $(\dagg_1\bar{h},[p])$ and $(\dagg_2,\bar{h},[p])$ with $S^\star(\dagg_1,\bar{h},[p]) = S^\star(\dagg_2,\bar{h},[p])$, if $\mathcal{R}_\gamma(\dagg_1,[p]) < \mathcal{R}_\gamma(\dagg_2,[p])$, then, there exists $N$ such that for $n_e \geq N$, $S_{\lambda,\gamma}(\dagg_1,\bar{h},[p]) < S_{\lambda,\gamma}(\dagg_2,\bar{h},[p])$. Furthermore for DAGs $\dagg_1,\dagg_2$ with $\dagg_1$ strict subgraph of $\dagg_2$, we have that $p~\degree[\text{moral}(\dagg_1)]+\|\dagg_1\|_{\ell_0} < p~\degree[\text{moral}(\dagg_2)]+\|\dagg_2\|_{\ell_0}$. Letting $\dagg_2 = \hat{\dagg}$ and $\dagg_1 =  \hat{\dagg}\setminus (i\to j)$,  we can conclude that the edge between the pair $(i,j)$ could have been removed to improve the score. Repeating this argument, we can conclude the desired result.

\end{proof}

\section{Selecting a cross-validated causal model}
\label{sec:supp_mat_cross_validation}
 The regularization parameters $(\lambda,\gamma)$ and the number of latent variables $\bar{h}$ in Algorithm~\ref{algo:equiv_class} and Algorithm~\ref{algo:structure_learning} are generally unknown. Here, we will propose an efficient cross-validation approach to select a causal model (and the associated interventional equivalence class). Our approach is based on the following reparameterization of the regularization parameters: $ \lambda_{B} := \lambda, \lambda_{I} := \lambda\gamma$, so that we must select $\bar{h},\lambda_B,\lambda_I$. This particular reparameterization and the nature of the procedures in  Algorithm~\ref{algo:equiv_class} and Algorithm~\ref{algo:structure_learning} lead to the following important simplifications: for a given $\bar{h}$, we can first do a 1-dimensional grid search for $\lambda_B$ to choose a DAG and another 1-dimensional grid search for $\lambda_I$ to choose an intervention set. Furthermore, building on the previous simplification, the regularization parameter $\lambda_B$ selects a DAG $\dagg$ from a finite set of available DAGs $\dagg_\text{set}$; these are all the candidate DAGs and all the pruned DAGs (according to the greedy backward deletion). Similarly, the regularization parameter $\lambda_I$ selects intervention targets from a finite collection of available sets $\cc{I}_\text{set}$; these are the collection of intervention targets according to the greedy backward deletion. Thus, choosing an optimal $\lambda_B,\lambda_I$ based on validation is equivalent to choosing an optimal DAG and optimal intervention targets from the sets $\dagg_\text{set}$ and $\cc{I}_\text{set}$. 
 
 With the observations above, we can propose an efficient cross-validation approach to select a causal model for both Algorithms~\ref{algo:equiv_class} and~\ref{algo:structure_learning}. For simplicity, we assume that the data is split into a training set and two test sets. The first test data will be used to determine the number of latent variables and the DAG (among the candidate set), and the second test data will be used to select the intervention set $\cc{I}$. The training data is parameterized by the covariance matrices $\hat{\Sigma}_e^{\text{train}}$ and mixture values $\hat{\pi}_e^{\text{train}}$ for every $e=1,2,\dots,m$; similarly, the test data is parameterized by $\hat{\Sigma}_e^{\text{test1}},\hat{\Sigma}_e^{\text{test2}}$ and mixture values $\hat{\pi}_e^{\text{test1}}$,$\hat{\pi}_e^{\text{test2}}$. We measure the likelihood of a given causal model $\hat{\Theta} = (\hat{B},\hat{\Gamma},\{\Omega_e,\Psi_e\}_{e=1}^m)$ on the first split of the test is: 
 \begin{equation*}
     \texttt{score-test}(\hat{\Theta}) := \sum_{e=1}^m \hat{\pi}_e^{\text{test1}}\ell(\hat{B},\hat{\Gamma},\hat{\Omega}_e,\hat{\Psi}_e;\hat{\Sigma}_e^{\text{test1}}).
 \end{equation*}
A similar measure can be defined to quantify the likelihood on the second split of the test data. 

We are now ready to state a cross-validated version of Algorithm~\ref{algo:equiv_class}, which is presented below. 
\FloatBarrier
\begin{algorithm}
\caption{Equivalence class of best scoring DAGs from a candidate set via \emph{cross-validated} \methName{}} 
\begin{algorithmic}[1]
\vspace{0.1in}
\STATE {\bf Input}: candidate DAG(s) $\dagg_\text{cand}$; training dataset and and two test datasets; max $\#$ of latent vars. $h_\text{max}$; initial intervention set $\cc{I} = [p]$ 
\vspace{.05in}
\STATE{\bf Causal parameters for each DAG and $\#$ latent vars.}: for each $\dagg \in \dagg_\text{cand}$ and $\bar{h} \leq \bar{h}_\text{max}$, supply training data as well as $\dagg$, $\bar{h}$, and $\cc{I}$ to Algorithm~\ref{algo:dag_score} to obtain the causal parameters $\hat{\Theta}_{\cc{I}}(\dagg,\bar{h})$ 
\vspace{0.05in}
\STATE{\bf Find an optimal DAG and $\#$ latent variables using test data}: use the first test dataset to obtain best DAG and number of latent variables $(\hat{\dagg}_\text{opt},\bar{h}_\text{opt}) = \arg\min_{\substack{\dagg \in \dagg_\text{cand}\bar{h} \leq \bar{h}_\text{max}}} \texttt{score-test}(\hat{\Theta}_{\cc{I}}(\dagg,\bar{h}))$ 
\vspace{0.05in}
\STATE{\bf Updating intervention targets $\cc{I}$ using test data}: initialize $\cc{I}_\text{set} = \{\cc{I}\}$ and 
\begin{enumerate}
\item[(a)] let $\{\hat{\Omega}_e\}_{e=1}^m$ be noise variance encoded in $\hat{\Theta}(\hat{\dagg}_\text{opt},\bar{h}_\text{opt})$
\item[(b)] estimate intervention strengths via $\xi_j := \frac{1}{m}\sum_{e=1}^m([\hat{\Omega}_e]_{j,j}-\frac{1}{m}\sum_{e=1}^m [\hat{\Omega}_e]_{j,j})^2$ for each $j \in [p]$
\item[(c)] remove the variable with weakest estimated intervention: $\cc{I} \leftarrow \cc{I} \setminus \{\arg\min_{j \in \hat{\cc{I}}}\xi_j\}$; add $\{\cc{I}\}$ to $\cc{I}_\text{set}$ and use Algorithm~\ref{algo:dag_score} to obtain the causal parameters $\hat{\Theta}_{\cc{I}}(\hat{\dagg}_\text{opt},\bar{h}_\text{opt})$
\item[(d)] obtain optimal intervention set: $\cc{I}_\text{opt} = \arg\min_{\cc{I} \in \cc{I}_\text{set}} \texttt{score-test}(\hat{\Theta}_\cc{I}(\hat{\dagg}_\text{opt},\bar{h}_\text{opt}))$ 
\end{enumerate}
\vspace{0.05in}
\vspace{0.05in}
\STATE{\bf Output:} equivalence class $\optimec(\hat{\dagg}_\mathrm{opt})$ 
\end{algorithmic} \label{algo:equiv_class_validation}
\end{algorithm}
\FloatBarrier
Here, we start with a candidate set of DAGs and a starting intervention set that is taken to be the full set of targets $[p]$. Step 2 of the algorithm scores all DAGs in the candidate set with the number of latent variables varied between $0,1,\dots,h_\text{max}$. Step 3 of the algorithm then computes the likelihood of each of these causal models on test data and chooses the best DAG and number of latent variables. For this selected DAG and the number of latent variables, Step 4 of the algorithm uses the second test set to choose the intervention targets. 

\FloatBarrier
\begin{algorithm}
\caption{{{}Improving "starting point" DAGs via cross-validated \methName{}}}
\begin{algorithmic}[1]
\vspace{0.1in}
\STATE {\bf Input}: candidate DAG(s) $\tilde{\dagg}_\text{cand}$; training dataset; max $\#$ of latent vars. $h_\text{max}$; initial intervention set $\cc{I} = [p]$ 
\vspace{.05in}
\STATE{\bf Backward deletion to remove spurious edges}: initialize $\dagg_\text{cand} = \tilde{\dagg}_\text{cand}$; for each ${\dagg} \in \tilde{\dagg}_\text{cand}$ and $\bar{h} \leq \bar{h}_{\text{max}}$:
\begin{enumerate}
    \item[(a)] supply data and $\hat{\cc{I}} = [p]$ to Algorithm~\ref{algo:dag_score} to obtain $\hat{\Theta}_{\cc{I}}(\dagg,\bar{h})$
    \item[(b)] let ${\dagg}$ be the DAG after deleting smallest edge in magnitude in ${\dagg}$; add $\dagg$ to $\dagg_{\text{cand}}$
    \item[(c)] repeat (a-b) until DAG $\dagg$ is empty
\end{enumerate}
\vspace{0.05in}
\STATE{\bf Output:} supply $\dagg_\text{cand}$ to Algorithm~\ref{algo:equiv_class_validation} to obtain an equivalence class $\optimec(\hat{\dagg}_\mathrm{opt})$
\end{algorithmic} \label{algo:structure_learning_valiation}
\end{algorithm}
\FloatBarrier
Here, as with Algorithm~\ref{algo:equiv_class_validation}, we start with a candidate set of DAGs and a starting intervention set that is taken to be the full set of targets $[p]$. For every DAG in the candidate set and the number of latent variables varied between $0,1,\dots,h_\text{max}$, step 2 scores the DAG and removes the weakest edge greedily until the DAG is empty. Each DAG along the path is included in the candidate set $\dagg_\text{cand}$. We then feed the candidate set $\dagg_\text{cand}$ to Algorithm~\ref{algo:equiv_class_validation} to obtain an output equivalence class.

\section{Additional synthetic experiments}

\subsection{Robustness to the strength of interventions on observed and latent variables}
\label{sec:additional_synthetic_robustness}
\begin{figure}[h!]
    \centering
    \includegraphics[scale = 0.6]{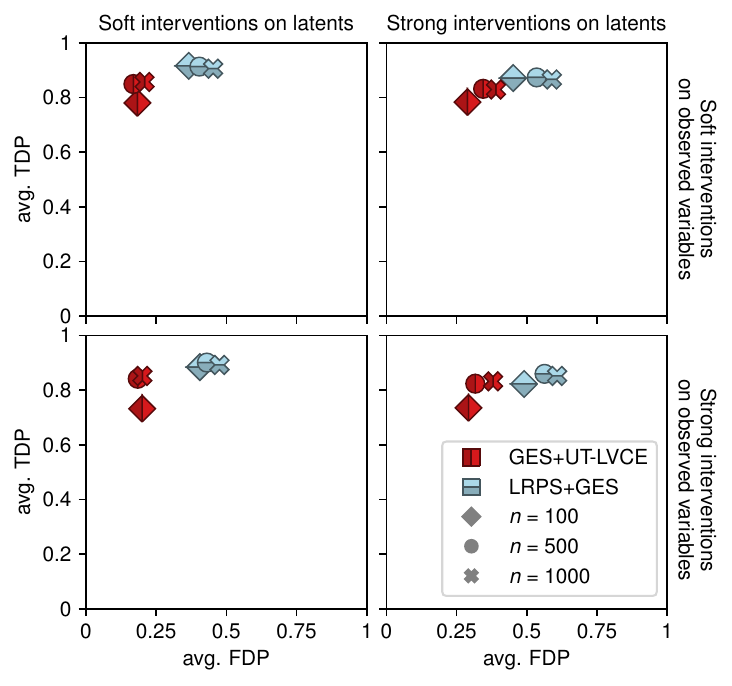}
    \caption{Performance of \methName{} as a structural learning procedure with GES initialization for different strengths of perturbations on the observed and latent variables as outlined in \eqref{eqn:settings_strengths}.}
        \label{fig:robustness}
\end{figure}

We explore the robustness of \methName{} as a structural learning algorithm to different strengths of perturbations on the observed variables and latent variables. In particular, we consider the setting described at the beginning of Section~\ref{sec:experiments_recovering} and different configurations for the strengths of interventions on observed and latent variables:
\begin{equation}
    \begin{aligned}
    &\text{Soft interventions on the observed variables: variance of }\delta_i^e \text{ in the interval }[3,6] \text{ for all }i \in \cc{I}^\star,\\
    &\text{Strong interventions on the observed variables: variance of }\delta_i^e \text{ in the interval }[6,12] \text{ for all }i \in \cc{I}^\star,\\
    &\text{Soft interventions on the latent variables: } \xi_i^e \text{ chosen uniformly and independently from } [0.1,1],\\
    &\text{Strong interventions on the latent variables: } \xi_i^e \text{ chosen uniformly and independently from } [1,5].
    \end{aligned}
    \label{eqn:settings_strengths}
\end{equation}
Figure~\ref{fig:robustness} shows the performance of \methName{} using GES starting points as well as LRpS \citep{marloes} for four settings: i) soft interventions on the observed and latent variables, i) soft interventions on the observed variables and hard interventions on the latent variables, iii) hard interventions on the observed variables and soft interventions on the latent variables, and iv) hard interventions on both the observed and the latent variables. In all settings, $|\cc{I}^\star| = 10$. We notice that the performance of \methName{} is robust across all the settings.

\subsection{Effect of graph density and number of latent variables on performance}
\label{sec:additional_synthetic_robustness_number_denseness}
\begin{figure}[h!]
    \centering
    \includegraphics[scale = 0.7]{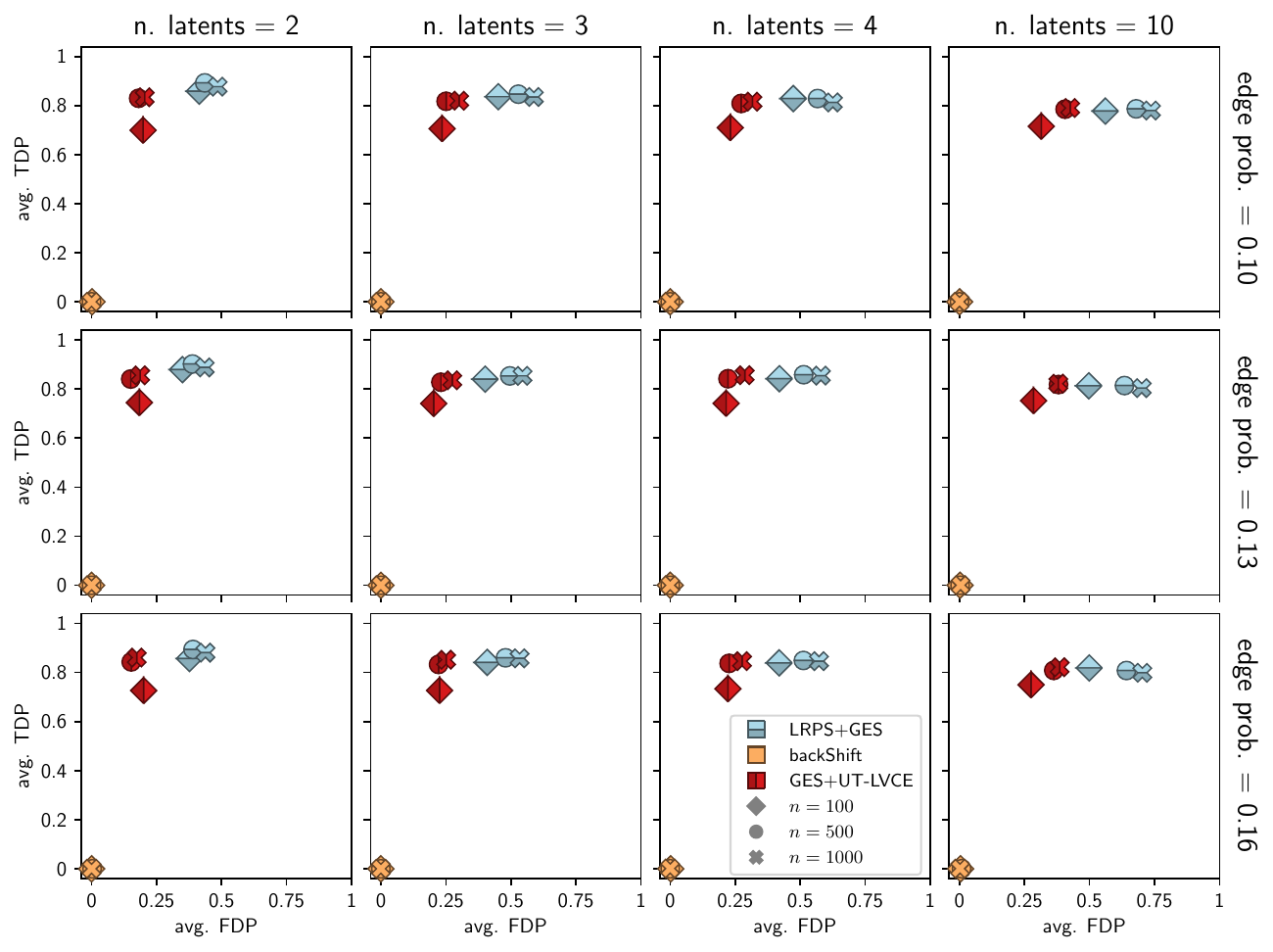}
    \caption{Performance of \methName{} as a structural learning procedure with GES initialization for different numbers of latent variables and edge probabilities in the data-generating graph.}
        \label{fig:latents}
\end{figure}

We explore the effect that different generating-graph densities and the number of latent variables have on the performance of \methName{} and the other methods. In particular, we consider the setting from the leftmost panel in \autoref{fig:comparisons}, and vary the edge probability of the data generating graph $(0.10, 0.13, 0.16)$ and the number of latent variables $(2,3,4,10)$. \autoref{fig:latents} shows the performance of \methName{} using GES starting points as well as LRpS \citep{marloes} and Backshift \citep{backshift}, for each combination of edge probability and number of latents. The performance of both methods slowly deteriorates as the number of latent variables increases.

\subsection{Goodness of GES input DAGs and performance of \methName{}}
\label{sec:additional_synthetic_GES_analysis}
\begin{figure}[h!]
    \centering
    \includegraphics[scale = 0.6]{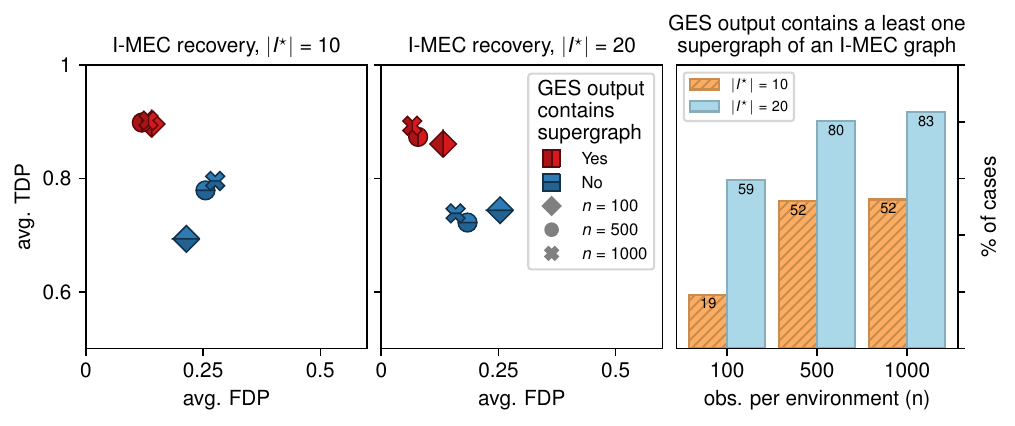}
    \caption{Left and middle: categorizing the performance of \methName{} into a setting where the GES input has a DAG that is a supergraph of a member of  $\simec({\dagg}^\star)$ and the complement setting; right: proportion of instances where the GES input has a DAG that is a supergraph of a member of  $\simec({\dagg}^\star)$.}
        \label{fig:binning}
\end{figure}

We explore how the performance of \methName{} is affected by the quality of the input GES DAGs. To that end, we consider the synthetic setting in Section~\ref{sec:experiments_comparison} for both $|\cc{I}^\star| \in \{10,20\}$. Among the $50$ randomly generated DAGs and $5$ runs, we bin the GES input solution into two categories: a category where there exists a DAG in the GES equivalence class that is a supergraph of a member of $\simec({\dagg}^\star)$, and another category where no DAG in the GES equivalence class is a supergraph of a member of $\simec({\dagg}^\star)$. Figure~\ref{fig:binning} shows the performance of \methName{} for these two categories. We observe that there is a substantial improvement in the performance of \methName{} if the GES initialization contains a DAG that is a supergraph of a member of $\simec({\dagg}^\star)$. Interestingly, we also observe that as the number of interventions increases, GES input DAGs are more likely to meet the aforementioned criteria. 

\subsection{Violations of Causal Dantzig and comparisons to \methName{}}
\label{sec:additional_synthetic_causal_dantzig}
\begin{figure}[h!]
    \centering
    \includegraphics[scale = 0.6]{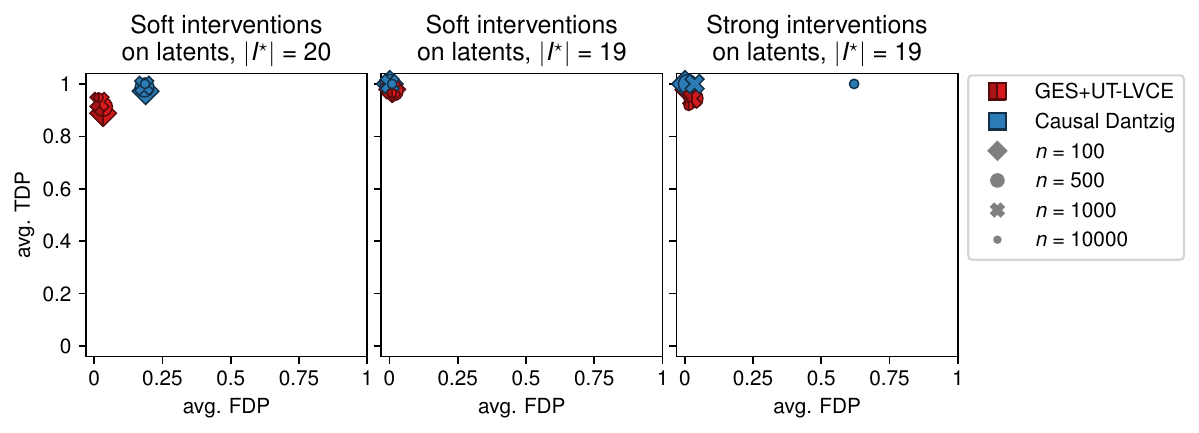}
    \caption{Performance of Causal Dantzig and \methName{} for different intervention strengths on the latent variables and for $|\cc{I}^\star| \in \{19,20\}$.}
        \label{fig:dantzig}
\end{figure}

We next explore the performance of Causal Dantzig \citep{Dantzig} under latent perturbations. Specifically, we consider soft interventions on the observed variables and both soft and hard interventions on the latent variables (see \eqref{eqn:settings_strengths}). Further, we consider the size of interventions to be $|\cc{I}^\star| \in \{19,20\}$. Figure~\ref{fig:dantzig} shows the performance of Causal Dantzig and \methName{} with GES initialization. Here, the performance is with respect to accurate recovery of the parental sets, and the average $\mathrm{FDP}$ and $\mathrm{TDP}$ are taken over $50$ DAGs and $5$ runs per DAG. We observe that when there is an intervention on the target variable or when there are strong interventions on the latent variables, Causal Dantzig performs poorly as compared to \methName{}. This is consistent with the fact that the assumptions for consistency with Causal Dantzig require interventions on all observed variables except the target variable as well as no interventions on the latent variables. 

\newpage
\section{Best scoring DAGs for the protein expressions dataset}
\label{sec:sachs_best_scoring}
We present the top three best scoring DAGs: `Dantzig 11', `Dantzig 9', and 'Dantzig 3' below:
\begin{figure}[h!]
    \centering
    \begin{subfigure}[b]{0.4\textwidth}
   \includegraphics[scale = 0.6]{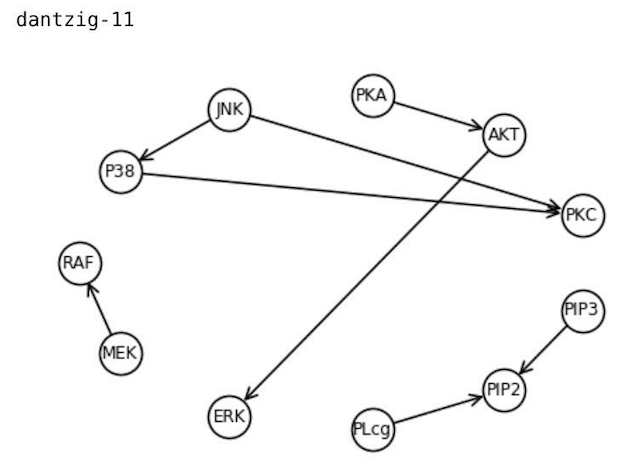}
   \end{subfigure}
   \begin{subfigure}[b]{0.4\textwidth}
     \includegraphics[scale = 0.6]{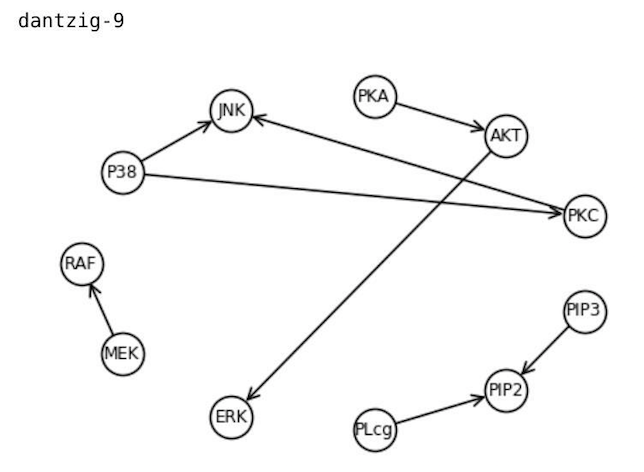}
   \end{subfigure}
   \begin{subfigure}[b]{0.4\textwidth}
      \includegraphics[scale = 0.6]{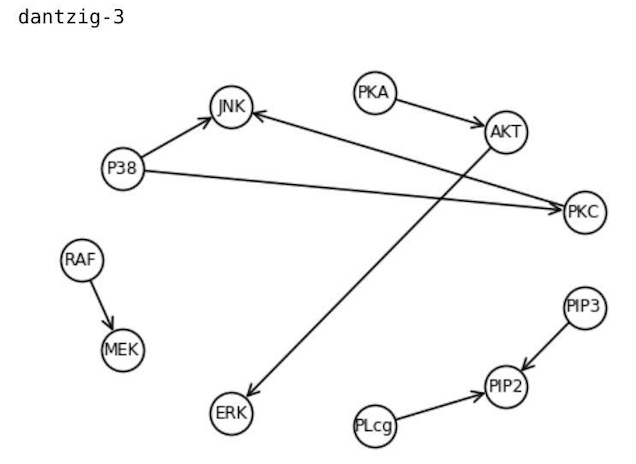}
   \end{subfigure}
    \caption{Top scoring causal models in the literature obtained by \methName{}.}
        \label{fig:dantzig_sachs}
\end{figure}